\documentclass[useAMS,usenatbib]{mnras}
\usepackage{graphicx,natbib,amssymb,amsmath,stmaryrd,makecell}
\usepackage{txfonts}
\usepackage{xcolor}
\usepackage{hyperref}
\usepackage{xspace,dsfont}
\usepackage{caption, lipsum, comment}

\newcommand{\astroway}[2]{\left[\frac{#1}{#2}\right]}

\newcommand{\Mpc}{{\rm Mpc}}

\newcommand{\kpc}{{\rm kpc}}

\newcommand{\expf}[1]{{{\rm e}^{#1}}}

\newcommand{\Planck}{{\it Planck}\xspace}

\newcommand{\vgh}{{\hat{\boldsymbol\gamma}}}

\newcommand{\id}{{\,\rm d}}

\newcommand{\beq}{\begin{equation}}   %

\newcommand{\eeq}{\end{equation}}   %

\newcommand{\beqa}{\begin{eqnarray}}   %

\newcommand{\eeqa}{\end{eqnarray}}   %

\newcommand{\beal}{\begin{align}}

\newcommand{\enal}{\end{align}}

\newcommand{\bspl}{\begin{split}}

\newcommand{\espl}{\end{split}}

\newcommand{\bsub}{\begin{subequations}}

\newcommand{\esub}{\end{subequations}}

\newcommand{\bmulti}{\begin{multline}}   %

\newcommand{\beqm}{\begin{mathletters}}   %

\newcommand{\eeqm}{\end{mathletters}}   %

\newcommand{\Ne}{N_{\rm e}}

\newcommand{\vek} [1]{\mbox{\boldmath${#1}$\unboldmath}}

\newcommand{\pot}[2]{#1 \times 10^{#2}}

\newcommand{\vXb}{\bar{\vek{X}}}

\newcommand{\changeJ}[1]{#1}

\newcommand{\Fb}{F_{\rm b}}
\newcommand{\Xe}{X_{\rm e}}

\newcommand{\Neb}{\bar{N}_{\rm e}}

\newcommand{\deltae}{\delta_{{\rm e}}}
\newcommand{\xie}{\xi_{{\rm e}}}
\newcommand{\sigmae}{\sigma_{{\rm e}}}

\newcommand{\Gammab}{\bar{\Gamma}}
\newcommand{\tauc}{\tau_{\rm c}}
\newcommand{\vomega}{{{\boldsymbol\omega}}}
\newcommand{\zetae}{\zeta_{\rm e}}

\newcommand{\LCDM}{$\Lambda$CDM\xspace}

\usepackage{bbm}
\def\iim{\mathbbm{i}}

\author[Chluba, Vasil \& Battye]
{Jens Chluba$^1$\thanks{E-mail:jens.chluba@manchester.ac.uk}, Geoffrey Vasil$^2$ and Richard Battye$^1$
\\
$^1$Jodrell Bank Centre for Astrophysics, School of Physics and Astronomy, The University of Manchester, Manchester M13 9PL, U.K.
\\
$^{2}$School of Mathematics and the Maxwell Institute for Mathematical Sciences,
University of Edinburgh, EH9 3FD, U.K.
}
\date{\vspace{-0mm}Accepted XXX. Received YYY; in original form ZZZ}

\title[It\^o calculus meets the Hubble tension]
{It\^o calculus meets the Hubble tension: Effects of small-scale electron density fluctuations on the CMB anisotropies}

\begin{document}

\maketitle

\begin{abstract}
In this work, we develop a novel formalism to include the effect of electron density fluctuations at ultra small scales (well below the sound horizon at last scattering) on the observed anisotropies of the Cosmic Microwave Background (CMB). We treat the electron field as an independent stochastic variable and obtain the required ensemble-averaged photon Boltzmann equations using It\^o calculus. 
Beyond changes to the average recombination history (which can be incorporated in the standard approach) our work identifies two new effects caused by the clumpiness of the medium. The first is a correction to the Thomson visibility function caused by correlations of the electron fluctuations along the line of sight, leading to an additional broadening of the visibility towards higher redshifts which causes extra damping and smearing of the CMB anisotropies. The second effect is a reduction of the effective scattering rate in the (pre-)recombination era that affects the photon transfer functions in a non-trivial manner.
These new effects are subdominant in \LCDM but can be significant in cosmologies with an early onset of structure formation (e.g., due to generation of enhanced small-scale power) as suggested by a number of indicators (e.g., the abundance of high redshift galaxies observed by JWST). 
We discuss the relevance of these new effects to the Hubble tension, finding that corrections which cannot be captured by simple modifications to the average recombination history arise. This highlights how important an understanding of the recombination process is in cosmological inference, and that a coordinated simulation and analysis campaign is required as part of the search for the origin of the various tensions in cosmology.
\end{abstract}


\section{Introduction}
Subtle cracks in the \LCDM model and our understanding of the Universe have emerged \citep[e.g.,][]{Abdalla2022tensions, DiValentino2025Tensions}. One of the most prominent issues is the Hubble tension (HT), a persistent discrepancy between the inferred value of the Hubble parameter, $H_0$, derived from early [e.g., Cosmic Microwave Background (CMB)]
and late [e.g., Supernovae (SN)] cosmological probes \citep[e.g.,][]{Verde2019, Knox2020, Ele2021HT, DiValentino2021}. The HT has stimulated much debate, thus
far without a clear resolution, remaining a major puzzle in modern cosmology. Possible explanations for the
HT range from mundane options (e.g., unaccounted systematic effects) to exotic new physics (e.g.,
early dark energy models) \citep[see][for broad review and model-comparisons]{DiValentino2021,NilsH0Olympics}. 
No matter what the solution to the HT might be, it is imperative to get to the bottom of it before we can be confident in our interpretation of the cosmological data or explore new physics beyond \LCDM. 

One of the {\it most conservative} avenues forward is through {\it modified recombination scenarios} \citep[e.g.,][]{Hart2020Hubble, Jedamzik2020Relieving, Sekiguchi2021PhRvD, Lee2023VFC, Lynch2024II, Mirpoorian2025ModRecI}.
The recombination history determines the number density of free electrons as a function of redshift and thus how photons and baryons decouple.
It is one of the crucial yet (mostly) theoretical ingredients in the interpretation of current and future
CMB data \citep[e.g.,][]{Hu1995, Chluba2006, Lewis2006}. For the analysis of CMB data from \Planck, much effort has gone into developing
detailed recombination codes such as {\tt CosmoRec} \citep{Chluba2010b}  and {\tt HyRec} \citep{Yacine2010c} that ensure an unbiased recovery of the key
cosmological parameters \citep{Fendt2009, Jose2010, Shaw2011, Planck2015params}. Without modern recombination treatments our inference of parameters relating to inflation physics (e.g., the spectral index of curvature perturbations, $n_{\rm S}$) would have been hampered \citep{Planck2015params}, highlighting the crucial role of the cosmological recombination model in cosmology.

However, the \LCDM recombination treatments assume standard radiative transfer and atomic physics in a uniform expanding medium to reach sub-percent precision in the recombination model. These assumptions {\it need not hold} and could therefore invalidate our interpretation of CMB observations. In
fact, one of \changeJ{the proposed} solutions to the HT allows for non-standard atomic physics during
the recombination era by introducing varying electron mass \citep{Hart2020Hubble, Sekiguchi2021PhRvD, Nils2025me}, a scenario that scored highly
in the $H_0$ Olympics \citep{NilsH0Olympics} and also recent model-comparisons \citep[e.g.,][]{Nils2025me, Calabrese2025extended}. Another model includes small-scale baryon density fluctuations violating
the assumption about the uniformity of the medium, causing an average acceleration of the recombination process \citep[e.g.,][]{Jedamzik2020Relieving, Galli:2021mxk, Thiele2021HT}. This
latter scenario is attractive because several observations hint towards an earlier onset of
structure formation and new physics hidden at small scales, linking to questions such as:
How did supermassive black holes form? What are the high redshift sources that JWST finds?
Are primordial black holes a part of the cosmic inventory? What causes the stochastic gravitational
wave background that was recently discovered? Are the ARCADE excess and EDGES measurement a signature of early structure formation? – {\it Could these all have one common primordial origin?}

Assuming that the small-scale Universe differs from \LCDM and indeed facilitates an {\it early onset of structure formation} (possibly starting even in the pre-recombination era) suggests that the free electron density also fluctuates. Modeling the processes that link the primordial physics to the exact realization of the electron density fluctuations is currently beyond our reach, but we can still ask how the presence of these fluctuations would affect the observed large-scale CMB anisotropies. To make progress, we shall assume that a huge separation of scales is present, with the CMB modes of interest having a wavelength that is much longer than that of the electron density perturbations. This is in line with the assumption that we cannot resolve the scales in the CMB on which the electron density fluctuations are present.
In this case, we can consider the electron fluctuations as an independent random variable with statistical properties that one has to specify. The CMB anisotropies then depend on the realization of the electron density fluctuations in our Universe and given that these fluctuations are on {\it very small scales}, we can perform an ensemble average over these fluctuations to obtain the predictions for the CMB anisotropies. 

To carry out the required ensemble average and propagate the effects through the photon Boltzmann hierarchy to the CMB power spectra we employ It\^o calculus. In simple words, along each line of sight we have a given realization of the electron density fluctuations. For a given large-scale CMB mode, this implies that the photon transfer functions receive random kicks on short timescales. On average, these kicks vanish, but correlations can build up and leave an overall net effect. The related stochastic differential equation problem can be reduced to a system of coupled ordinary differential equations using It\^o calculus.

{\it Which effects are expected?} -- To determine the fluctuations we first have to define the ensemble-averaged recombination history. This average recombination history itself can already depart from that in standard \LCDM. For instance, clumpiness of the medium is indeed expected to lead to an average acceleration of the recombination process \citep[e.g.,][]{Jedamzik:2013gua,
Jedamzik2020Relieving}. However, the CMB anisotropies {\it do not} directly depend on the free electron fraction but rather integrated quantities such as the Thomson visibility function and photon damping scale. As such, additional corrections arise due to time-time correlations that are not captured by modifying the recombination history alone. As we show here, this causes an additional broadening of the visibility towards higher redshifts and hence additional damping and smearing. In addition, we find that the corrections to the scattering rates affect the various transfer functions differently, leading to modifications that have to be computed with a Boltzmann code.

Using It\^o calculus, we find evolution equations for an infinite hierarchy of coupled moments of the cosmological variables with the electron density field. We implement the new Boltzmann system into the anisotropy module of {\tt CosmoTherm} \citep{Chluba2011therm, kite_spectro-spatial_2023-III} to illustrate the effects and validate various approximations. We show, however, that the problem can be simplified if the corrections remain small. The problem then depends on a model for the statistical properties of the electron density fluctuations. We discuss various cases for illustration and show that the new effects should be relevant to the HT. However, at this stage new theoretical studies are required to better understand the possible link to physical scenarios that generate the electron density fluctuations. This work is thus the starting point for \changeJ{a broader} investigation that requires a combination of numerical simulations in the pre-recombination era with non-standard small-scale physics.

The paper is structured as follows: In Sect.~\ref{sec:prelim} we sketch the problem and how one can think of the electron density fluctuations at small scales as an independent stochastic field. In Sect.~\ref{sec:electron_fluct} we recap the problem in terms of the perturbation variables and explain the main origin of the new effects using a perturbative treatment akin to the derivation of the Langevin equation for Brownian motion. This already reveals the main correction, which we then generalize using It\^o calculus (see Sect.~\ref{sec:ito_formalism}). It also yields a corrected Boltzmann hierarchy that is applicable when the corrections remain sufficiently small. In Sect.~\ref{sec:LN_delta_b_fluctuations}, we discuss some of the simple ways to propagate baryon density fluctuations to electron density fluctuations, which we then use in Sect.~\ref{sec:effects_T_Cell} for our illustrations. We close with a discussion of the limitations and some future directions in our summary section.

\vspace{-3mm}
\section{Preliminary ingredients}
\label{sec:prelim}
The goal of this paper is to compute the solutions for the CMB power spectra in the presence of fluctuations in the free electron fraction. We start by setting up some of the key assumptions about the electron fluctuations. For our purpose, we can assume that the free electron density, $\Ne$, is a random variable, with certain realizations along each line of sight that we now define more carefully.

\subsection{Thinking of $\Ne$ as a random variable}
We assume that the electron density fluctuates at very small scales, with wavenumbers {\it much larger} than for the scales we observe in the CMB. To fully propagate all the effects to the CMB power spectra one would have to consider mode-coupling and real-space radiative transfer effects as well as details of the perturbed recombination processes, which automatically lead to correlations with the standard perturbation variables such as the baryon density. 
To simplify matters and assess which leading order effects might appear, we will instead think of the problem as an isotropic random field with correlations along the line of sight.
This simplification is justified as long as a large separation of scales is present.
We can furthermore anticipate that the mapping from primordial physics to the relevant electron density fields may be caused by a highly non-linear process implying that a statistical approach is generally more promising.

Within each conformal-time slice, the field is then characterized by a probability distribution $P(\eta, \Ne)$ but with negligible variations at the large scales of interest to us for the CMB anisotropy observations. For now, we shall not specify the precise process that generates the field. However, we demand it to have the following properties
\begin{align}
\label{eq:Ne_av_def_Pe}
&\int P(\eta, \Ne) \id \Ne =1,
\qquad 
\int \Ne P(\eta, \Ne) \id \Ne=\Neb(\eta).
\end{align}
The average recombination history, $\Neb(\eta)$, can in principle depart {\it non-perturbatively} from the standard recombination history of the homogeneous Universe. In Sect.~\ref{sec:LN_delta_b_fluctuations}, we will consider electron density fluctuations generated from log-normal baryon density fluctuations, which clearly operates in the non-perturbative regime. 

We also define the moments of $\Ne$ and the electron density fluctuations, $\deltae=\Delta \Ne/\Neb=\Ne/\Neb-1$, as
\begin{subequations}
\begin{align}
\Neb^{(m)}(\eta)
&=\int \Ne^m P(\eta, \Ne) \id \Ne,
\\
\delta_{\rm e}^{(m)}(\eta)
&=\int \left(\frac{\Ne}{\Neb}-1\right)^m P(\eta, \Ne) \id \Ne\equiv \int \deltae^m P(\eta, \deltae) \id \deltae, 
\end{align}
\end{subequations}
with $\Neb^{(0)}=\delta_{\rm e}^{(0)}=1$, $\Neb^{(1)}=\Neb$, and $\delta_{\rm e}^{(1)}=0$ by construction. 
Along the line of sight, we furthermore allow $\left<\delta_{\rm e}(\eta)\,\delta_{\rm e}(\eta')\right>~\neq~0$. 
For a given electron density field, correlations naturally appear (even in the Gaussian limit) given that the size of ionization bubbles directly imply a certain time-correlation in the isotropic limit.
Formally, this can be written as
\begin{align}
\label{eq:cov_def_PDF}
C_{\rm e}(\eta, \eta')=\left<\delta_{\rm e}(\eta)\,\delta_{\rm e}(\eta')\right>
&=\int  \deltae \deltae' P(\eta, \eta', \deltae, \deltae') \id \deltae\!\id \deltae',
\end{align}
where we used the joint probability of $\delta_{\rm e}$ and $\delta_{\rm e}'$, such that
\begin{align}
P(\eta, \deltae)
&=\int  P(\eta, \eta', \deltae, \deltae') \id \deltae'\id \eta'.
\end{align}
We shall omit higher order correlators, although these could also become important in the non-Gaussian limit.

\subsection{Ornstein-Uhlenbeck process for $\deltae$ (Gaussian case)}
To gain a simpler understanding, we can specify the corrrelation kernel $C_{\rm e}(\eta, \eta')$ explicitly by thinking of $\deltae$ as generated by an Ornstein-Uhlenbeck (OU) process \citep{OUprocess1930}:
\begin{align}\label{eq:OU_def}
\id \deltae
&=-\alpha(\eta) \, \deltae \id \eta + \sigma(\eta)\id W.
\end{align}
Here, $\alpha(\eta)$ defines the time correlation length and $\sigma(\eta)\id W$ determines a Wiener process with standard deviation, $\sigma(\eta)$. The OU process naturally allows us to account for time-time correlations, which decay on a timescale $\Delta \eta_{\rm c} = 1/\alpha$. The model parameters can in principle be chosen independently, however, for our application to cosmology certain constraints will become relevent.

Defining $\gamma=\int^\eta_0 \alpha(\eta') \id \eta'$, Eq.~\eqref{eq:OU_def} has the formal solution
\begin{align}
\deltae(\eta) 
&= \deltae(0)\,\expf{-\gamma} +  \int_0^\eta \expf{\gamma'-\gamma}\,\sigma(\eta')\id W(\eta').
\end{align}
One can simply think of the integral over $\id W$ as a discrete sum over an ordered sequence of random realizations of the unit variance field, $W(\eta)$.
The mean of the field vanishes, $\mathbb{E}\left[\deltae\right]=0$. 
Using $\mathbb{E}\left[\id W_1\id W_2\right]=\delta_{\rm D}(\eta_1-\eta_2) \id \eta_1\!\id\eta_2$ with Dirac-$\delta$, $\delta_{\rm D}(x)$, yields
\begin{align}
\mathbb{E}\left[\deltae\,\deltae'\right]
&= 
\mathbb{E}\left[\deltae^{2}(0)\right]  \expf{-(\gamma+\gamma')}  + 
\expf{-(\gamma+\gamma')} \int_{0}^{\min(\eta,\eta')} \sigma_{1}^{2} \, \expf{2\gamma_{1}} \id \eta_{1}.
\end{align}
Assuming that $\alpha$ and $\sigma$ are roughly constant over the width of the correlation kernel, we then find the correlation function, 
\begin{align}
\label{eq:corel_delta2e}
C_{\rm e}(\eta, \eta')
&\approx \mathbb{E}\left[\deltae^{2}(0)\right]  \expf{-(\gamma+\gamma')} + \frac{\sigma^2}{2\alpha}\left[\expf{-|\gamma-\gamma'|}-\expf{-(\gamma+\gamma')}\right].
\end{align}
The statistical invariance, $\mathbb{E}\left[\deltae^{2}(0)\right] \approx \sigma^{2}/(2\alpha)=\sigmae^2(\eta)$, defines the electron density variance at $\eta$. Hence 
\begin{align}
\label{eq:corel_delta2e_alt}
C_{\rm e}(\eta, \eta')
&\approx  \frac{\sigma^2}{2\alpha} \expf{-|\gamma-\gamma'|}.
\end{align}
As the time-separation increases, the correlation quickly drops on a timescale $\Delta \eta_{\rm c}= 1/\alpha$. In the derivations using It\^o calculus the dependence on the initial conditions is directly clarified (see Sect.~\ref{sec:ito_formalism}), while we use the correlation kernels in Eq.~\eqref{eq:corel_delta2e} and Eq.~\eqref{eq:corel_delta2e_alt} for our perturbative treatment (Sect.~\ref{sec:pert_sol}).


\subsection{Ornstein-Uhlenbeck process for $1+\deltae$ (Log-normal case)}
The OU process for $\deltae$ does not ensure the positivity of the total electron density field. This can be remedied by specifying a log-normal OU process for $1+\deltae$. 
Assuming that $\xi \in \mathcal{N}(0,1)$ is a standard Gaussian random variable with zero mean and unit variance, we define 
\begin{align}
    1+\deltae(\xi) = \frac{\expf{\xi \sqrt{\ln(1+\sigmae^{2} )} }}{\sqrt{1+\sigmae^{2}}}.
\end{align}
This implies $\mathbb{E}\left[\deltae\right]  = 0$ and $\mathbb{E}\left[\deltae^{2}\right]  = \sigmae^{2}$, as required.

For a bi-variate standard Gaussian distribution with correlation coefficient $|c| \le 1$, we may write
\begin{align}
    f(\xi_{1},\xi_{2}) = \frac{\expf{-\frac{\xi_{1}^{2} -2 c \, \xi_{1}\xi_{2}+\xi_{2}^{2} }{2(1-c^{2})}}}{2\pi \sqrt{1-c^{2}}},
\end{align}  
with $\mathbb{E}\left[ \xi_{i} \right] =  0$, $  \mathbb{E}[ \xi_{i} \, \xi_{j}] = c + (1-c)\, \delta_{ij}$ and $c(\gamma, \gamma') = \expf{-|\gamma - \gamma'|}$ for a standard stationary Ornstein-Uhlenbeck process.
Therefore, 
\begin{align}
\mathbb{E}\left[\deltae(\xi_{i}) \right] =0,\quad
    \mathbb{E}\left[\deltae(\xi_{1})\,\deltae(\xi_{2})\right]\, = \, (1+\sigmae^{2})^{c(\gamma, \gamma')} - 1. 
\end{align}
For $\sigmae\ll 1$, we have $\mathbb{E}\left[\deltae(\xi_{1})\,\deltae(\xi_{2})\right]\approx \sigmae^{2} \expf{-|\gamma - \gamma'|}$, as in the Gaussian case, Eq.~\eqref{eq:corel_delta2e_alt}. For large values of $\sigmae$, the correlation drops more steeply than for the Gaussian setup. This also reduces the characteristic correlation time as we show next.

\subsection{Correlation time}
We can now define the correlation time, which determines the typical time it takes for the correlation to drop by one e-fold. We shall use the integral definition
\begin{align}
\Delta \eta_{\rm c}=\frac{\int_0^\infty C_{\rm e}(\eta, \eta+\Delta \eta) \id \Delta \eta}{\sigmae^{2}},
\end{align}
which is well-motivated for a top-hat correlation function. 
The correlator in the Gaussian version of the OU process, Eq.~\eqref{eq:corel_delta2e_alt}, yields
\begin{align}
\Delta \eta^{\rm G}_{\rm c}=\int_0^\infty  \expf{-|\alpha \Delta \eta|}\id \Delta \eta=\frac{1}{\alpha},
\end{align}
as anticipated. For the log-normal version, we obtain
\begin{align}
\Delta \eta^{\rm LN}_{\rm c}
&=\frac{1}{\sigmae^2}\int_0^\infty[(1+\sigmae^{2})^{\expf{-|\alpha \Delta \eta|}} - 1]\id \Delta \eta
\nonumber\\
&=\frac{1}{\alpha}\frac{{\rm Li}(1+\sigmae^2)-\ln[\ln(1+\sigmae^2)]-\gamma_{\rm E}}{\sigmae^2},
\end{align}
where ${\rm Li}(z)$ denotes the logarithmic integral and $\gamma_{\rm E}$ the Euler constant. The correlation time therefore obeys $\Delta \eta^{\rm LN}_{\rm c}\lesssim \Delta \eta^{\rm G}_{\rm c}$.

With these preliminary ingredients we can perform a perturbative treatment of the problem akin to how the Langevin equation for Brownian motion is obtained (see Sect.~\ref{sec:pert_sol}). We stress that the Ornstein-Uhlenbeck process only provides a simple model for including time-time correlations. The Gaussian model is a useful starting point. However, it suffers from unphysical negative total density regions no matter how conservatively we set the latent OU parameters $\alpha$ and $\sigma$. In physical systems, we furthermore expect these parameters to depend on each other in subtle ways. These dependencies have to be determined using detailed simulations for the growth of structures in various early eras. 

\section{Modeling the effects of electron density fluctuations on the CMB}
\label{sec:electron_fluct}
In this section, we explain how the effect of small-scale electron density fluctuations along the line of sight can be accounted for in the CMB power spectra using a perturbative treatment. We start with the photon Boltzmann equations to show how the effects propagate at second order in the electron density field. This is then generalized in the next section that employs It\^o calculus. 

\subsection{Modified photon Boltzmann equation}
\label{sec:main_equations}
To approach the problem, we first explicitly write the photon Boltzmann equation for the temperature field:
\begin{align}
&\!\!\frac{\partial \Theta}{\partial \eta}+\vgh\cdot \nabla\,\Theta
=-\frac{\partial \Phi}{\partial \eta}-\vgh\cdot \nabla\,\Psi-\Gamma \left[\Theta-\Theta_0-\frac{\Theta_2}{10}\,-\vgh \cdot {\boldsymbol\varv}_{\rm b}
\right].
\end{align}
Here, the Thomson scattering rate is $\Gamma=a \Ne \sigma_{\rm T}$. The perturbation variables, all with their common denomination as clarified below, are real space functions, i.e., $\Theta=\Theta(\eta, \vek{x}, \vgh)$, $\Phi=\Phi(\eta, \vek{x})$, $\Psi=\Psi(\eta, \vek{x})$, ${\boldsymbol\varv}_{\rm b}={\boldsymbol\varv}_{\rm b}(\eta, \vek{x})$ and $\Ne=\Ne(\eta)=\Neb(\eta)[1+\deltae(\eta)]$, with fluctuations $\deltae=\Ne/\Neb-1$. Transforming to Fourier space gives
\begin{align}
\label{eq:Theta_evol_start_F}
&\!\!\frac{\partial \Theta}{\partial \eta}+\iim k \chi\Theta
=-\frac{\partial \Phi}{\partial \eta} -\iim k \chi\Psi-\bar{\Gamma}(1+\deltae)\!\left[\Theta - \Theta_0\!-\frac{\Theta_2}{10}-\chi\varv_{\rm b}
\right],
\end{align}
where $\bar{\Gamma}=a \Neb \sigma_{\rm T} $ is the average scattering rate, $\chi=\vgh\cdot \hat{\vek{k}}$ and $\boldsymbol\varv_{\rm b}=\hat{\vek{k}}\,\varv_{\rm b}$ for the velocity modes in Fourier space. 
Note that here $\Theta=\Theta(\eta, k, \chi)$ and $\deltae=\deltae(\eta)$, and similar for all other perturbation variables. We have neglected any spatial dependence of $\deltae$, treating it as a pure random field in time. Without this approximation, a convolution integral would appear in the last term, making the problem untractable with the methods we use here. However, due to the strong separation of scales, this approximation seems well-justified.

We can next obtain the photon Boltzmann hierarchy and also the baryon velocity equations in Fourier space, by performing the Legendre transform of the photon Boltzmann equations in the usual way. 
To write the full set of evolution equations in conformal Newtonian gauge, closely following \citet{Ma1995}, for convenience we introduce the following definitions
\bsub
\begin{align}
\label{eq:definition}
\delta \rho_{\rm m}&=\rho_{\rm c}\,\delta_{\rm c}+\rho_{\rm b}\,\delta_{\rm b},
\\
\delta \rho_{{\rm r}, \ell}&=4[\rho_\gamma\,\Theta_\ell+\rho_\nu\,\Theta_{\nu, \ell}],
\end{align}
\esub
where $\rho_{\rm c}$, $\rho_{\rm b}$, $\rho_\gamma$ and $\rho_\nu$ are the background energy densities of cold dark matter, baryons, photons and neutrinos, respectively. We furthermore denote the dark matter and baryon density perturbations in Fourier space as $\delta_{\rm c}$ and $\delta_{\rm b}$. The photon and neutrino temperature transfer functions at multipole $\ell$ are given by $\Theta_\ell$ and $\Theta_{\nu, \ell}$.

For the Newtonian potentials $\Phi$ and $\Psi$ we use the equations
\bsub
\begin{align}
\label{eq:Potentials}
k^2 \Phi + 3 \mathcal{H}(\partial_\eta \Phi-\mathcal{H} \Psi)&=4 \pi G a^2 [\delta \rho_{\rm m}+\delta \rho_{\rm r, 0}],
\\
k^2 (\Phi+\Psi)&=-8 \pi G a^2 \delta \rho_{\rm r, 2},
\end{align}
\esub
where $\mathcal{H}=a H = a^{-1} \partial_\eta a$ is the conformal time Hubble parameter. Note that $\Phi = - \Phi^{\rm Ma}=-\Phi^{\rm Baumann}$ in our convention of the metric.

For the dark matter and baryon density and velocity perturbations, the latter denoted by $\varv_{\rm c}$ and $\varv_{\rm b}$, we then have
\bsub
\begin{align}
\label{eq:dm_equations}
\partial_\eta \delta_{\rm c}&=-3\partial_\eta \Phi-k \varv_{\rm c},
\\
\partial_\eta \varv_{\rm c}&= k\Psi-\mathcal{H} \varv_{\rm c},
\\
\partial_\eta \delta_{\rm b}&=-3\partial_\eta \Phi-k\varv_{\rm b},
\\[-1.5mm]
\partial_\eta \varv_{\rm b}&= k\Psi-\mathcal{H} \varv_{\rm b}
+\frac{3\bar{\Gamma}}{R}(1+\delta_{\rm e})\left[\Theta_1-\frac{\varv_{\rm b}}{3}\right].
\end{align}
\esub
The photon hierarchy is given by
\bsub
\begin{align}
\label{eq:Theta_equations}
\partial_\eta \Theta_0&=-k \Theta_1-\partial_\eta\Phi,
\\
\partial_\eta \Theta_1&=\frac{k}{3}\Theta_0-\frac{2 k}{3}\Theta_2+\frac{k}{3}\Psi-\bar{\Gamma}(1+\delta_{\rm e})\left[\Theta_1-\frac{\varv_{\rm b}}{3}\right],
\\
\partial_\eta \Theta_2&=\frac{2k}{5}\Theta_1-\frac{3 k}{5}\Theta_3-\frac{9}{10}\bar{\Gamma}(1+\delta_{\rm e})\,\Theta_2,
\\
\partial_\eta \Theta_{\ell\geq 3}&=\frac{k \ell}{2\ell+1}\Theta_{\ell -1}-\frac{k (\ell+1)}{2\ell+1}\Theta_{\ell +1}-\bar{\Gamma}(1+\delta_{\rm e})\,\Theta_\ell.
\end{align}
\esub
The neutrino equations are omitted here but take their usual form (without any dependence on the electron density fluctuations). We omitted polarization terms, but will include them later.

\vspace{-3mm}
\subsection{Naive understanding based on visibility function averaging}
\label{sec:naive}
The Thomson visibility function, $g(\eta)$, plays a crucial role in the formation of the CMB anisotropies. It directly appears in the line of sight approach \citep{Seljak1997} and is given by $g(\eta)=\id f/\id \eta=- \exp(-\tau)\,\id\tau/\id \eta$ where $f(\eta)=\exp(-\tau)$ and $\tau=\int_\eta^{\eta_0} \Gamma \id \eta'$. For $\Gamma=\bar{\Gamma}(1+\deltae)$, we can already understand when corrections will appear. The key features can be understood by deriving the ensemble averages of $g$ and $f$ assuming a multi-variate Gaussian probability for the density perturbations. 

We start by defining a realization of electron perturbations in infinitesimal $\Delta \eta$ steps along the line of sight, $\vek{\delta}_{\rm e}=(\delta_{{\rm e}, 0}, \delta_{{\rm e}, 1}, \ldots, \delta_{{\rm e}, N})$. We then assume that the probability distribution of correlations between different $\delta_{{\rm e}, i}$ is given by a multi-variate Gaussian
\begin{align}
\label{eq:Prob}
\mathcal{P}(\vek{\delta}_{\rm e})
&=\frac{1}{\sqrt{(2\pi)^N |\vek{C}|}}\exp\left(-\frac{1}{2}\vek{\delta}^T_{\rm e}\, \vek{C}^{-1}\,\vek{\delta}_{\rm e}\right),
\end{align}
where $C_{ij}$ is the covariance matrix. Let us first consider the averaging of $f$. With $\bar{f}=\expf{-\bar{\tau}}$ and $\bar{\tau}=\int_\eta^{\eta_0} \bar{\Gamma} \id \eta'$, we then have to calculate
\begin{align}
f(\eta)
&=\bar{f}(\eta)\,\exp\left[- \int_\eta^{\eta_0} \id \eta' \, \bar{\Gamma}(\eta') \, \delta_{\rm e}(\eta')\right]
\nonumber 
\\
&\approx \bar{f}(\eta)\, \exp\left[-\Delta \eta \,\bar{\vek{\Gamma}}(\eta) \cdot \vek{\delta}_{\rm e}(\eta) \right]
\end{align}
where $\bar{\vek{\Gamma}}(\eta)$ and $\vek{\delta}_{\rm e}(\eta)$ are the discrete vectors evaluated at the various $\eta_i$ with components set to zero at $\eta_i<\eta$. To obtain the ensemble average, we then simply have to evaluate the integral
\begin{align}
\left<f\right>&=\bar{f}(\eta)\int \exp\left[-\Delta \eta \,\bar{\vek{\Gamma}}(\eta) \cdot \vek{\delta}_{\rm e}(\eta) \right]\,\mathcal{P}(\vek{\delta}_{\rm e})\id \delta_{\rm e}^N
\nonumber \\[-1mm]
&=
\bar{f}(\eta)\int \frac{1}{\sqrt{(2\pi)^N |\vek{C}|}}\,\exp\left[-\Delta \eta \,\bar{\vek{\Gamma}}(\eta) \cdot \vek{\delta}_{\rm e}(\eta) -\frac{1}{2}\vek{\delta}^T_{\rm e}\, \vek{C}^{-1}\,\vek{\delta}_{\rm e} \right]\id \delta_{\rm e}^N
\nonumber 
\\
\label{eq:corr_f}
&=\bar{f}(\eta)\,\exp\left[\frac{\Delta \eta^2}{2} \,\bar{\vek{\Gamma}}^T(\eta) \,\vek{C} \,\bar{\vek{\Gamma}}(\eta)\right].
\end{align}
Here, we have completed the square using
\begin{align}
&\frac{1}{2}\vek{\delta}^T_{\rm e}\, \vek{C}^{-1}\,\vek{\delta}_{\rm e} +\Delta \eta \,\bar{\vek{\Gamma}}(\eta) \cdot \vek{\delta}_{\rm e}(\eta) +  K \equiv \frac{1}{2}(\vek{\delta}_{\rm e}-\vek{\mu})^T\, \vek{C}^{-1}\,(\vek{\delta}_{\rm e}-\vek{\mu})
\nonumber \\
&\qquad =\frac{1}{2}\vek{\delta}_{\rm e}^T\, \vek{C}^{-1}\,\vek{\delta}_{\rm e}-\vek{\mu}^T\, \vek{C}^{-1}\,\vek{\delta}_{\rm e}+\frac{1}{2}\vek{\mu}^T\, \vek{C}^{-1}\,\vek{\mu}
\end{align}
which implies $\vek{\mu}=-\Delta \eta \,\vek{C} \,\bar{\vek{\Gamma}}(\eta)$ and $K=\frac{\Delta \eta^2}{2} \,\bar{\vek{\Gamma}}^T(\eta) \,\vek{C} \,\bar{\vek{\Gamma}}(\eta)$. 

The result in Eq.~\eqref{eq:corr_f} shows one of the most important features of the problem. For a diagonal covariance $C_{ij}=\sigma_i^2 \delta_{ij}$ (where $\sigma_{i}$ is bounded by the size of $\deltae$ and is independent of $\Delta \eta$), we have 
\begin{align}
\label{eq:def_fav}
\left<f\right>=\bar{f}(\eta)\,\exp\left[\sum_{i}\frac{\Delta \eta^2}{2} \sigma_i^2\,\bar{\Gamma}_i^2 \right]\rightarrow \bar{f}(\eta)
\end{align}
for $\Delta \eta \rightarrow 0$. Without correlations between $\eta$ slices, there is no new effect beyond changing the average recombination history. This limit was considered in previous approaches \citep{Jedamzik2020Relieving, Galli:2021mxk}. 
However, physically there is a correlation across some finite coherence length, $\Delta \eta_{\rm c}$, e.g., related to the finite physical size of electron density halos.
Thus, for non-vanishing off-diagonal correlation, we anticipate a correction even as $\Delta \eta \rightarrow 0$, essentially leading to the replacement $\Delta \eta^2 \rightarrow 2\Delta \eta\,\Delta \eta_{\rm c}$ in Eq.~\eqref{eq:def_fav}.

To prove this more formally, let us define 
\begin{align}
C_{ij}&=
\begin{cases}
\sigma_i^2 &\text{for $|i-j|\leq N_{\rm c}\equiv \Delta \eta_{\rm c}/\Delta \eta$}
\\
0 &\text{otherwise}.
\end{cases}
\end{align}
This mimics an off-diagonal covariance matrix that also explicitly depends on the chosen resolution, $\Delta \eta$. For large $\Delta \eta$, the matrix becomes essentially diagonal, while for $\Delta \eta \rightarrow 0$ many neighboring time bins can give the same contribution.

With this definition, we can then evaluate the fluctuation-induced optical depth correction
\begin{align}
   \Delta \tau&= \frac{\Delta \eta^2}{2} \,\bar{\vek{\Gamma}}^T(\eta) \,\vek{C} \,\bar{\vek{\Gamma}}(\eta) 
    = \frac{\Delta \eta^{2}}{2} \sum_{i=1}^{N} \sum_{j=i - N_{\rm c} \ge 0 }^{i + N_{\rm c} \le N}  \Gammab_{i}\, C_{ij}\, \Gammab_{j}
    \\[-1mm]
    \nonumber 
    &\approx \frac{\Delta \eta^{2}}{2}  \sum_{i=1}^{N}  2 N_{\rm c}\Gammab_{i}^2\sigma_i^2
    =\sum_{i=1}^{N}  \Delta \eta \,\Delta \eta_{\rm c} \Gammab_{i}^2 \sigma_i^2 \rightarrow \int \Delta \eta_{\rm c} \Gammab_{i}^2 \sigma_i^2 \id \eta,
\end{align}
where we assumed $\Gammab_{i}\approx \Gammab_{j}$ over the typical correlation length $\Delta \eta_{\rm c}$. The factor of two arises because both past and future correlations contribute to the final sum, leading to a cancellation of one factor of $\Delta \eta$. We also see that when $\Delta \eta_{\rm c} = 0$, we recover $\left<f\right>\rightarrow \bar{f}(\eta)$, but more generally we have
\begin{align}
\label{eq:corr_f_limit}
\left<f\right>&=
\bar{f}(\eta)\,\exp\left[\int_\eta^{\eta_0} \Delta \eta_{\rm c}(\eta')\,\sigma_{\rm e}^2(\eta') \,\bar{\Gamma}^2(\eta')\id \eta' \right],
\end{align}
with an optical depth correction, $\Delta \tau=\int_\eta^{\eta_0} \Delta \eta_{\rm c}(\eta')\,\sigma_{\rm e}^2(\eta') \,\bar{\Gamma}^2(\eta')\id \eta'$.

Alternatively, the final result in Eq.~\eqref{eq:corr_f} can be converted into a double integral as
\begin{align}
\label{eq:corr_f_fin}
\left<f\right>&=\bar{f}(\eta)\,\exp\left[\frac{1}{2}\int_\eta^{\eta_0} \id \eta' \int_\eta^{\eta_0} \id \eta'' \,\bar{\Gamma}(\eta') \,C(\eta',\eta'') \,\bar{\Gamma}(\eta'')\right].
\end{align}
For vanishing correlation between $\eta$ slices we have $C(\eta',\eta'')=0$, and therefore no correction. Assuming a Gaussian probability for the correlation in $\eta$, we have the normalization condition
$$\int C(\eta,\eta') \id \eta' = 2\Delta \eta_{\rm c} \sigma^2_{\rm e}(\eta)\,\delta_{\rm D}(\eta'-\eta),$$ where $\delta_{\rm D}(x)$ denotes the Dirac-$\delta$ function and we considered both past and future correlations.
This then implies Eq.~\eqref{eq:corr_f_limit}. 

\vspace{-3mm}
\subsubsection{Correction to the visibility function}
To compute the averaged visibility function, we have to consider 
\begin{align}
\left<\deltae f\right>
&=
\bar{f}(\eta)\int \frac{\delta_{{\rm e},N}}{\sqrt{(2\pi)^N |\vek{C}|}}\,\exp\left[-\Delta \eta \,\bar{\vek{\Gamma}}(\eta) \cdot \vek{\delta}_{\rm e}(\eta) -\frac{1}{2}\vek{\delta}^T_{\rm e}\, \vek{C}^{-1}\,\vek{\delta}_{\rm e} \right]\id \delta_{\rm e}^N
\nonumber \\ 
&=\bar{f}(\eta)\exp\left[\frac{\Delta \eta^2}{2} \,\bar{\vek{\Gamma}}^T\!(\eta) \,\vek{C} \,\bar{\vek{\Gamma}}(\eta)\right]
\left\{-\Delta \eta \,\vek{C}\,\bar{\vek{\Gamma}}(\eta)\right\}_N,
\end{align}
where we used the mean value after completing the square. 
Taking the limit $\Delta \eta\rightarrow  0$ then implies $\left<\deltae f\right>\rightarrow -\Delta \eta_{\rm c}(\eta)\, \sigma_{\rm e}^2(\eta)\,\bar{\Gamma}(\eta)\left< f\right>$, where $\left\{\Delta \eta \,\vek{C}\,\bar{\vek{\Gamma}}(\eta)\right\}_N\rightarrow \Delta \eta_{\rm c} \sigma_{\rm e}^2(\eta)\,\bar{\Gamma}(\eta)$ without any extra factor of two as only future\footnote{The vector $\vek{\delta}_{\rm e}$ is ordered reversely, with $\delta_{{\rm e}, N}$ being the first entry at $\eta\leq \eta_0$.} correlations from $\eta'\geq \eta$ contribute. This then implies the average visibility
\begin{align}
\label{eq:vis_Gaussian_derivations}
\langle g(\eta)\rangle\approx \bar{\Gamma}(\eta)[1-\bar{\Gamma}(\eta) \, \Delta \eta_{\rm c}(\eta)\, \sigma_{\rm e}^2(\eta)]\left< f(\eta)\right>\equiv
\partial_\eta \langle f(\eta)\rangle
\end{align}
as could have been naively expected. 
Knowing how the visibility function links to the photon Boltzmann equation immediately suggests that one can assume $\Gammab(1+\deltae)\rightarrow \Gammab[1-\Gammab\Delta \eta_{\rm c} \sigma_{\rm e}^2]$. However, we will show below that this is too simplistic and that differences arise for dipole and quadrupole terms even at leading order.

\subsubsection{New model parameters}
\label{sec:model_pars_new}
The derivations above identify two new parameters in the presence of electron density fluctuations. One is the optical depth across the coherence length, 
\begin{align}
\label{eq:tau_c}
\tauc=\Gammab \Delta \eta_{\rm c}, 
\end{align}
the other is the variance of the density fluctuations, $\sigmae^2$, in each time-slice. These are combined to the effective parameter
\begin{align}
\label{eq:zeta_e}
\zeta_{\rm e}=\tauc\sigmae^2,
\end{align}
which should not exceed unity for physical reasons. A Gaussian treatment of the problem furthermore demands $\sigmae^2 \lesssim 0.1$, a limitation that is overcome for a log-normal field, with the log-normal variance set to $\sigma_{\rm e, LN}^2=\ln(1+\sigmae^2)$ allowing any value of  $\sigmae^2=\langle \deltae^2\rangle$.

The physical reason for requiring $\zeta_{\rm e} \le 1$ is that we know scattering damps photon perturbations. Indeed, $\zeta_{\rm e} > 1$ would lead to exponential \textsl{growth} of photon fluctuations by scattering events, which would be in tension with observations. 
We were unable to prove that the regime $\zeta_{\rm e} > 1$ can be excluded for physical systems in cosmology. We expect this to require conditions and geometric settings specific to a particular model. Methods used in random heterogeneous materials \citep{TorquatoBook} may allow making progress in this respect, however, it is clear that $\zeta_{\rm e} > 1$ could generally occur (see Appendix~\ref{sec:example_spheres} for a simple example).

Using Eq.~\eqref{eq:vis_Gaussian_derivations} and the definition of $\zetae$,  the corrections to the visibility then enter as 
\begin{align}
\langle g(\eta)\rangle\approx \bar{\Gamma}(\eta)[1-\zetae(\eta)]\,\exp\left(-\int_{\eta}^{\eta_0} \bar{\Gamma}(\eta')[1-\zetae(\eta')] \id \eta' \right).
\end{align}
Let us consider the limiting cases. For $\zetae \ll 1$, no effect appears, meaning that either the coherence length is extremely small while $\sigmae^2$ remains moderate or that the amplitude of the fluctuations becomes negligible and everything is well-captured by the average scattering process. As $\zetae\rightarrow 1$, fewer scattering happen than predicted by the average rate. If the increase of $\zetae$ is due to increasing $\sigmae^2$, this is equivalent to a very clumpy Universe, with many voids and few high-density peaks, such that the effective scattering rate is significantly smaller. As $\zetae$ approaches unity, scatterings can be neglected and photons primarily free-stream through the Universe, while only occasionally getting stuck in dense regions.

We stress that a simple Gaussian treatment {\it cannot} be used to cover the full range of scenarios in this picture. It is also clear that in physical models, $\tauc$ and $\sigmae^2$ cannot be varied arbitrarily nor independently. However, one can expect the main effect of electron density fluctuations to reduce the effective average scattering rate in the models we use. This means that the visibility function can change shape and broaden significantly, hence affecting the CMB anisotropies in a non-trivial way.

\begin{figure}
    \centering
    \includegraphics[width=\columnwidth]{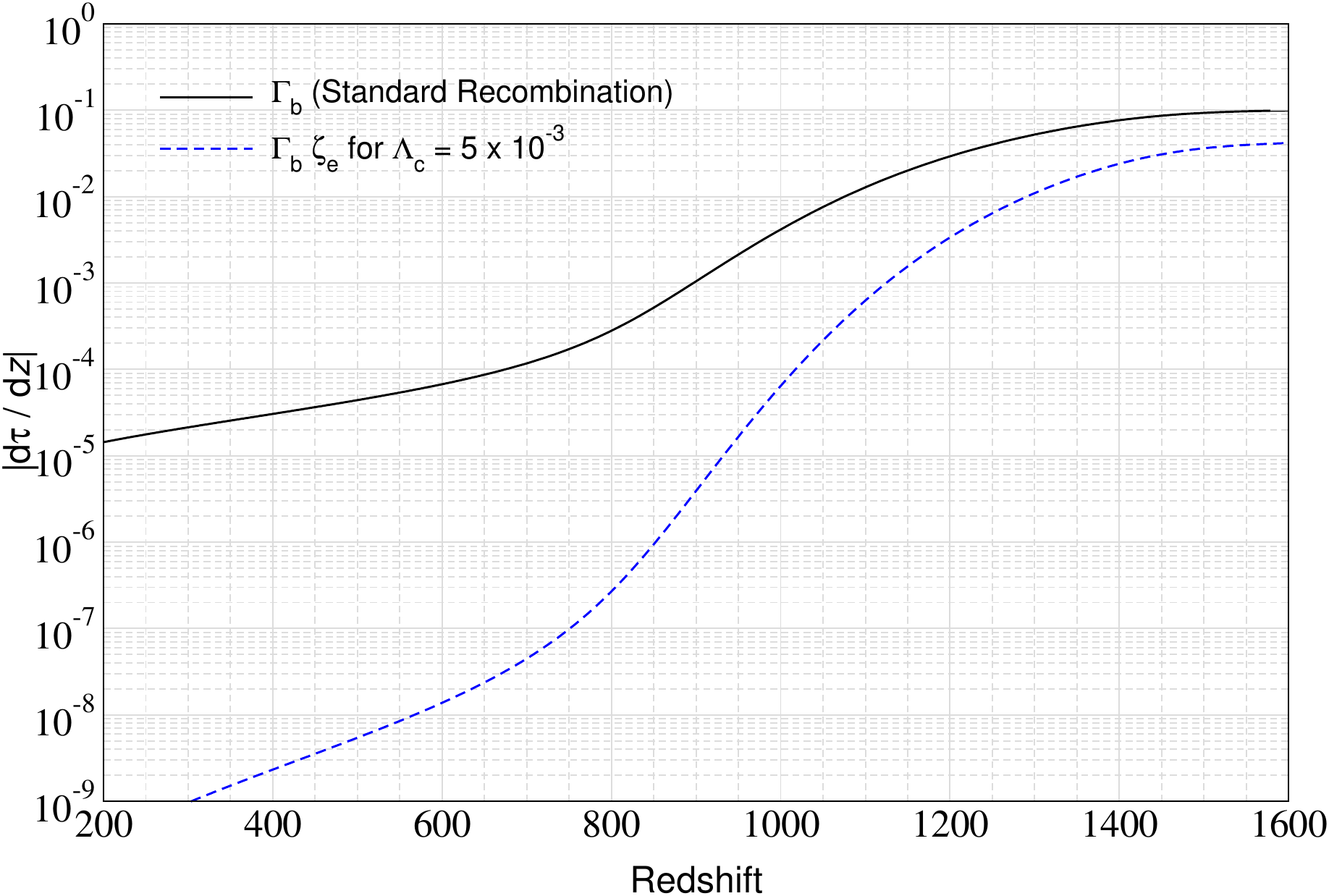}
    \\
    \caption{Differential optical depth contributions. The solid black line is for $\Gamma_{\rm b}\equiv \Gammab$, while the blue line shows $\Gammab\zeta_{\rm e}=\Gammab^2 r_{\rm s}\,\Lambda_{\rm c}$ for $\Lambda_{\rm c}=\pot{5}{-3}$.}
    \label{fig:Gammab_mods}
\end{figure}

\begin{figure}
    \centering
\includegraphics[width=\columnwidth]{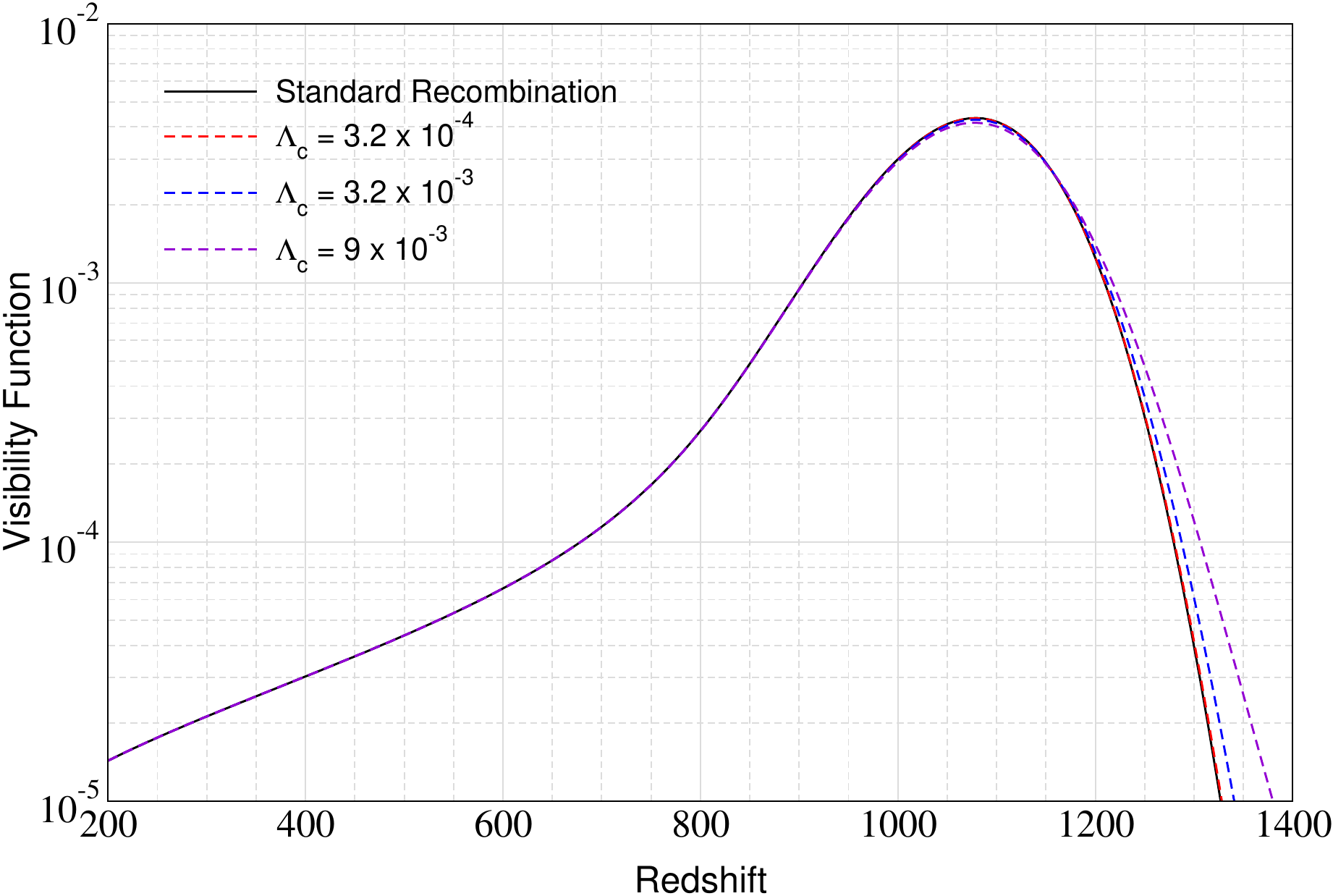}
\\
    \caption{Modifications to the Thomson visibility function for various values of $\Lambda_{\rm c}$. Electron density variations cause a broadening of the visibility function and reduction of the maximal last-scattering probability.}
    \label{fig:vis_mods}
\end{figure}

\vspace{-3mm}
\subsection{Illustrating the visibility corrections}
\label{sec:estimate_Gamma}
Even if additional corrections arise from changes to the transfer functions, we will see below that visibility function corrections already capture one of the main effects of electron density perturbations along the line of sight on the CMB power spectra. To parametrize things more explicitly, let us describe the model in terms of the sound horizon, $r_{\rm s}$, and scattering rate, $\Gammab$:
\begin{align}
\zeta_{\rm e}(\eta)=\Gammab\,\Delta \eta_{\rm c} \sigma_{\rm e}^2=\Gammab\, r_{\rm s}\Lambda_{\rm c}
\end{align}
with $\Lambda_{\rm c}(\eta)=\Delta \eta_{\rm c}\sigma^2_{\rm e}/r_{\rm s}$. We shall normalize everything at $z=1100$. In this way, we can write
\begin{align}
\Lambda_{\rm c}(\eta)
&=\pot{1.8}{-3}\,
\astroway{\Delta \eta_{\rm c}}{10\,\kpc}\,
\astroway{\sigma_{\rm e}}{5}^2\,\hat{\Lambda}_{\rm c}(\eta),
\end{align}
where $\hat{\Lambda}_{\rm c}(\eta)$ is a free function normalized to unity at $z=1100$. For reference, the sound horizon is $r_{\rm s}\simeq 142\,\Mpc$ at $z=1100$ and one expects the perturbations to be present on smaller scales.

In Fig.~\ref{fig:Gammab_mods}, we illustrate the scattering rate as a function of redshift. We assumed constant $\Lambda_{\rm c}$ for illustration. Due to the $\Gammab^2$ weighting of the correction, for constant $\Lambda_{\rm c}$ the main contribution arises before recombination. 
In Fig.~\ref{fig:vis_mods}, we show a few examples for the Thomson visibility. For increasing $\Lambda_{\rm c}$, the main effects are a broadening of the visibility function towards high redshifts with an associated reduction of the last-scattering probability at $z\simeq 1100$. This leads to extra damping and smearing of the CMB power spectra. 

For the computations we assumed $(\Delta \eta_{\rm c}, \sigma_{\rm e})=(5 \,\kpc, 3)$, $(50 \,\kpc, 3)$ and $(50 \,\kpc, 5)$ at $z=1100$, covering scales of large galaxies and their surrounding and electron density enhancements of a few times the average. 
One can also think about the effect in terms of an optical depth correction factor, $\zeta_{\rm e}(\eta)=\tauc \sigma_{\rm e}^2$, where $\tauc=\Gammab\,\Delta \eta_{\rm c}$ is the optical depth across the coherence length. In Fig.~\ref{fig:Gammab_mods}, we have $\tauc \approx \pot{2}{-3}$ at $z=1100$ for $\sigmae\simeq 3$, giving rise to a percent-level correction around that redshift.

The chosen examples already suggest a noticeable effect on the CMB power spectra would be expected. However, we will demonstrate that a modification to the visibility function does not capture the full effect and that additional corrections arise affecting the transfer functions due to extra source terms. 

\subsubsection{Link to the power spectrum of electron density fluctuations}
\label{sec:power_spectrum}
To understand the link between $\sigmae^2$ and $\Delta\eta_{\rm c}$ a bit better, let us consider a power spectrum of electron fluctuations at small scales, $P_{\rm e}(k, \eta)$ for a Gaussian random field. This directly defines the two-point correlation functions of the field as
\begin{align}
\xi(r_{ij}, \eta_i)= \left<\deltae(\vek{r}_i) \,\deltae(\vek{r}_j)\right>= \int \id \ln k \,\Delta^2_{\rm e}(k, \eta_i) \, j_0(k r_{ij})
\end{align}
with $r_{ij}=|\vek{r}_i-\vek{r}_j|$, $\Delta^2_{\rm e}(k, \eta_i)=\frac{k^3}{2\pi^2}\,P_{\rm e}(k, \eta_i)$ and spherical Bessel function $j_0(x)$. This can be converted into the time-time correlation matrix by setting $r_{ij}=|\eta_i-\eta_j|$ under the assumption of isotropy. For zero lag, one recovers the variance
\begin{align}
\sigmae^2(\eta_i)\equiv \xi(r_{ij}=0, \eta_i)= \int \id \ln k \,\Delta^2_{\rm e}(k, \eta_i),
\end{align}
which defines the effective coherence scale of the field as
\begin{align}
\label{eq:Detac_power}
%
\Delta\eta_{\rm c}(\eta_i)= \frac{\int \xi(r_{ij} , \eta_i)\id r_{ij}}{\sigmae^2(\eta_i)}
=\frac{\pi}{2}\frac{\int \id \ln k \,\Delta^2_{\rm e}(k, \eta_i)/k}{\int \id \ln k \,\Delta^2_{\rm e}(k, \eta_i)},
\end{align}
where we used $\int j_0(k r_{ij}) \id r_{ij} =\pi/[2k]$. This expression tells us how far from $\eta_i$ the two-point correlation function remains effectively constant.\footnote{This statement is obvious if $\xi(r_{ij} , \eta_i)$ is a top-hat.} 
This implies
\begin{align}
\zeta_{\rm e}(\eta)
\approx 
 \Gammab(\eta)\int \xi(r , \eta)\id r=\Gammab(\eta)\int \frac{\pi\id k}{2 k^2} \,\Delta^2_{\rm e}(k, \eta), \label{zeta Gamma PS eq}
\end{align}
where we assumed that $\Gammab$ remains roughly constant across the coherence scale. This assumption can be softened by instead considering the power spectrum of scattering rate fluctuations.

Assuming a single-mode power spectrum at $k_{\rm s}$, we naturally have $\Delta\eta_{\rm c}=\pi/[2 k_{\rm s}]=\lambda_{\rm s}/4$, where $\lambda_{\rm s}$ is the comoving wavelength of the mode. For a log-normal power spectrum $\Delta^2_{\rm e}(k, \eta_i)$ with mean $k_{\rm s}$ and $k$-space variance $\sigma^2_{\rm s}$ [see Eq.~\eqref{eq:delta_b} for explicit form for log-normal $1+\deltae$] one finds $\Delta\eta_{\rm c}=\lambda_{\rm s}\,\expf{\sigma^2_{\rm s}}/4$. 

These examples show that for a Gaussian random field the two parameters $\sigma_{\rm e}^2$ and $\Delta\eta_{\rm c}$ are essentially independent, with the typical amplitude of the fluctuations setting $\sigma_{\rm e}^2$ and the scale-distribution (i.e., the shape of the power-spectrum) determining $\Delta\eta_{\rm c}$. Assuming some initial perturbations, gravitational collapse will generally lead to increasing peak heights, i.e., $\sigma_{\rm e}^2$, and decreasing coherence length $\Delta\eta_{\rm c}$. For ionizing radiation, the coherence length can in addition change dramatically and depart from the coherence length of baryons or dark matter. Modeling all these details is beyond the scope of this work and will require dedicated simulations.

\subsection{Perturbative treatment}
\label{sec:pert_sol}
The system in Sect.~\ref{sec:main_equations} can be written as a matrix equation. Defining the vector $\vek{X}=(\Phi, \Psi, \delta_{\rm c}, \ldots, \varv_{\rm b}, \Theta_0, \Theta_1, \ldots, \Theta_N)$, where we truncate at some suitable $\ell_{\rm max}=N$, we can then write
\begin{align}
    \partial_{\eta} \vek{X} = \vek{A} \,\vek{X} - \Gammab (1+\deltae)\, \vek{B}\,\vek{X},
\end{align}
where $\vek{A}$ and $\vek{B}$ are matrices to define the system. We note that $\vek{B}$ is a diagonal matrix with off-diagonal terms only between $\Theta_1$ and $\varv_{\rm b}$. Written in this from, we can determine the transfer functions of the system for various initial conditions and source terms. 

Setting $\deltae=0$, we can obtain the usual solution around the average recombination history from the system
\begin{align}
    \partial_{\eta} \vXb  &= \vek{A} \,\vXb 
    - \Gammab \,\vek{B} \,\vXb .
\end{align}
The corresponding transfer functions are related to an initial value problem and shall be denoted as $\hat{\vek{X}}(\eta)=\hat{\bar{\vek{X}}}(0, \eta)$. We assume adiabatic initial conditions. 
Given $\bar{\vek{X}}$, we can next compute the correction, $\delta \vek{X}=\vek{X}-\bar{\vek{X}}$, caused by electron density fluctuations, which follows from the system
\begin{align}
    \partial_{\eta} \delta\vek{X} &= 
    \vek{A} \,\delta\vek{X} - \Gammab (1+\deltae)\, \vek{B}\, \delta\vek{X}
    - \Gammab\,\deltae \vek{B}\,\bar{\vek{X}}.
\end{align}
We observe a non-zero sourcing $\propto \deltae \vek{B}\,\bar{\vek{X}}$, which will give rise to fast-varying fluctuations that survive at second order in $\deltae$.  

If we linearize the problem in terms of $\deltae$ and $\delta\vek{X}$, we find the first and second order corrections using
\bsub
\begin{align}
    \partial_{\eta} \delta\vek{X}^{(1)} &= \vek{A} \,\delta\vek{X}^{(1)}
    - \Gammab\, \vek{B} \, \delta\vek{X}^{(1)}
    - \Gammab\, \deltae \vek{B}\,\bar{\vek{X}},
\\
    \partial_{\eta} \delta\vek{X}^{(2)} &= \vek{A} \,\delta\vek{X}^{(2)}
    - \Gammab\, \vek{B} \, \delta\vek{X}^{(2)}
    - \Gammab\, \deltae \vek{B}\,\delta\vek{X}^{(1)}.
\end{align}
\esub
Denoting the {\it transfer function matrix} that solves the evolution from $\eta$ to $\eta'$ as $\hat{\vek{X}}(\eta, \eta')$, we can write the formal solutions as
\begin{align}
\nonumber
    \delta\vek{X}^{(1)}(\eta') &= 
    \hat{\vek{X}}(\eta, \eta') \, \delta\vek{X}^{(1)}(\eta) - 
    \int_{\eta}^{\eta'}\hat{\vek{X}}(\eta_1, \eta')
    \,\Gammab_1 \delta_{\rm e, 1} \vek{B}_1\bar{\vek{X}}_1
    \id \eta_1,
\\
\nonumber
\delta\vek{X}^{(2)}(\eta') &= 
\hat{\vek{X}}(\eta, \eta') \, \delta\vek{X}^{(2)}(\eta) -
\!\int_{\eta}^{\eta'}\!\!\hat{\vek{X}}(\eta_1, \eta')
    \,\Gammab_1 \delta_{\rm e, 1} \vek{B}_1 \delta\vek{X}^{(1)}_1
    \id \eta_1,
\end{align}
where $\vek{B}_1=\vek{B}(\eta_1)$ and so forth.
Note that the first and second order transfer function matrix is the same as for the unperturbed system because the relevant linear operator $L[\delta\vek{X}]=\partial_\eta\delta\vek{X}-[\vek{A}-\Gammab \vek{B}]\,\delta\vek{X}$ is identical. 
Assuming $\eta'=\eta+\Delta \eta$ with $\Delta \eta \gg \Delta \eta_{\rm c}= 1/\alpha$ but $\Delta \eta$ short enough to avoid changes in $\vek{B}$, $\bar{\vek{X}}$ and $\hat{\vek{X}}(\eta, \eta')\approx \mathds{1}$, we can then write the average change of the solution as
\begin{align}
    \left<\delta\vek{X}^{(1)}(\eta, \eta') \right>_{\Delta \eta}&\approx 0,
\nonumber
\\
\nonumber
\left<\delta\vek{X}^{(2)}(\eta, \eta') \right>_{\Delta \eta} &\approx \Gammab^2\,\frac{\vek{B}^2 \bar{\vek{X}}}{\Delta \eta}\! 
\int_{\eta}^{\eta'}\!\!\id \eta_1 
\int_{\eta}^{\eta_1}\!\!\id \eta_2 
\,\left<\delta_{\rm e, 1}\delta_{\rm e, 2} \right>
\approx 
\frac{\Gammab^2 \sigma^2}{2\alpha^2} 
\,\vek{B}^2 \bar{\vek{X}},
\end{align}
where we can use Eq.~\eqref{eq:corel_delta2e} or Eq.~\eqref{eq:corel_delta2e_alt} for $\Delta \gamma=\alpha \Delta \eta \gg 1$.\footnote{In the derivation, one finds terms that are suppressed by $\Delta \gamma^{-1}\expf{-\Delta \gamma}$ and $\Delta \gamma^{-1}$, which one can drop.} This means that we can write the evolution equation for the correction as
\begin{align}
\label{eq:corr_Theta_second_av}
    \partial_{\eta} \left<\delta\vek{X}^{(2)}\right> &\approx  \vek{A} \,\left<\delta\vek{X}^{(2)}\right>
    -\Gammab\,\vek{B}\left<\delta\vek{X}^{(2)}\right>
    +\Gammab\,\tauc\sigma_{\rm e}^2 \,\vek{B}^2 \bar{\vek{X}},
\end{align}
where we replaced $\tauc=\Gammab/\alpha$, $\sigma_{\rm e}^2\equiv \sigma^2/[2\alpha]$. This equation can be used to compute the transfer function $\big<\delta\hat{\vek{X}}^{(2)}(\eta)\big>=\big<\delta\hat{\vek{X}}^{(2)}(0, \eta)\big>$ for the full evolution from the initial moment.

We note that the procedure above is akin to a Langevin treatment of Brownian motion, where perturbations vary and decorrelate much faster than the evolution from the background system. 
The term $\propto\vek{B}^2 \bar{\vek{X}}$ excites a growing mode as long as $\bar{\vek{X}}$ does not change much. However, once $\bar{\vek{X}}$ begins to vary, the evolution is regulated by the terms $\vek{M}=\vek{A}-\Gammab\vek{B}$, such that the system is solvable with a finite correction. We could alternatively solve the simpler system
\begin{align}
\label{eq:final_system_simplified_I_per}
\frac{\!\id \!\left<\vek{X}\right>}{\id \eta} &\approx \vek{A}\left<\vek{X}\right>
-\Gammab\,\vek{B}\left<\vek{X}\right>+
\Gammab \,\tauc\sigma_{\rm e}^2 \,\vek{B}^2 \left<\vek{X}\right>
\end{align}
to directly incorporate the corrections. Note that this expression is limited to small $\zeta_{\rm e}=\tauc\sigma_{\rm e}^2=\Gammab\,\Delta \eta_{\rm c}\sigma_{\rm e}^2$, which can be violated at early times even if $\zeta_{\rm e}< 1$ at $z\simeq 1100$. We will derive higher order terms using It\^o-calculus (Sect.~\ref{sec:ito}) but even these do not eliminate the issue fully, highlighting a limitation of the approach. 

\subsection{Expected changes to the CMB power spectra}
\label{sec:CMB_Power}
Following the usual steps, the CMB temperature power spectrum evaluated today at $\eta_0$ is related to the integrals
\begin{align}
\label{eq:Cell_def}
C_{\ell}
&=
4 \pi \int \id \ln k \,\Theta^2_\ell(\eta_0, k)\, \Delta^2_{\rm R}(k)
\end{align}
over the photon transfer functions. Here, we introduced the primordial scalar power spectrum, $\Delta^2_{\rm R}(k)$. The latter appears since we have to compute the ensemble average over the initial conditions at large scales, which relates to terms $$\left<\bar{\Theta}^*_\ell(\eta_0, \vek{k}_1) \,\bar{\Theta}_\ell(\eta_0, \vek{k}_2) \right>=(2\pi)^3\delta^{(3)}(\vek{k}_1-\vek{k}_2)\, \bar{\Theta}^2_\ell(\eta_0, k) \,2\pi^2 k^{-3}\,\Delta^2_{\rm R}(k).$$ As above, we assume that due to the presence of small-scale electron density fluctuations no new directional dependence will be introduced. However, we have to independently average over realizations of the electron density fluctuations in the time-coordinate. The transfer functions are then given by $\Theta_\ell(\eta_0, k)=\bar{\Theta}_\ell(\eta_0, k)+\delta \Theta^{(1)}_\ell(\eta_0, k)+\delta \Theta^{(2)}_\ell(\eta_0, k)$, where the $\delta \Theta^{(m)}_\ell(\eta_0, k)$ stem from the electron density fluctuations. To understand how various terms appear, it is easiest to consider the average of $\left<\vek{X}(\eta_0, k) \cdot\vek{X}(\eta_0, k) \right>_{\rm e}$ over the electron density realizations. Because all corrections to the transfer functions are initially sourced by the same scalar perturbations (and the $\deltae$ field is scale-independent), no mode-coupling occurs. We then have
\begin{align}
&\left<\vek{X}(\eta_0, k) \cdot\vek{X}(\eta_0, k) \right>_{\rm e}
\approx \bar{\vek{X}}(\eta_0, k)\cdot\bar{\vek{X}}(\eta_0, k)
\!+\!2\bar{\vek{X}}(\eta_0, k) \cdot \left<\delta \vek{X}^{(1)}(\eta_0, k) \right>_{\rm e}
\nonumber\\
&\qquad\; +\left<\delta \vek{X}^{(1)}(\eta_0, k)\cdot \delta \vek{X}^{(1)}(\eta_0, k) \right>_{\rm e}+2\bar{\vek{X}}(\eta_0, k) \cdot \left<\delta \vek{X}^{(2)}(\eta_0, k) \right>_{\rm e}.
\nonumber
\end{align}
The first term is due to the average evolution, with a recombination history given by $\Neb$.
The second term averages out, since $\delta \vek{X}^{(1)}(\eta_0, k)$ is linear in $\deltae$. The last term can be directly computed using Eq.~\eqref{eq:corr_Theta_second_av}, and takes the form
\begin{align}
\left<\delta \vek{X}^{(2)}\right>_{\rm e}
&\approx 
\int_{0}^{\eta_0}
\id \eta_1
\,\bar{\Gamma}_{1} \zeta_{\rm e, 1}\,\hat{\vek{X}}(\eta_1, \eta_0)\,\vek{B}^2_1\bar{\vek{X}}_1,
\end{align}
where we used the formal transfer function matrix for the background evolution, $ \hat{\vek{X}}(\eta_1, \eta_0)$, from $\eta_1$ to $\eta_0$. This term accounts for corrections from correlations along the line-of sight.

The remaining term can be evaluated as 
\begin{align}
\label{eq:gen_aver_first_2_hard}
\left<\delta \vek{X}^{(1)}\cdot \delta \vek{X}^{(1)} \right>_{\rm e}
&=
\int_{0}^{\eta_0}
\id \eta_1
\int_{0}^{\eta_0}
\id \eta_2
\left<\delta_{\rm e, 1}\delta_{\rm e, 2}\right>  
\\[1mm] \nonumber 
&\qquad \times
[\bar{\Gamma}_{1}\hat{\vek{X}}(\eta_1, \eta_0)\vek{B}_1\bar{\vek{X}}_1]\cdot 
[\bar{\Gamma}_{2}\hat{\vek{X}}(\eta_2, \eta_0)\vek{B}_2\bar{\vek{X}}_2]
\nonumber\\\nonumber
&\approx
2 \int_{0}^{\eta_0}
\id \eta_1\,\bar{\Gamma}_{1} \zeta_{\rm e, 1}\,|\hat{\vek{X}}(\eta_1, \eta_0)\,\vek{B}_1\bar{\vek{X}}_1|^2,
\end{align}
which is of similar order as the correction from $\big<\delta \vek{X}^{(2)}\big>_{\rm e}$ and in contrast accounts for corrections from time-correlations between two independent modes.
Here, we have used Eq.~\eqref{eq:corel_delta2e_alt}
\begin{align}
\label{eq:second_type}
&\int_0^{\eta_0}\id \eta_2
F(\eta_1, \eta_2)\,\left<\delta_{\rm e, 1}\delta_{\rm e, 2}\right> 
\approx F(\eta_1, \eta_1)\,\int_{\eta_1-\Delta \eta}^{\eta_1+\Delta \eta}\id \eta_2
\left<\delta_{\rm e, 1}\delta_{\rm e, 2}\right>
\nonumber \\[1mm]
&\qquad \qquad \approx 
F(\eta_1, \eta_1)\,\frac{\sigma^2(\eta_1)}{\alpha^2(\eta_1)} \, \left[1-\expf{-\Delta \gamma}\right]\approx F(\eta_1, \eta_1)\,\frac{\sigma^2(\eta_1)}{\alpha^2(\eta_1)}
\end{align}
with $\Delta \gamma=\alpha\Delta \eta\gg 1$, realizing that $\left<\delta_{\rm e, 1}\delta_{\rm e, 2}\right>$ is only non-zero in a very short interval of $\eta_2$ around $\eta_1$. 
%
Putting things together, we have
\begin{align}
\label{eq:Final_corrections}
\left<\vek{X}\cdot\vek{X} \right>_{\rm e}
&\approx \bar{\vek{X}}\cdot\bar{\vek{X}}
+\left<\delta \vek{X}^{(1)}\cdot \delta \vek{X}^{(1)}\right>_{\rm e}+2\bar{\vek{X}} \cdot \left<\delta \vek{X}^{(2)}\right>_{\rm e}
\nonumber\\
&\approx \left|\bar{\vek{X}}+\left<\delta \vek{X}^{(2)}\right>_{\rm e}\right|^2
+\left<\delta \vek{X}^{(1)}\cdot \delta \vek{X}^{(1)}\right>_{\rm e}
\end{align}
for small $\left<\delta \vek{X}^{(2)}(\eta_0, k) \right>_{\rm e}$. This means we can simply compute the first contribution in the second line using the combined evolution as suggested in Eq.~\eqref{eq:final_system_simplified_I_per}. For the second term, a more complicated problem is encountered, requiring a detailed consideration of each sourcing event, which we will only discuss in passing.

We will see that Eq.~\eqref{eq:Final_corrections} indeed captures the leading order term as long as $\tauc \sigma_{\rm e}^2$ remains small. 
This limit can in principle be violated at high redshifts, as we can already anticipate from our estimates in Sect.~\ref{sec:estimate_Gamma}. However, we show below how to better stabilize the problem by including higher order contributions. This reveals additional corrections in powers $\tauc^2\sigma_{\rm e}^2 \vek{B}^2$. We can also generalize to log-normal distributions of $1+\deltae$ using It\^o calculus.

\section{Generalized treatment using It\^o calculus}
\label{sec:ito_formalism}
In this section we employ It\^o calculus to obtain a generalization of the Boltzmann hierarchy for the Gaussian and log-normal Ornstein-Uhlenbeck processes. For our models, this should yield the same result as a treatment based on the Stratonovich integral, which would account for correlations with the future solution realization of the field that are absent in our setup.

\subsection{It\^o formalism}
\label{sec:ito}
We now derive the evolution equations using It\^o calculus, thinking of $\deltae$ as a correlated stochastic process. To obtain the final answer, we have to average the result over all realizations of this field, which then allows us to obtain evolution equations for the averaged problem.
Formally, we then have a coupled problem of the form
\begin{align}
\id \vek{X} &= [\vek{M}  - \Gammab\deltae \vek{B}]\,\vek{X}\id \eta,
\nonumber 
\\
\id \deltae&=-\alpha \deltae \id \eta + \sigma \id W,
\end{align}
where $\Gammab$, $\alpha$, $\sigma$, $\vek{M}=\vek{A}-\Gammab \vek{B}$ and $\vek{B}$ are all slowly varying and $\deltae$ is treated as an independent variable. 
We now consider any function $F(\eta, \vek{X}, \deltae)$ and write the total differential as
\begin{align}
\id F &= \partial_\eta F \id \eta +  \id \vek{X} \cdot \nabla_{\vek{X}}  F  + \id \deltae \, \partial_{\deltae} F.
\end{align}
We are interested in computing $\left<\!\id F \right>$. Since $\id \deltae$ contains a stochastic driving term, we pick up a new term according to It\^o calculus
\begin{align}
\left<\!\id F\right>
&\equiv
\left<F(\eta+\id \eta, \vek{X}+\id \vek{X}, \deltae+\id \deltae) - F(\eta, \vek{X}, \deltae)\right>=\id \left<F\right>
\\     
\nonumber
&
\simeq 
\left<\partial_\eta F  +  [\vek{M} - \Gammab\,\deltae\vek{B}]\,\vek{X}\cdot \nabla_{\vek{X}} F
-\alpha \deltae \partial_{\deltae} F+\frac{\sigma^2}{2}\,\partial^2_{\deltae} F \right>\!\id \eta.
\end{align}
The only unusual term is $\frac{\sigma^2}{2}\,\partial^2_{\deltae} F$, which follows from the fact that one encounters terms $\propto \left<\!\id W \id W'\right> \rightarrow \id \eta$.

Taking the formal limit $\id \eta \rightarrow 0$, we find that
\begin{align*}
\frac{\!\id \!\left<F\right>}{\id \eta} &= \left<\partial_\eta F  + [\vek{M} - \Gammab\,\deltae \vek{B}]\vek{X}\cdot \nabla_{\vek{X}} F
+\alpha \left(\sigmae^2\,\partial^2_{\deltae} F -\deltae \partial_{\deltae} F\right)\right>,
\end{align*}
where we used $\sigmae^2=\sigma^2/[2\alpha]$ as before. 
With this expression we can generate the evolution equations for any of the required quantities averaged over the electron fluctuations. In particular, we can derive a solution for $\left<\vek{X} \cdot\vek{X} \right>$.
Introducing $\left<\delta \vek{X}\right>=\left<\vek{X}\right>-\bar{\vek{X}}$, we then have
\begin{align}
\label{eq:gen_aver_ALL}
&\left<\vek{X} \cdot\vek{X} \right>
\approx \bar{\vek{X}}\cdot\left(\bar{\vek{X}}
+2\left<\delta \vek{X}\right>\right)+\left<\delta \vek{X}\cdot \delta \vek{X}\right>.
\end{align}
Note that here $\left<\delta \vek{X} \right>$ will automatically include higher order corrections in terms of $\deltae$ and thus does not vanish. We will focus on this term in what follows.
Thus, for now all we need is an equation for $\left<\vek{X}\right>$ which together with $\partial_{\eta} \bar{\vek{X}} = \vek{M} \bar{\vek{X}}$ solves the problem.
Considering $F_p=\deltae^p \vek{X}$ and introducing the moment vectors $\vek{\kappa}_p=\left<\deltae^p \vek{X}\right>$, we obtain the infinite system 
\begin{align}
\frac{\!\id \vek{\kappa}_{p}}{\!\id \eta} &= \vek{M} \vek{\kappa}_{p} -\Gammab  \vek{B} \,\vek{\kappa}_{p+1}
-\alpha \,p \, \vek{\kappa}_{p}+\alpha p\,(p-1)\,\sigmae^2\, \vek{\kappa}_{p-2}
\end{align}
using the It\^o formalism. Truncating at second order, we then have to solve the coupled system of equations
\bsub
\label{eq:final_system_2}
\begin{align}
    \frac{\!\id \vek{\kappa}_0}{\id \eta} &=\vek{A}\vek{\kappa}_0-\Gammab\vek{B}\left[\vek{\kappa}_0+\vek{\kappa}_1\right],
\\[-0.5mm]
\frac{\!\id \vek{\kappa}_1}{\!\id \eta} &=\vek{A}\vek{\kappa}_1
-\Gammab\vek{B}\left[\vek{\kappa}_1+\vek{\kappa}_2\right]-\alpha \vek{\kappa}_1,
\\[-0.5mm]
\frac{\!\id \vek{\kappa}_2}{\!\id \eta} &=\vek{A}\vek{\kappa}_2-\Gammab\vek{B} \vek{\kappa}_2-2\alpha \vek{\kappa}_2
+2\alpha\sigmae^2\vek{\kappa}_0
\end{align}
\esub
to obtain the solution for $\vek{\kappa}_0=\left<\vek{X}\right>$. We have separated the term $\propto \Gammab$ which can become large at early time. Corrections are sourced by the term $\propto \sigmae^2\left<\vek{X}\right>\approx \sigmae^2\bar{\vek{X}}$. Without this variance from auto-correlation in the $\deltae$ field, there would not be any corrections. However, without time-correlations (related to the terms $\propto \alpha$) there would not be any propagation of effects on $\left<\vek{X}\right>$. The time-correlations are therefore crucial for any effect to appear, as we already understood from the discussion in Sect.~\ref{sec:naive}.

\vspace{-4mm}
\subsubsection{Consistency with perturbative treatment}
We now compare to the perturbative treatment we presented in Sect.~\ref{sec:pert_sol}. In order to do this, we again perform a short time integration of the system given by Eq.~\eqref{eq:final_system_2}. Assuming that at $\eta$ initially we have $\left<\vek{X}\right>_0=\bar{\vek{X}}$, all higher order correlators also vanish at $\eta$. 

There are three time-scales in the problem. The correlations evolve on the shortest time-scale, $\Delta \eta_{\rm c}=1/\alpha$. We also have the scattering time-scale $\eta_{\rm sc}=\Gammab^{-1}$, which can become short at early times, driving the system into the tight-coupling regime. Finally, we have the time-scale on which the large-scale anisotropies evolve, which is related to the wavenumber of the mode, $\eta_k=k^{-1}$, which, as we shall see below, can be ignored for the scales under consideration. 
Since the leading order term is $\propto \Gammab^2 \sigmae^2/\alpha$, we expect the correction to be a series in the parameter $\tauc=
\Gammab/\alpha$, which is the optical depth across the coherence time. Introducing $\id \gamma = \alpha \id \eta$, we rewrite Eq.~\eqref{eq:final_system_2} as
\bsub
\label{eq:final_system_2_eps}
\begin{align}
\frac{\!\id \vek{\kappa}_0}{\id \gamma} &=\tauc\left\{\vek{A}_{\Gammab}\vek{\kappa}_0-\vek{B}\left[\vek{\kappa}_0+\vek{\kappa}_1\right]\right\},
\\[-1mm]
\frac{\!\id \vek{\kappa}_1}{\!\id \gamma} &=\tauc\left\{\vek{A}_{\Gammab}\vek{\kappa}_1
-\vek{B}\left[\vek{\kappa}_1+\vek{\kappa}_2\right]\right\}-\vek{\kappa}_1,
\\[-1mm]
\frac{\!\id \vek{\kappa}_2}{\!\id \gamma} &=\tauc\left\{\vek{A}_{\Gammab}\vek{\kappa}_2-\vek{B} \vek{\kappa}_2\right\}-2\vek{\kappa}_2
+2\sigmae^2\vek{\kappa}_0,
\end{align}
\esub
where we use $\vek{A}_{\Gammab}=\vek{A}/\Gammab$.
Inserting the Ansatz $\vek{\kappa}_m=\sum_{k=0}^\infty \tauc^m\,\vek{\kappa}^{(m)}_p$, we then can solve order by order in $\tauc$. 
We note that at finite order of $\tauc$, we are bound to $\tauc < 1$; however, one can take the formal limit to large $\tauc$ by considering higher order terms.

Reorganising the system then yields
\bsub
\label{eq:final_system_2_eps_orders}
\begin{align}
\frac{\!\id \vek{\kappa}^{(m)}_0}{\id \gamma} &=\vek{M}_{\Gammab}\vek{\kappa}^{(m-1)}_0-\vek{B}\vek{\kappa}^{(m-1)}_1,
\\[-1mm]
\frac{\!\id \vek{\kappa}^{(m)}_1}{\!\id \gamma} &=
\vek{M}_{\Gammab}\vek{\kappa}^{(m-1)}_1-\vek{B}\vek{\kappa}^{(m-1)}_2
-\vek{\kappa}^{(m)}_1,
\\[-1mm]
\frac{\!\id \vek{\kappa}^{(m)}_2}{\!\id \gamma} &=\vek{M}_{\Gammab}\vek{\kappa}^{(m-1)}_2-2\vek{\kappa}^{(m)}_2
+2\sigmae^2\vek{\kappa}^{(m)}_0,
\end{align}
\esub
with $\vek{M}_{\Gammab}=\vek{M}/\Gammab$.
At zeroth order we find
\begin{align}
\label{eq:sol_0}
\vek{\kappa}^{(0)}_0 &=\vXb, \qquad \vek{\kappa}^{(0)}_1=0, \qquad \vek{\kappa}^{(0)}_2=\sigmae^2 \left[1-\expf{-2\gamma}\right] \vXb.
\end{align}
At first order, it is
\bsub
\label{eq:sol_1}
\begin{align}
\vek{\kappa}^{(1)}_0 &=\gamma \,\vek{M}_{\Gammab}\vXb, 
\quad \vek{\kappa}^{(1)}_1=-\sigmae^2 \,\left[1-2\expf{-\gamma}+\expf{-2\gamma}\right]\vek{B}\vXb
\\
\vek{\kappa}^{(1)}_2&=\sigmae^2 \gamma\left[1-\expf{-2\gamma}\right] \vek{M}_{\Gammab}\vXb.
\end{align}
\esub
At second order, we then have
\bsub
\label{eq:sol_2}
\begin{align}
\vek{\kappa}^{(2)}_0 &=\frac{\gamma^2}{2} \,\vek{M}_{\Gammab}^2\vXb+\sigmae^2 \gamma\,\left[1-\frac{3-4\expf{-\gamma}+\expf{-2\gamma}}{2\gamma}\right]\vek{B}^2\vXb, 
\\[-1mm]
\vek{\kappa}^{(2)}_1&=-\sigmae^2 \,\left[1-2\gamma\,\expf{-\gamma}-\expf{-2\gamma}\right]\vek{M}_{\Gammab}\vek{B}\vXb
\nonumber\\
&\qquad
-\sigmae^2\gamma\left[1+\expf{-2\gamma}-\frac{1-\expf{-2\gamma}}{\gamma}\right]\vek{B}\vek{M}_{\Gammab}\vXb
\\[-1mm]
\vek{\kappa}^{(2)}_2&=\sigmae^2 \frac{\gamma^2}{2}\left[1-\expf{-2\gamma}\right] \vek{M}_{\Gammab}^2\vXb
\nonumber \\[-1mm]
&\qquad+\sigmae^4 \gamma\,(1-\expf{-2\gamma})\left[1-\frac{2(1-\expf{-2\gamma})}{\gamma}\right]\vek{B}^2\vXb.
\end{align}
\esub
We can now drop all terms $\expf{-p\gamma}$ and $1/\gamma$ as we are considering the limit of large $\gamma$. Collecting terms, we then find
\begin{align}
\vek{\kappa}_{0}&\approx 
\left(1+\frac{\gamma}{\alpha}\vek{M} +\frac{\gamma^2}{2\alpha^2} \,\vek{M}^2\right)\vXb
+\tauc^2\sigma^2_{\rm e}\gamma\,\vek{B}^2\vXb.
\end{align}
Replacing the factors $\gamma=\alpha\Delta \eta $, we can go back to computing the average change over the considered short time-interval, $\Delta \eta$. Dropping higher order terms in $\Delta \eta$, we finally obtain
\begin{align}
\frac{\Delta\!\left<\vek{X}\right>}{\Delta \eta}&\approx 
\vek{M} \vXb +\frac{\Gammab^2\sigma^2_{\rm e}}{\alpha}\,\vek{B}^2 \vXb.
\end{align}
The first term captures the usual evolution, while the second is the correction from the electron density fluctuations. We could thus again have solved the simpler transfer problem
\begin{align}
\label{eq:final_system_simplified_I}
\frac{\!\id \!\left<\vek{X}\right>}{\id \eta} &\approx \vek{M}\!\left<\vek{X}\right>+
\frac{\Gammab^2\sigmae^2}{\alpha}\vek{B}^2\!\left<\vek{X}\right>.
\end{align}
Computationally, this captures the main effects on the transfer functions and visibility and is consistent with Eq.~\eqref{eq:final_system_simplified_I_per}. This is an important simplification, since one essentially has to solve a similar problem as for the standard CMB anisotropies. 

\vspace{-3mm}
\subsubsection{Solutions to infinite orders in the $\left<\deltae^p \delta\vek{X}\right>$ moments}
We can also express the infinite hierarchy in terms of 
$\vek{\mu}_{p} = \left<\deltae^{p} \delta \vek{X} \right>$, where $\delta \vek{X}=\vek{X}-\vXb$. 
With this definition, we obtain
\begin{align}
    \frac{\!\id \vek{\mu}_{p}}{\id \gamma} &= \tauc \left\{\vek{M}_{\Gammab} \vek{\mu}_{p} -\vek{B} \left[\vek{\mu}_{p+1}+\lambda_{p+1}\vXb\right]\right\}
    -p \, \vek{\mu}_{p}+p\,(p-1)\,\sigmae^2\, \vek{\mu}_{p-2},
    \nonumber \\
\frac{\!\id \lambda_{p}}{\id \gamma}
&=-p \, \lambda_{p}+p\,(p-1)\,\sigmae^2\, \lambda_{p-2},
\end{align}
where $\lambda_{p}=\left<\deltae^p\right>$ are the moments of $\deltae$, which can be solved independently. 

Using the initial conditions for a Gaussian, it is easy to show that all the moments, $\lambda_{p}$, remain independent of time with: 
\bsub
\begin{align}
\lambda_{0}&=1, 
\quad\lambda_{2m-1}=0, 
\quad\lambda_{2m}=(2m-1)!!\,\sigma_{\rm e}^{2m},
\end{align}
\esub
where we used $\big<\deltae^2\big>\equiv \sigma^2_{\rm e}=\sigma^2/[2\alpha]$. We again insert the Ansatz $\vek{\mu}_p=\sum_{m=0}^\infty \tauc^m\,\vek{\mu}^{(m)}_p$ and then rewrite the system as
\begin{align}
\frac{\!\id \vek{\mu}^{(m)}_{p}}{\id \gamma} &= -p \, \vek{\mu}^{(m)}_{p}+p\,(p-1)\,\sigmae^2\, \vek{\mu}^{(m)}_{p-2}
\nonumber \\
&\qquad\qquad +\vek{M}_{\Gammab} \vek{\mu}^{(m-1)}_{p} -\vek{B} \left[\vek{\mu}^{(m-1)}_{p+1}+(1-\delta_{k0})\lambda_{p+1}\,\vXb \right].
\end{align}
The formal solution then is
\begin{align}
\vek{\mu}^{(m)}_{p}(\gamma) &=
\expf{-p \gamma}
\,\vek{\mu}^{(m)}_{p,0}
+p(p-1)\,\sigmae^2\int_0^\gamma \expf{p(\gamma'-\gamma)}\,\vek{\mu}^{(m)}_{p-2}(\gamma')\id \gamma'
\\ \nonumber 
&\!\!\!\!\!\!\!\!\!\!\!\!\!\!\!\!
+\int_0^\gamma \expf{p(\gamma'-\gamma)}\,\left[\vek{M}_{\Gammab}\vek{\mu}^{(m-1)}_{p}(\gamma')
-\vek{B} \vek{\mu}^{(m-1)}_{p+1}(\gamma')\right]\id \gamma'
-(1-\delta_{k0})\,\mathcal{I}_{p}\vek{B}\vXb ,
\\ \nonumber
\mathcal{I}_p&=\int_0^\gamma \expf{p(\gamma'-\gamma)}\,\hat{\lambda}_{p+1}\id \gamma'
=
\begin{cases}
0 &\text{for even $p$}
\\
\frac{p!!}{p}\,\sigmae^{p+1}(1-\expf{-p\gamma}) &\text{for odd $p$},
\end{cases}
\end{align}
where we have assumed that all coefficients vary on much longer timescales than the considered time-step, $\Delta \eta$. 
The recursion can be started with $p=m=0$ and then solved raising $p$ up to the required order followed by raising $m$ to restart the loop. Since we already have the background solution, we can assume that at zeroth order in $\sigma_{\rm e}$ all $\vek{\mu}_{p}$ vanish initially, i.e., $\vek{\mu}^{(m)}_{p, 0}=0$. Due to the absence of sources at this order, we then find $\vek{\mu}^{(0)}_{p}(\gamma)=0$. At first order in $\sigma_{\rm e}$, we obtain
\begin{align*}
\vek{\mu}^{(1)}_{2m}(\gamma) &= 0, &\vek{\mu}^{(1)}_{2m-1}(\gamma)  &= 
-(2m-1)!!\,\sigmae^{2m}\,\left[1-\expf{-\gamma}\right]\vek{B}\vXb.
\end{align*}
At second order, we have
\begin{align*}
\vek{\mu}^{(2)}_{0}(\gamma) &= \sigmae^2\left[\gamma-1-\expf{-\gamma}\right]\vek{B}^2\vXb,
\\
\vek{\mu}^{(2)}_{1}(\gamma) &= -\sigmae^2\left[1-\expf{-\gamma}\right]\vek{B}\vXb
+
\sigmae^2\left[1-(1+\gamma)\expf{-\gamma}\right]\vek{M}_{\Gammab}\vek{B}\vXb,
\\
\vek{\mu}^{(2)}_{2}(\gamma) &=\sigmae^4\left[\gamma -\expf{-\gamma}+\expf{-2\gamma}\right]\vek{B}^2\vXb,
\\
\vek{\mu}^{(2)}_{2m-1}(\gamma) &= (2m-1)!!\,\sigmae^{2m-2}\,\vek{\mu}^{(2)}_{1}(\gamma).
\end{align*}
For the higher orders in $\tauc$, we carry out the computations using {\tt Mathematica}. Since we are only interested in $\vek{\mu}_{0}(\gamma)$, which we then use to compute the average change, $\vek{\mu}_{0}(\gamma)/\Delta \eta$, over the interval $\Delta \eta$, we focus on this quantity. We recognize that all terms $\propto \expf{-p \gamma}$ can be dropped, since we assume that $\gamma=\alpha\Delta \eta\gg 1$. We then also encounter terms that are higher order in $\Delta\eta$, which can be neglected once we take the limit $\Delta \eta\rightarrow 0$. Similarly, we find terms that are suppressed by $1/\gamma$ relative to the leading order, which we also drop. This then leads to the average change of $\vek{\mu}_{0}(\eta)$ over the interval $\Delta \eta$
\begin{align}
\label{eq:final_term_Gaussian}
\frac{\Delta \vek{\mu}_{0}(\eta)}{\Delta \eta}&\approx 
\left[
1-\frac{\Gammab^2\sigma^2_{\rm e}}{\alpha^2}\vek{B}^2
+\frac{\Gammab^4\sigma^4_{\rm e}}{2\alpha^4}\vek{B}^4
-\frac{\Gammab^6\sigma^6_{\rm e}}{6\alpha^6}\vek{B}^6\right]\frac{\Gammab^2\sigma^2_{\rm e}}{\alpha(1-\Gammab/\alpha)}\,\vek{B}^2\,\vXb
\nonumber\\
&=\frac{\Gammab\zetae}{(1-\tauc)}\exp\left[-\tauc\zetae\,\vek{B}^2\right]\vek{B}^2 \vXb
.
\end{align}
In the last step we used $\tauc=\Gammab/\alpha$, $\zetae=\tauc \sigmae^2=\frac{\Gammab\sigma^2_{\rm e}}{\alpha}$ and took the formal limit to all orders in $\tauc$. We analytically confirmed terms up to $12^{\rm th}$ order in $\tauc$. This expression has better convergence properties than the leading order terms. Still, we find instabilities for certain parameter combinations as we will discuss below.

Physically, exponential suppression of the correction can be caused by two effects. Firstly, the amplitude of the fluctuations can become large, while $\tauc < 1$, in which case photons no longer frequently encounter high density regions, such that the corrections are suppressed. Alternatively, $\tauc$ can become large, such that scattering effects dominate over distances shorter than $\Delta \eta_{\rm c}$. In this case, tight coupling occurs, and we no longer depend on the density fluctuations. We, therefore, recommend dropping the factor $1/(1-\tauc)$, which was obtained assuming $\tauc < 1$, to allow for a smooth transition between all extremes.

We also note that we dropped terms $\propto \vek{M}$. To leading order, these appear as 
\begin{align}
\frac{\Delta \vek{\mu}_{0}(\eta)}{\Delta \eta}\Bigg|_{\vek{M}}\approx  \frac{\Gammab^2\sigmae^2}{\alpha^2}[\vek{B}\vek{M}-\vek{M}\vek{B}]\,\vek{B}\vXb
=\frac{\Gammab^2\sigmae^2}{\alpha^2}[\vek{B}\vek{A}-\vek{A}\vek{B}]\,\vek{B}\vXb
\end{align}
but because $\vek{A}$ is suppressed by a factor of the wavenumber, these enter as background evolution corrections, which we neglect.

\vspace{-3mm}
\subsection{It\^o formalism for log-normal $\deltae$ field}
\label{sec:ito_log_normal}
We now derive the evolution equations using It\^o calculus when $\deltae$ is related to a log-normal stochastic process. In many ways, this is more realistic for applications in cosmology and also naturally ensures that $1+\deltae > 0$. 
Formally, we have a problem of the form
\begin{align}
\id \delta \vek{X} &= \left\{[\vek{M} - \Gammab\,\deltae \vek{B}]\,\delta\vek{X} - \Gammab\,\deltae \vek{B}\vXb\right\}\!\id \eta,
\nonumber 
\\
\id\xie&=\alpha [\mu-\xie] \id \eta + \sigma \id W,
\end{align}
for the correction $\delta\vek{X}=\vek{X}-\vXb$. As before, all coefficients are assumed to vary slowly in time. We also assume that $1+\deltae = \expf{\xie}$ follows a log-normal distribution, driven by the independent stochastic variable $\xie$ with parameters $\alpha$, $\mu$ and $\sigma$. Here, $\mu$ is used to accommodated for the differences between the mean and median value of the distribution. 

We now again consider any function $F(\eta, \vek{X},\xie)$ and follow the same steps as above, yielding
\begin{align}
    \frac{\!\id \!\left<F\right>}{\id \gamma} &= \left<\partial_\gamma F  +  \tauc\left\{[\vek{M}_{\Gammab} - \deltae \vek{B}]\,\delta\vek{X} - \deltae \vek{B}\vXb\right\}\cdot \nabla_{\delta \vek{X}} F \right>
    \nonumber \\
    &\qquad\qquad\qquad + \left<(\mu-\xie) \partial_{\xie} F+\sigmae^2\,\partial^2_{\xie} F \right>, 
    \label{eq:SDE_lognormal}
\end{align}
where we have immediately written everything in terms of $\gamma$ and $\tauc$. Let us first set $F=\xie^p$ to give
\begin{align}
    \frac{\!\id \rho_p}{\id \gamma} &= 
    -p\, \rho_{p}+p \mu\,\rho_{p-1}+p(p-1)\sigmae^2\,\rho_{p-2}.
\end{align}
We enforce $\rho_0= 1$ and $\rho_1=\mu=-\sigmae^2/2$. The latter choice ensures that $\big<1+\deltae\big>=1$ for a log-normal distribution. By demanding that $\id \rho_p/\id \gamma=0$ for all $p$, we then have the simple recursion
\begin{align}
\rho_{p} &= -\frac{\sigmae^2}{2}\,\rho_{p-1}+(p-1)\sigmae^2\,\rho_{p-2}.
\end{align}
This implies $\rho_2=\sigmae^2+\sigmae^4/4$, $\rho_3=-3\sigmae^4/2-\sigmae^6/8$ and so on.

We again want to compute $\big<\delta\vek{X}\big>$. With $\vek{\mu}_p=\big<\deltae^p \delta\vek{X}\big>$, we also have to solve a second system for moments of the form $\vomega_{pm}=\big<\xie^p \deltae^m \delta\vek{X}\big>$, which leads to a complicated sourcing structure that is not favored. 
Thus, it seems most convenient to instead use moments $\big<\xie^p \delta\vek{X}\big>$ and then insert $\deltae=\sum_{m=1}^\infty \frac{\xie^m}{m!}$. This implies
\begin{align}
\label{eq:log_normal_system}
    \frac{\!\id \vomega_{p}}{\id \gamma} &= \tau_{\rm c}\left\{\vek{A}_{\Gammab} \,\vomega_{p} - \sum_{m=0}^\infty\frac{\vek{B}\vomega_{p+m}}{m!}- \Lambda_{p}\vek{B}\vXb\right\}
    \nonumber \\[1mm]
    &\qquad\qquad -p\vomega_{p} + p\mu\vomega_{p-1}+p(p-1)\sigmae^2\,\vomega_{p-2}.
\end{align}
The density moments $\Lambda_{p}=\big<\deltae\xie^p\big>=\big<\expf{\xie} \xie^p\big>-\big<\xie^p\big>$ can be computed using the recursion
\begin{align}
\Lambda_{p} &= \frac{\sigmae^2}{2}\,\Lambda_{p-1}+(p-1)\sigmae^2\,\Lambda_{p-2}+\sigmae^2\rho_{p-1}.
\nonumber 
\end{align}
We obtained this expression by realizing that $\big<\expf{\xie}\xie^p\big>$ follows the same recursion as $\big<\xie^p\big>$ but with $\mu=\sigmae^2/2$. The first few values are $\Lambda_{2m}=0$ and $\Lambda_1=\sigmae^2$, $\Lambda_3=3\sigmae^4+\frac{1}{4}\sigmae^6$, $\Lambda_5=15\sigmae^6+\frac{5}{2}\sigmae^8+\frac{1}{16}\sigmae^{10}$.

The system in Eq.~\eqref{eq:log_normal_system} is not easy to solve as it in principle requires infinite order in the moments. 
We again use an Ansatz $\vomega_p=\sum_{m=0}^\infty \tau_{\rm c}^m\,\vomega^{(m)}_p$. This yields
\begin{align}
\label{eq:log_normal_system_orders}
\frac{\!\id \vomega^{(m)}_{p}}{\id \gamma} &= \vek{A}_{\Gammab} \,\vomega^{(m-1)}_{p} - \sum_{t=0}^\infty\frac{\vek{B}\vomega^{(m-1)}_{p+t}}{t!} - (1-\delta_{k0})\Lambda_{p}\vek{B}\vXb
    \nonumber \\
    &\qquad\qquad -p\left(\vomega^{(m)}_{p} +\frac{\sigmae^2}{2}\vomega^{(m)}_{p-1}\right)+p(p-1)\,\sigmae^2\,\vomega^{(m)}_{p-2},
    \nonumber 
\end{align}
where we have inserted $\mu=-\sigmae^2/2$ as required. Although $\Lambda_{2m}=0$, due to the terms $\sum_{t=0}^\infty\vek{B}\vomega^{(m-1)}_{p+t}/t!$ we now also obtain contributions with odd powers of $\vek{B}$. The intermediate expressions are not very illuminating. In the end, we find
\begin{align}
\frac{\Delta \vek{\mu}_{0}(\eta)}{\Delta \eta}&\approx 
\frac{\Gammab \tauc (\expf{\sigmae^2}-1)\,\expf{-\frac{\sigmae^2}{4}}}{(1-\tauc)}\exp\left[-\tauc\sigmae^2(2\vek{B}+\tauc\,\vek{B}^2)\right]\vek{B}^2\,\vXb,
\end{align}
to capture the leading order dependencies. We derived corrections up to $12^{\rm th}$ order in $\tauc$, but were unable to find a general pattern beyond the expression given above. For our studies, we shall use the simplified expression, even if it is less clear than for Gaussian $\deltae$ field. For small $\sigmae$ and $\tauc$ both approaches agree. We also note that to fix the variance $\langle \deltae^2 \rangle$ to the same as in the Gaussian approach we can use $\sigmae^2=\ln(1+\langle \deltae^2 \rangle)$ for the log-normal case.

\subsection{System for $F=\vek{X}\vek{X}^T$}
So far we have dropped the term $\left<\delta \vek{X}\cdot \delta \vek{X}\right>$ in Eq.~\eqref{eq:gen_aver_ALL}. In reality one expects a correction that is similar to the one from $\bar{\vek{X}}\cdot \left<\delta \vek{X}\right>$; however, it is not as easy to solve for this contribution. As Eq.~\eqref{eq:gen_aver_first_2_hard} indicates, we essentially need the full transfer function library for sourcing of perturbations $\propto \vek{B} \vXb$ from any redshift. This becomes a computational challenge, which we defer. 

However, we can already write down a more general hierarchy using It\^o calculus. Instead of solving for $\left<\delta \vek{X}\cdot \delta \vek{X}\right>$, we can directly find a system to solve for $\vek{X}\cdot\vek{X}$. Attempting $F_p=\deltae^p\,\vek{X}\cdot\vek{X}$ quickly reveals that it is hard to close the system. However, by considering the much larger problem for $F_p=\deltae^p\,\vek{X}\vek{X}^T$ (i.e., the full correlation matrix of all observables), we can show that
\begin{align}
\label{eq:final_system_II}
    \frac{\!\id \!\left<\vek{X}\vek{X}^T\right>}{\id \eta} &=2\vek{M}\left<\vek{X}\vek{X}^T\right>+2\Gammab\vek{B}\left<\deltae \vek{X}\vek{X}^T\right>,
\\\nonumber
\frac{\!\id \!\left<\deltae\vek{X}\vek{X}^T\right>}{\id \eta} &=2\vek{M}\left<\deltae \vek{X}\vek{X}^T\right>+2\Gammab\vek{B}\left<\deltae^2 \vek{X}\vek{X}^T\right>-\alpha \left<\deltae \vek{X} \vek{X}^T\right>,
\\\nonumber
\frac{\!\id \!\left<\deltae^2\vek{X}\vek{X}^T\right>}{\id \eta} &=2\vek{M}\left<\deltae^2 \vek{X} \vek{X}^T\right>-2\alpha \left<\deltae^2 \vek{X}\vek{X}^T\right>+\sigma^2\left<\vek{X}\vek{X}^T\right>.
\end{align}
Computationally, this is a significantly larger problem as it involves a system of $N\times N$ equations, where $N$ is the single transfer problem, which can become challenging. 

We can further simply the problem by carrying out the short time-scale average. Following the same steps as before, we have
\begin{align}
\label{eq:final_system_small_Deta_sol_II}
\left<\vek{X}\vek{X}^T\right> &\approx 
\bar{\vek{X}} \bar{\vek{X}}^T+
\frac{\Gammab^2 \vek{B}^2\bar{\vek{X}} \bar{\vek{X}}^T\sigma^2}{\alpha^2}\left[1-\frac{3}{2\xi}+\frac{2\expf{-\xi}}{\xi}-\frac{\expf{-2\xi}}{2\xi} \right]\Delta \eta,
\nonumber \\
\left<\deltae\vek{X}\vek{X}^T\right>&\approx
\left<\deltae\vek{X}\vek{X}^T\right>_0\expf{-\xi}
+
\frac{\Gammab \vek{B}\bar{\vek{X}} \bar{\vek{X}}^T\sigma^2}{\alpha^2}\left[1-\expf{-\xi}\right]^2,
\\ \nonumber
\left<\deltae^2\vek{X} \vek{X}^T\right>&\approx \left<\deltae^2\vek{X}\vek{X}^T\right>_0\expf{-2\xi}
+\frac{\bar{\vek{X}}\bar{\vek{X}}^T\!\sigma^2}{2\alpha}\left[1-\expf{-2 \xi}\right],
\end{align}
such that $\Delta \left<\vek{X}\vek{X}^T\right>\approx \frac{\Gammab^2 \sigma^2}{\alpha^2}\,\Delta \eta\,\vek{B}^2\left<\bar{\vek{X}} \bar{\vek{X}}^T\right>$. 

Instead of solving the coupled system for various moments, Eq.~\eqref{eq:final_system_II}, we could therefore also simply solve the reduced problem
\begin{align}
\label{eq:final_system_simplified}
\frac{\!\id \!\left<\vek{X}\vek{X}^T\right>}{\id \eta} &\approx 2\vek{M}\left<\vek{X}\vek{X}^T\right>+
2 \Gammab \zetae \,\vek{B}^2 \left<\vek{X}\vek{X}^T\right>
\end{align}
to capture all the corrections in one go. The final power spectrum contributions are then given by $\left<\vek{X}\cdot\vek{X}\right>={\rm Tr}(\vek{X} \vek{X}^T)$. However, here we will only consider the corrections caused by $\vXb\cdot \left<\delta \vek{X}\right>$. We also note that higher order terms may have to be considered in a more complete approach.

\subsection{Corrected Boltzmann hierarchy and line-of-sight approach}
\label{sec:Boltz_simp}
We are now nearly done with connecting the effects back to the computation of the CMB power spectra. Evaluating Eq.~\eqref{eq:final_term_Gaussian}, we find the modified evolution equations 
\begin{align}
\label{eq:equations_sim}
\partial_\eta \varv_{\rm b}&= k\Psi-\mathcal{H} \varv_{\rm b}
+\frac{3\bar{\Gamma}}{R}\left\{1-f_1\left(\tauc, \sigmae\right)\right\} \left[\Theta_1-\frac{\varv_{\rm b}}{3}\right],
\\ \nonumber
\partial_\eta \Theta_1&=\frac{k}{3}\Theta_0-\frac{2 k}{3}\Theta_2+\frac{k}{3}\Psi-\bar{\Gamma}\left\{1-f_1\left(\tauc, \sigmae\right)\right\} \left[\Theta_1-\frac{\varv_{\rm b}}{3}\right],
\\ \nonumber
\partial_\eta \Theta_2&=\frac{2k}{5}\Theta_1-\frac{3 k}{5}\Theta_3-
\frac{9}{10}\bar{\Gamma}\left\{1-f_2\left(\tauc, \sigmae \right)\right\}\,\Theta_2,
\\ \nonumber
\partial_\eta \Theta_{\ell\geq 3}&=\frac{k \ell}{2\ell+1}\Theta_{\ell -1}-\frac{k (\ell+1)}{2\ell+1}\Theta_{\ell +1}-\bar{\Gamma}\left\{1-f_3\left(\tauc, \sigmae\right)\right\} \Theta_\ell,
\end{align}
with $\tauc=\Gammab/\alpha$ and the functions
\bsub
\label{eq:equations_sim_G}
\begin{align}
\label{eq:equations_sim_Ga}
f^{\rm G}_{\rm e}\left(\tau_{*}, \sigmae\right)&=\tau_{*} \, \sigmae^2
\exp\left(-\tau_{*}^2\,\sigmae^2\right),
\\
\label{eq:equations_sim_Gb}
f_1\left(\tauc, \sigmae\right)
&=f^{\rm G}_{\rm e}\left(\frac{1+R}{R}\tauc, \sigmae\right),
\\
f_2\left(\tauc, \sigmae\right)
&=f^{\rm G}_{\rm e}\left(\frac{9}{10}\tauc, \sigmae\right),
\\
f_3\left(\tauc, \sigmae\right)&=f^{\rm G}_{\rm e}\left(\tauc, \sigmae\right),
\end{align}
\esub
valid for Gaussian $\deltae$ field. Instead for a log-normal $\deltae$ field, we just have to replace $f^{\rm G}_{\rm e}$ with
\begin{align}
\label{eq:equations_sim_LN}
f^{\rm LN}_{\rm e}\left(\tau_{*}, \sigmae\right)&=
\tau_{*}\, (\expf{\sigmae^2}-1)\,\expf{-\frac{\sigmae^2}{4}}
\exp\left(-(2+\tau_{*})\, \tau_{*} \,\sigmae^2\right)
\end{align}
in the definitions of $f_1$, $f_2$ and $f_3$.
We confirmed these expressions up to $\sigmae \simeq 2-3$ but expect them to give reasonable results even for higher values. We also note that the expressions for the Gaussian limit in principle are only valid for $\sigmae<1$. Thinking of $\sigmae$ as a phenomenological parameter, both approaches should allow to gain some physical insight into the problem. However, we will demand that all $f_i<1$ to avoid unphysical growth of perturbations.

The new system can be readily solved using the standard solver with a model for $\sigmae^2$ and $\tauc$. We note that the corrections {\it do not} simply appear as a correction to the visibility function in the naive sense, as one cannot capture the effects by simply rescaling $\bar{\Gamma}$. In particular, the leading order term $(1-\zeta_{\rm e})$ is not sufficient. 
Defining $\Gamma_{\rm e}=\bar{\Gamma}\left\{1-f_3(\tauc, \sigmae)\right\}$, we can rewrite Eq.~\eqref{eq:Theta_evol_start_F} as
\begin{align}
\label{eq:Theta_evol_start_F_corrected}
\frac{\partial \Theta}{\partial \eta}+\iim\,k\,\chi\,\Theta
&=-\frac{\partial \Phi}{\partial \eta}-\iim\,k\,\chi\,\Psi-\Gamma_{\rm e}\left[\Theta-\Theta_0-\frac{\Theta_2}{10}\,-\chi \,\varv_{\rm b}
\right]\nonumber\\
&\!\!\!\!\!\!\!
+\Gammab\,
\left[f_1-f_3
\right]
\left(\Theta_1-\chi \,\varv_{\rm b}\right)
+ \frac{9}{10}\Gammab\,
\left[f_2-f_3\right]
\Theta_2.
\end{align}
Note that here the angle dependence is still present in all variables, i.e., $\Theta_\ell=\Theta_\ell(\chi)\propto (2\ell + 1) P_\ell(\chi)$, even if we do not distinguish the functions. This shows that we have new photon source terms in addition to visibility function and transfer function corrections. Collecting terms, this leads to the modified line of sight solution 
\begin{align}
\label{eq:Theta_los}
\Theta_\ell
&=\int_0^{\eta_0} 
\id\eta \, g_{\rm e}(\eta)\,\Bigg\{
\left(
\Theta_0+\Psi
\right)\,j_\ell+ \varv_{\rm b}
j^{(1, 0)}_\ell+ \frac{\Theta_2}{2} j^{(2, 0)}_\ell
\Bigg\}
\\ \nonumber 
&\qquad
+\int_0^{\eta_0}\id\eta \, f_{\rm e}(\eta)\,\Bigg\{
\left(
\frac{\partial \Psi}{\partial \eta}-\frac{\partial \Phi}{\partial \eta}+
3\Gammab  \Theta_{1} 
\left[f_1-f_3 \right]
\right)
\,j_\ell
\\ \nonumber 
&\qquad\qquad
- 
\Gammab \varv_{\rm b} 
\left[f_1-f_3\right]
j^{(1, 0)}_\ell
+ \frac{9}{2}\Gammab \Theta_2
\left[f_2-f_3\right] j^{(2, 0)}_\ell
\Bigg\},
\end{align}
where the spherical Bessel functions $j_\ell(x)$ have argument $x=k \Delta \eta$ with $\Delta \eta=\eta_0-\eta$ and we introduced the functions $j^{(1, 0)}_\ell(x)=\partial_x j_\ell(x)$ and $j^{(2, 0)}_\ell(x)=\frac{1}{2}\left[3 \partial^2_x j_\ell(x)+j_\ell(x)\right]$. The modified visibility function, $g_{\rm e}=\Gamma_{\rm e} \expf{-\tau_{\rm e}}$, and $f_{\rm e}=\expf{-\tau_{\rm e}}$ are furthermore evaluated using $\partial_\eta \tau_{\rm e} =\Gamma_{\rm e}$.

\vspace{-2mm}
\subsubsection{Adding polarization terms}
\label{sec:polarization}
So far, we have neglected the effect of polarization. Extending the matrix $\vek{B}$ to account for those yields the corresponding modifications to the system. For the temperature multipoles, only the equation for the quadrupole is modified to become \citep[e.g.,][]{Seljak1996, Hu1997}
\bsub
\begin{align}
\label{eq:equations_sim_Theta_2}
\partial_\eta \Theta_2&=\frac{2k}{5}\Theta_1-\frac{3 k}{5}\Theta_3
-\bar{\Gamma}\left\{1-f_3\right\}\,\Theta_2
\nonumber\\
&\qquad\qquad\qquad 
+\frac{\bar{\Gamma}}{10}\!\left\{1-f^{\rm P}_2\right\}[\Theta_2+\Theta^{\rm P}_0+\Theta^{\rm P}_2],
\\
f_2^{\rm P}\left(\tauc, \sigmae\right)&=\frac{10}{7}f^{S}_{\rm e}\left(\tauc, \sigmae\right)-\frac{3}{7}
f^{S}_{\rm e}\left(\frac{3}{10}\tauc, \sigmae\right),
\end{align}
\esub
where one can again use the functions, $f^{S}_{\rm e}$, for the Gaussian or log-normal driving scenarios. 

Similarly, for the polarization hierarchy we have
\begin{align}
\label{eq:equations_sim_Theta_P}
\partial_\eta \Theta^{\rm P}_0&=-k \Theta^{\rm P}_1-\bar{\Gamma}\left\{1-f_3\right\} \Theta^{\rm P}_{0}
+\frac{\bar{\Gamma}}{2}\!\left\{1-f^{\rm P}_2\right\}[\Theta_2+\Theta^{\rm P}_0+\Theta^{\rm P}_2],
\nonumber\\
\partial_\eta \Theta^{\rm P}_1&=\frac{k}{3}\Theta^{\rm P}_0-\frac{2 k}{3}\Theta^{\rm P}_2-\bar{\Gamma}\left\{1-f_3\right\} \Theta^{\rm P}_{1},
\nonumber\\
\partial_\eta \Theta^{\rm P}_2&=
\frac{2k}{5}\Theta^{\rm P}_1-\frac{3 k}{5}\Theta^{\rm P}_3
-\bar{\Gamma}\left\{1-f_3\right\}\,\Theta^{\rm P}_2
\nonumber\\
&\qquad\qquad\qquad 
+\frac{\bar{\Gamma}}{10}\!\left\{1-f^{\rm P}_2\right\}[\Theta_2+\Theta^{\rm P}_0+\Theta^{\rm P}_2],
\nonumber\\
\partial_\eta \Theta^{\rm P}_{\ell\geq 3}&=\frac{k \ell}{2\ell+1}\Theta^{\rm P}_{\ell -1}-\frac{k (\ell+1)}{2\ell+1}\Theta^{\rm P}_{\ell +1}-\bar{\Gamma}\left\{1-f_3\right\}\Theta^{\rm P}_{\ell},
\end{align}
which summarizes all the new terms. Setting all the $f_i=0$ reduces the system to the usual unperturbed case.

To derive the terms for the line-of-sight approach, we write the modified transfer equation with polarization terms
\begin{align}
\label{eq:Theta_evol_start_F_corrected_P}
&\frac{\partial \Theta}{\partial \eta}+\iim\,k\,\chi\,\Theta
=-\frac{\partial \Phi}{\partial \eta}-\iim k \chi\Psi-\Gamma_{\rm e}\left[\Theta-\Theta_0-\frac{\Theta_2+\Theta_0^{\rm P}+\Theta_2^{\rm P}}{10}-\chi\varv_{\rm b}
\right]\nonumber\\
&\;\;\qquad 
+\Gammab\,
\left[f_1-f_3
\right]
\left(\Theta_1-\chi\,\varv_{\rm b}\right)
- \Gammab\,
\left[f^{\rm P}_2-f_3\right]\frac{\Theta_2+\Theta_0^{\rm P}+\Theta_2^{\rm P}}{10}.
\end{align}
Comparing with Eq.~\eqref{eq:Theta_evol_start_F_corrected} and Eq.~\eqref{eq:Theta_los} then yields
\begin{align}
\label{eq:Theta_los_P}
\Theta_\ell
&=\int_0^{\eta_0} 
\id\eta \, g_{\rm e}(\eta)\,\Bigg\{
\left(
\Theta_0+\Psi
\right)\,j_\ell+ \varv_{\rm b}
j^{(1, 0)}_\ell+ \frac{\Theta_2+\Theta_0^{\rm P}+\Theta_2^{\rm P}}{2} j^{(2, 0)}_\ell
\Bigg\}
 \nonumber  \\ 
&\qquad
+\int_0^{\eta_0}\id\eta \, f_{\rm e}(\eta)\,\Bigg\{
\left(
\frac{\partial \Psi}{\partial \eta}-\frac{\partial \Phi}{\partial \eta}+
3\Gammab  \Theta_{1} 
\left[f_1-f_3 \right]
\right)
\,j_\ell
\\ \nonumber 
&\qquad\quad
- 
\Gammab \varv_{\rm b} 
\left[f_1-f_3\right]
j^{(1, 0)}_\ell
-\frac{\Gammab}{2} \left[\Theta_2+\Theta_0^{\rm P}+\Theta_2^{\rm P}\right]
\left[f^{\rm P}_2-f_3\right] j^{(2, 0)}_\ell
\Bigg\}
\end{align}
for the line of sight solution. 

Similarly, for the $E$-mode polarization terms we have 
\begin{align}
\label{eq:Theta_los_P_E}
\Theta^{E}_\ell
&=\int_0^{\eta_0} 
\id\eta \left\{g_{\rm e}-f_{\rm e}\Gammab (f_2^{\rm P}-f_3)\right\}\frac{\sqrt{6}(\Theta_2+\Theta_0^{\rm P}+\Theta_2^{\rm P})}{2} j^{(2, 2)}_\ell
\end{align}
with $j^{(2, 2)}_\ell(x)=\sqrt{\frac{3}{8}\frac{(\ell+2)!}{(\ell-2)!}}\,\frac{j_\ell(x)}{x^2}$. Comparing to \citet{Hu1997} we find $P^{(0)}_2=(\Theta_2+\Theta_0^{\rm P}+\Theta_2^{\rm P})/10$ and $E_\ell=-\sqrt{6}(2\ell + 1)\Theta^{\rm P}_\ell$. With these definitions we can reproduce the results of {\tt CLASS} to high precision using {\tt CosmoTherm} \citep{kite_spectro-spatial_2023-III}.

\vspace{-2mm}
\subsection{Tight coupling limit}
In the standard tight coupling limit, one can show that well inside the horizon the monopole follows the equation \citep[e.g.,][]{BaumannBook}
\begin{align}
\label{eq:equations_sim_TC}
\partial^2_\eta \Theta_0
+\frac{k^2 c_{\rm s}^2}{\Gammab}\left[\alpha_2+\frac{R^2}{1+R}\right]
\partial_\eta\Theta_0+ k^2 c_{\rm s}^2\Theta_0&\approx 0
\end{align}
with photon sound speed $c_{\rm s}^2=1/[3(1+R)]$ and $\alpha_2=8/9$ without and $\alpha_2=16/15$ with polarization terms \citep{Weinberg1971, Kaiser1983}.\footnote{We derive a more general tight-coupling equation in Appendix~\ref{app:TC_derivation} that also includes the potential sources and Hubble frictions terms, however, these are not relevant to the following discussion.}

To incorporate the corrections from the electron fluctuations, we can recognize that the term $\alpha_2$ stems from the tight coupling solution of the photon quadrupole. With the scattering corrections included, this means $\alpha_2\rightarrow \alpha_2/(1-f_2)$ without polarization effects included.\footnote{The case with polarization terms has a more complicated dependence on $f_2^{\rm P}$ and $f_3$ which is omitted here.} Similarly, the term $\propto R^2/(1+R)$ stems from the tight coupling limit of the photon dipole and thus is modified to $R^2/(1+R)\rightarrow R^2/[(1+R)(1-f_1)]$. Put together, this yields the modified photon diffusion scale (without polarization corrections)
\begin{align}
\label{eq:equations_sim_TC_kD}
\frac{1}{k^2_{\rm D}}\approx 
\int_0^\eta\,\frac{\id \eta'}{
6(1+R)\Gammab}
\left[\frac{8}{9(1-f_2)}+\frac{R^2}{(1+R)(1-f_1)}\right],
\end{align}
meaning that $k^{-2}_{\rm D}$ is expected to increase, implying {\it more} damping at a fixed value of $k$. We also note that at the relative contributions from shear viscosity and heat conduction are modified differently.

\vspace{-2mm}
\section{Recombination history with small-scale baryon density fluctuations}
\label{sec:LN_delta_b_fluctuations}
After having developed the framework to include electron density fluctuations in the computation of the CMB anisotropies, we can start computing the effects explicitly. To build a more realistic model, we first study how the electron recombination history is modified in the presence of baryon density fluctuation. For this, we compute the recombination problem in a separate Universe approach using the {\tt CosmoRec} module of {\tt CosmoSpec} \citep{Chluba2010b, Chluba2016CosmoSpec}, for a modified baryon density, $N_{\rm b}=\Fb\,\bar{N}_{\rm b}$. We only change the baryon density locally, but assume that the modification is time-independent and also does not affect the average expansion history. In the perturbative limit, this means that the average recombination history can be obtained using 
\begin{align}
    \label{eq:av_Ne_Fb}
\Ne(z, \Fb)&= N^{\rm st}_{\rm e}(z)+N^{\rm st, (1)}_{\rm e}(z)\, \Delta \Fb+\frac{1}{2}\,N^{\rm st, (2)}_{\rm e}(z) \,\Delta \Fb^2
\nonumber\\
&\qquad
+\frac{1}{6}\,N^{\rm st, (3)}_{\rm e}(z) \,\Delta \Fb^3
+\frac{1}{24}\,N^{\rm st, (4)}_{\rm e}(z) \,\Delta \Fb^4+\ldots
\end{align}
with $N^{\rm st, (k)}_{\rm e}(z)=\partial^k_{\Fb}\Ne(z, \Fb)\big|_{\Fb=1}$ and $\Delta F_{\rm b}=\Fb-1$. Here, we expanded around the standard ionization history, $N^{\rm st}_{\rm e}(z)$, with $\Fb=1$. We demand that $\big< F_{\rm b} \big>=1$, by construction. This means that upon ensemble averaging we have the moment expansion
\begin{align}
    \label{eq:av_Ne}
\big< \Ne\big>&=N^{\rm st}_{\rm e}(z)+\frac{1}{2}\,N^{\rm st, (2)}_{\rm e}(z) \,\big<\delta_{\rm b}^2 \big>
\nonumber \\
&\qquad \qquad +\frac{1}{6}\,N^{\rm st, (3)}_{\rm e}(z) \,\big<\delta_{\rm b}^3 \big>+\frac{1}{24}\,N^{\rm st, (4)}_{\rm e}(z) \,\big<\delta_{\rm b}^4 \big>+\ldots
\end{align}
with $\Delta F_{\rm b}=\delta_{\rm b}$. The leading correction arises due to the second moment of the $\delta_{\rm b}$-field. However, due to the non-linear nature of the recombination process higher order moments contribute significantly. 

To demonstrate the main effects, we build a library of recombination histories in the range $F_{\rm b}=[10^{-3}, 10^3]$. We assume that $F_{\rm b}=1+\delta_{\rm b}$ follows a log-normal distribution with fixed $\sigma_{\rm b}$:
\begin{align}
    \label{eq:delta_b}
P(\delta_{\rm b}, \sigma_{\rm b})&=\frac{1}{\sqrt{2\pi}\sigma_{\rm b} (1+\delta_{\rm b})}\exp\left[-\frac{[\ln (1+\delta_{\rm b})+\sigma_{\rm b}^2/2]^2}{2\sigma_{\rm b}^2}\right].
\end{align}
The distribution is normalized as $\int P(\delta_{\rm b}, \sigma_{\rm b}) \id \delta_{\rm b}=1$ and has zero mean, $\int P(\delta_{\rm b}, \sigma_{\rm b}) \, \delta_{\rm b} \id \delta_{\rm b}=0$. Using the library of recombination histories, we can then specify a model for $\sigma_{\rm b}(z)$ to mimic the effects of evolving density fluctuation, and compute $\big<\Ne\big>$ using an equation similar to Eq.~\eqref{eq:Ne_av_def_Pe}:
\begin{align}
\label{eq:Ne_av_def_Pb}
\Neb(z)&=\int \Ne(z, F_{\rm b}) \, P(\delta_{\rm b}, \sigma_{\rm b}) \id \delta_{\rm b},
\end{align}
where $\sigma_{\rm b}$ can depend on redshift.
This can then be used to formulate an approximate version in terms of $\deltae$, as we show below.

\begin{figure}
    \centering
    \includegraphics[width=\columnwidth]{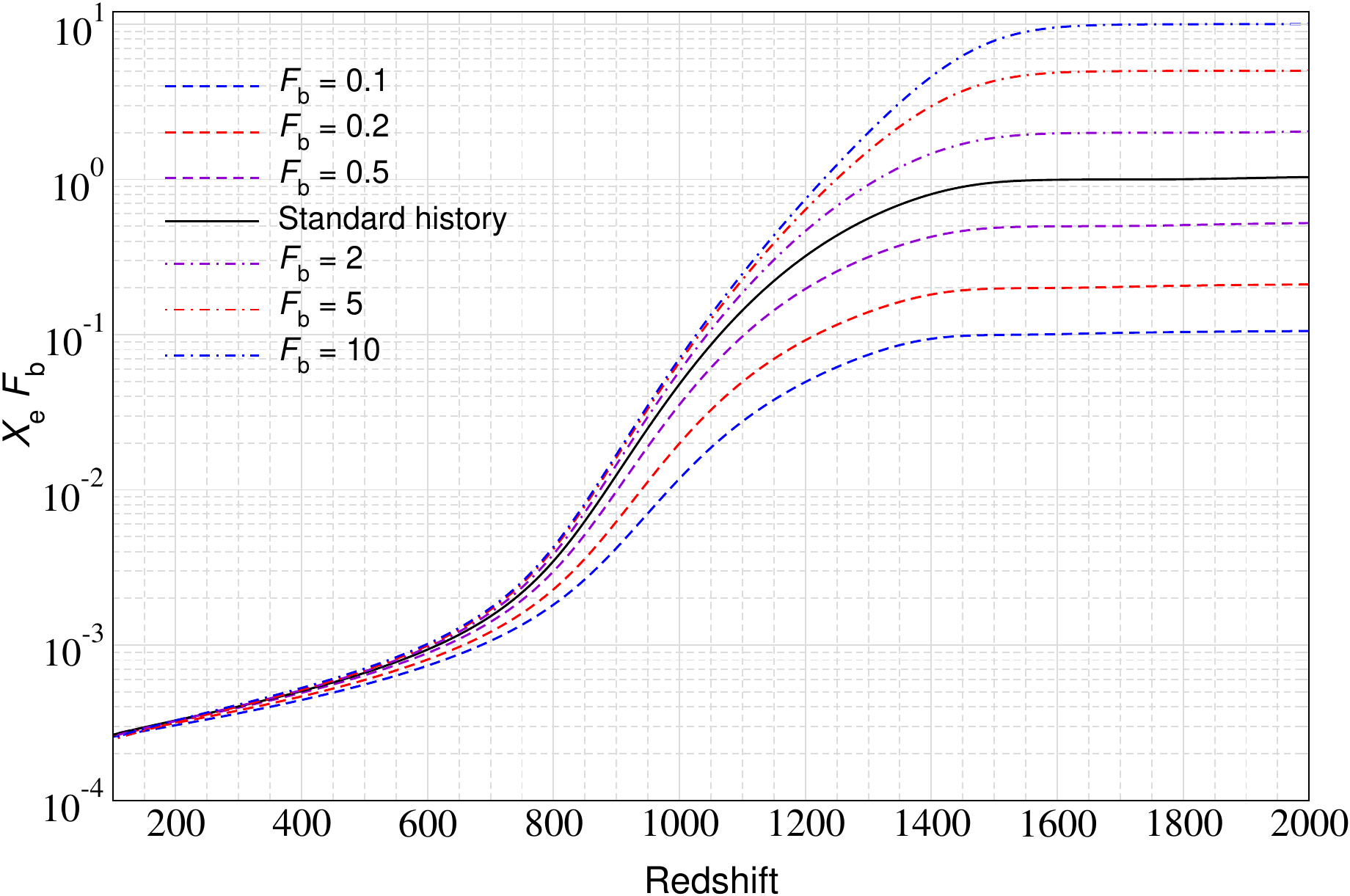}
    \caption{Dependence of the recombination history on $\Fb$. The weighted free-electron fraction is shown $\Xe^*=\Xe \Fb$. At each redshift, the electron faction can be averaged over a distribution $P(\delta_{\rm b}, \sigma_{\rm b})$ to obtain the average recombination history.}
    \label{fig:Fb_Xe}
\end{figure}
In Fig.~\ref{fig:Fb_Xe}, we illustrate the dependence of the recombination history on $\Fb$. The main effect manifests itself in a shifting and stretching of the ionization history in redshift, with the Universe recombining later for $\Fb<1$, and earlier for $\Fb> 1$. It is interesting to note that in all cases, the recombination process ends with roughly the same freeze-out fraction. This means that even for large fluctuations in $F_{\rm b}$, the responses in $\deltae$ are expected to be marginal, i.e., $\deltae\simeq 0$, at $z\lesssim 500$. On the other hand, in the pre-recombination era we can see that $\deltae\simeq \delta_{\rm b}$ As expected. We can, therefore, obtain a rough mapping between $\deltae$ and $\delta_{\rm b}$ (see Sect.~\ref{sec:deltae_mapping}).

\begin{figure}
    \centering
    \includegraphics[width=\columnwidth]{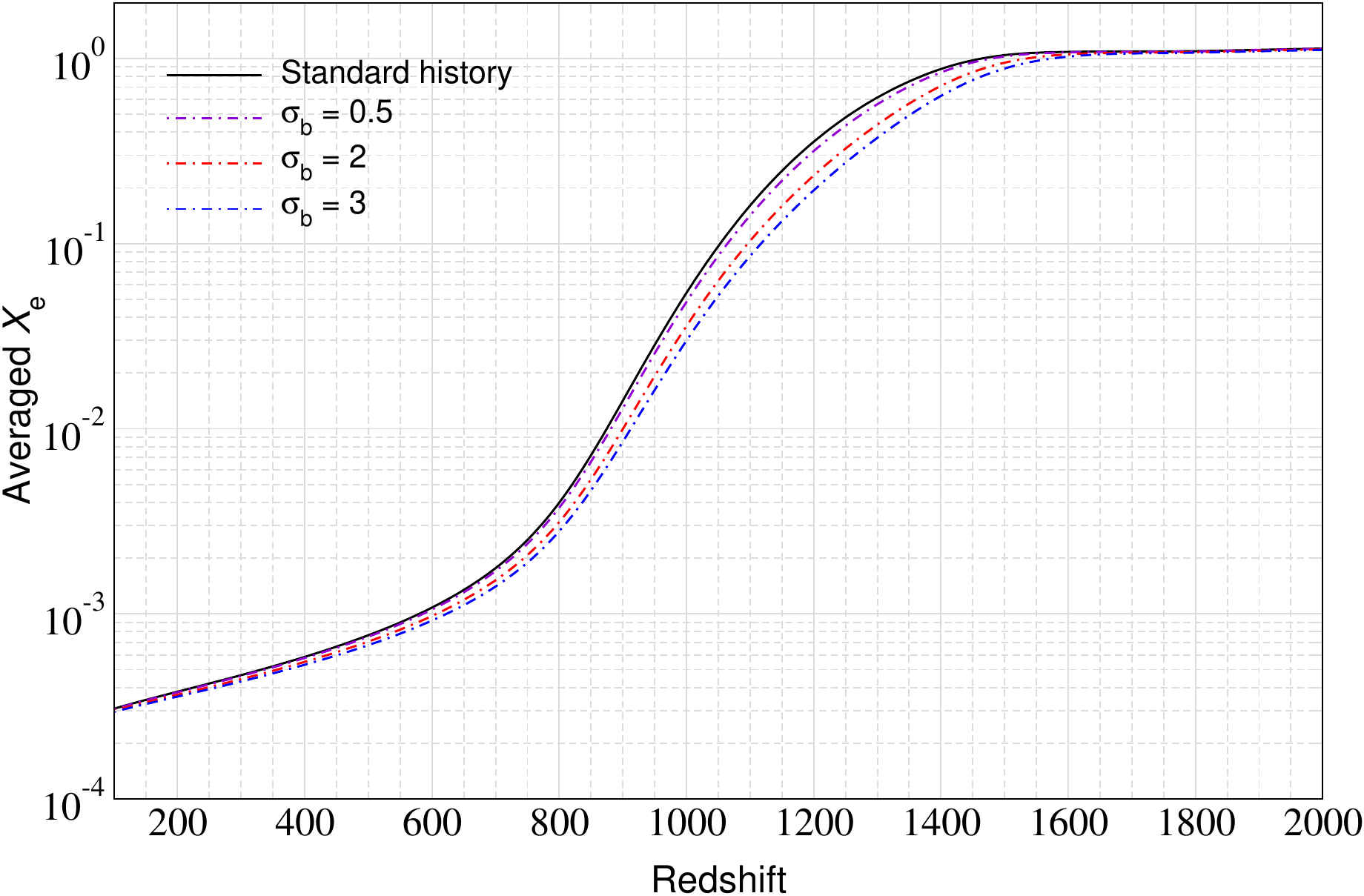}
    \\[2mm]
    \includegraphics[width=\columnwidth]{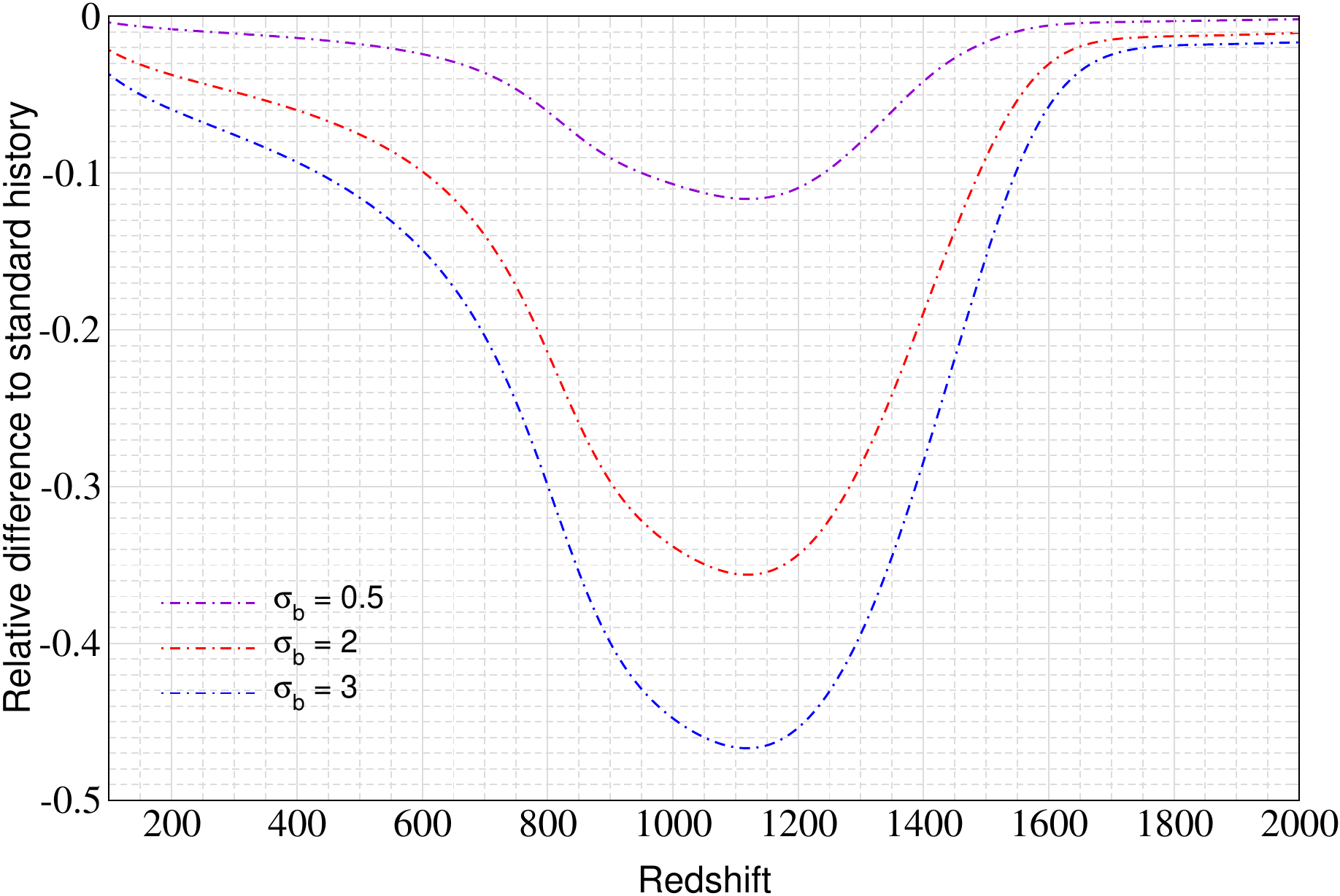}
    \caption{Average recombination history (upper panel) and relative difference (lower panel) for various values of $\sigma_{\rm b}$. Baryon density perturbations lead to an acceleration of recombination around $z\simeq 1100$.}
    \label{fig:Av_Xe}
\end{figure}
In Fig.~\ref{fig:Av_Xe}, we illustrate the average recombination history and the relative difference with respect to the standard history for various choices of $\sigma_{\rm b}$. The recombination process is on average accelerated in the presence of baryon fluctuations. However, due to the differences in the responses of $\Ne$, both the late and early time solutions are affected less. At $z\gg 1100$, one has $\deltae \approx \delta_{\rm b}$, which implies on average there is no effect. However, the variance $\big<\deltae^2\big> \approx \big<\delta_{\rm b}^2\big>$ is largest at early times, while it becomes very small at late times for this particular set of parameters, as we illustrate now.

\subsection{Simple mapping between $\delta_{\rm b}$ and $\deltae$}
\label{sec:deltae_mapping}
Given $P(\delta_{\rm b}, \sigma_{\rm b})$, what can we say about $P(\deltae)$ and its moments? For our setup, we can already anticipate that even for constant $\sigma_{\rm b}$, at late times, the corresponding electron response is very small, consistent with $\big<\deltae^2\big>\ll 1$. In contrast, at early times one will have $P(\deltae) \approx P(\delta_{\rm b}, \sigma_{\rm b})$, since $\deltae\approx \delta_{\rm b}$ for a fully ionized medium (cf. Fig.~\ref{fig:Fb_Xe}). We can quantify the mapping from $\delta_{\rm b}$ to $\deltae$ by simply plotting a histogram of $P(\delta_{\rm b}, \sigma_{\rm b})\,\Fb$ as a function of $1+\deltae$.
\begin{figure}
\centering
\includegraphics[width=\columnwidth]{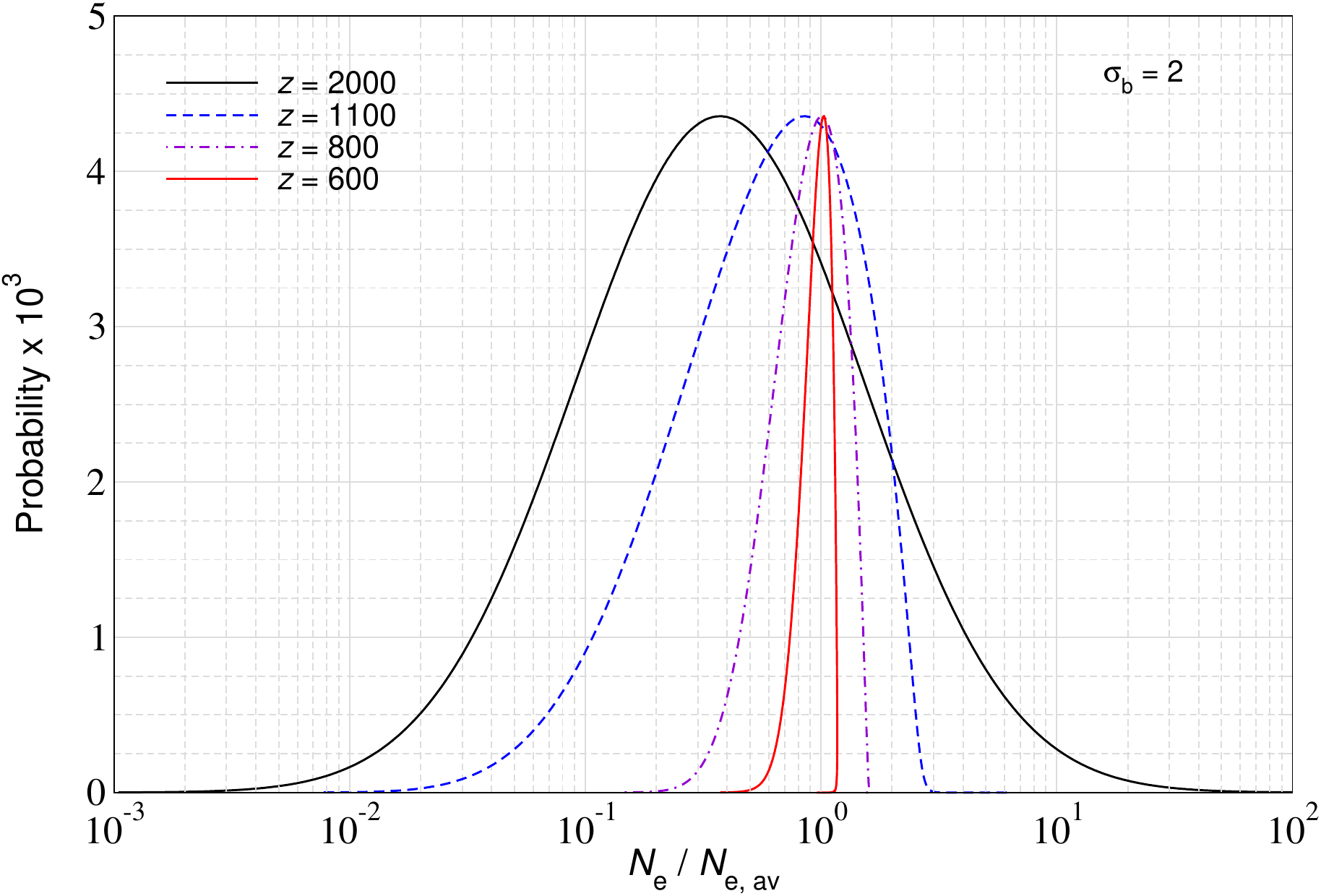}
\caption{Electron density probability distribution at various redshifts and fixed baryon density distribution. Note that here $\Ne/N_{\rm e, av}$ is essentially $1+\deltae$.}
\label{fig:PDF}
\end{figure}
The result is shown in Fig.~\ref{fig:PDF} for fixed baryon density distribution. At high redshifts one essentially recovers the distribution for $\Fb$, while at lower redshifts the distribution becomes much narrower even if the baryon density distribution has the same variance. This is caused by the smaller variation of the $\Ne$ response for a given change in $\Fb$. We note, however, that our modeling of the $\Ne$ responses in the separate Universe approach is extremely simplistic and one can expect much larger variations for more realistic models. Nevertheless, these insights can guide our explorations in the context of CMB anisotropies and their modification due to electron density fluctuations.

\begin{figure}
\centering
\includegraphics[width=\columnwidth]{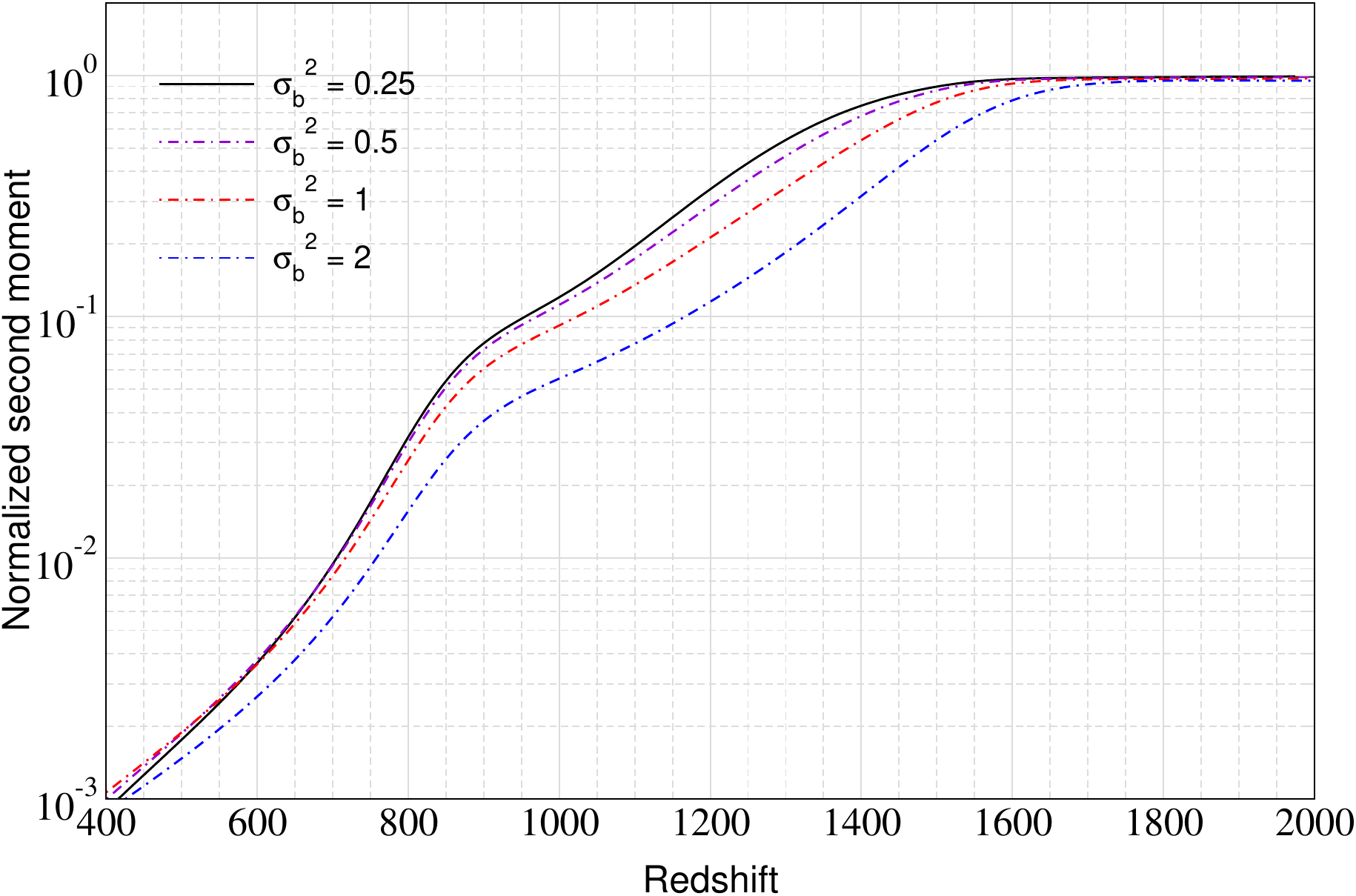}
\caption{Normalized electron density moment, $\big<\deltae^2\big>/[\expf{\sigma_{\rm b}^2}-1]$ as a function of redshift and for various values of $\sigma_{\rm b}$.}
\label{fig:Moment2}
\end{figure}
To further explore the properties of $P(\deltae)$ given the baryon density fluctuations, we consider the moments, $\big<\deltae^m\big>$. At each redshift, these can be directly computed given the model for $P(\delta_{\rm b}, \sigma_{\rm b})$. Assuming, the log-normal distribution, we have $\big<\delta_{\rm b}^2\big>=\expf{\sigma_{\rm b}^2}-1$. Normalizing $\big<\deltae^2\big>$ by this, we find the redshift evolution as given in Fig.~\ref{fig:Moment2}. Early on, $\big<\deltae^2\big>\simeq \big<\delta_{\rm b}^2\big>$ as expected, while around and after recombination the electron density responses reduce as also anticipated from the discussion surrounding Fig.~\ref{fig:PDF}.

\begin{figure}
\centering
\includegraphics[width=\columnwidth]{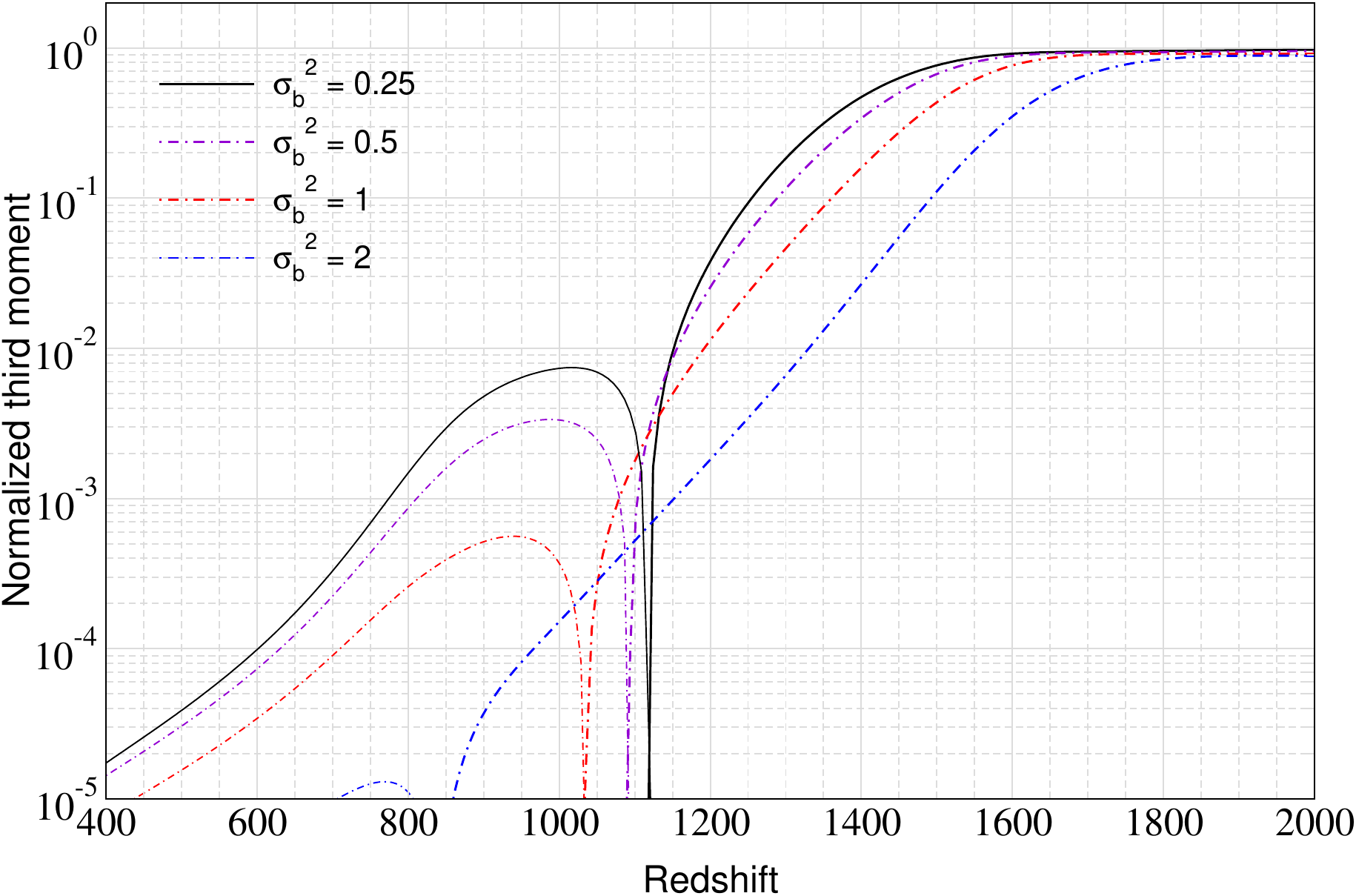}
\caption{Normalized electron density moment, $\big<\deltae^3\big>/[\expf{3\sigma_{\rm b}^2}-3\expf{\sigma_{\rm b}^2}+2]$ as a function of redshift and for various values of $\sigma_{\rm b}$. The thinner line style indicates negative moments.}
\label{fig:Moment3}
\end{figure}
In Fig.~\ref{fig:Moment3}, we also illustrate the dependence of $\big<\deltae^3\big>$ on redshift. This time we normalize by $\big<\delta_{\rm b}^3\big>=\expf{3\sigma_{\rm b}^2}-3\expf{\sigma_{\rm b}^2}+2$. For the electrons, the third moment even changes sign at late times due to non-linearities in the response function. We checked that at early times, we can roughly reproduce the scaling of the third moment by simply using a log-normal distribution for $\deltae$ with the second moment as computed from the full distribution. 

Even if not exact, the setup described here provides a leading order model that we shall use below. In the context of the Hubble tension, baryon variance at the level of $\sigma_{\rm b}^2\simeq 0.2-0.5$ was considered in previous works \citep[e.g.,][]{Thiele2021HT, Galli:2021mxk}, although only the effects on the average recombination history were included.

\section{Modifications to the photon transfer functions and CMB power spectra}
\label{sec:effects_T_Cell}
For $\deltae=0$, the coupled system of ordinary differential equations can be solve numerically using standard Boltzmann solver such as {\tt CAMB} \citep{CAMB} or {\tt CLASS} \citep{CLASSCODE}. Below we implement a modified system that allows us to capture the corrections from the electron fluctuations numerically. We will use the anisotropy module of {\tt CosmoTherm} for these computations \citep{Chluba2011therm, kite_spectro-spatial_2023-III}. The calculations carried out here are all meant to illustrate the main effects whereas a detailed parameter analysis using data will be carried out separately.

The main parameters of the model are $\tauc$ and $\sigmae^2$, which generally can be functions of $\eta$. We will implement the simplified Boltzmann hierarchy (BH) discussed in Sect.~\ref{sec:Boltz_simp} and also illustrate the results for the transfer functions using the moment hierarchy from It\^o calculus given in Sect.~\ref{sec:ito}. This shows that for certain parameter choices the simplified treatment becomes insufficient; however, it is easier to use for explorations and remains accurate in the perturbative regime. 

\subsection{Effects on the photon transfer functions}
\label{sec:effects_Transfer}
We start by illustrating the effects on the photon transfer functions for various choices of the parameters. For reference, the optical depth across the sound horizon is $\tau_{\rm s}=\Gammab r_{\rm s}\approx 9.8$ at $z\simeq 1100$ for the standard recombination history.

\begin{figure}
\centering
\includegraphics[width=\columnwidth]{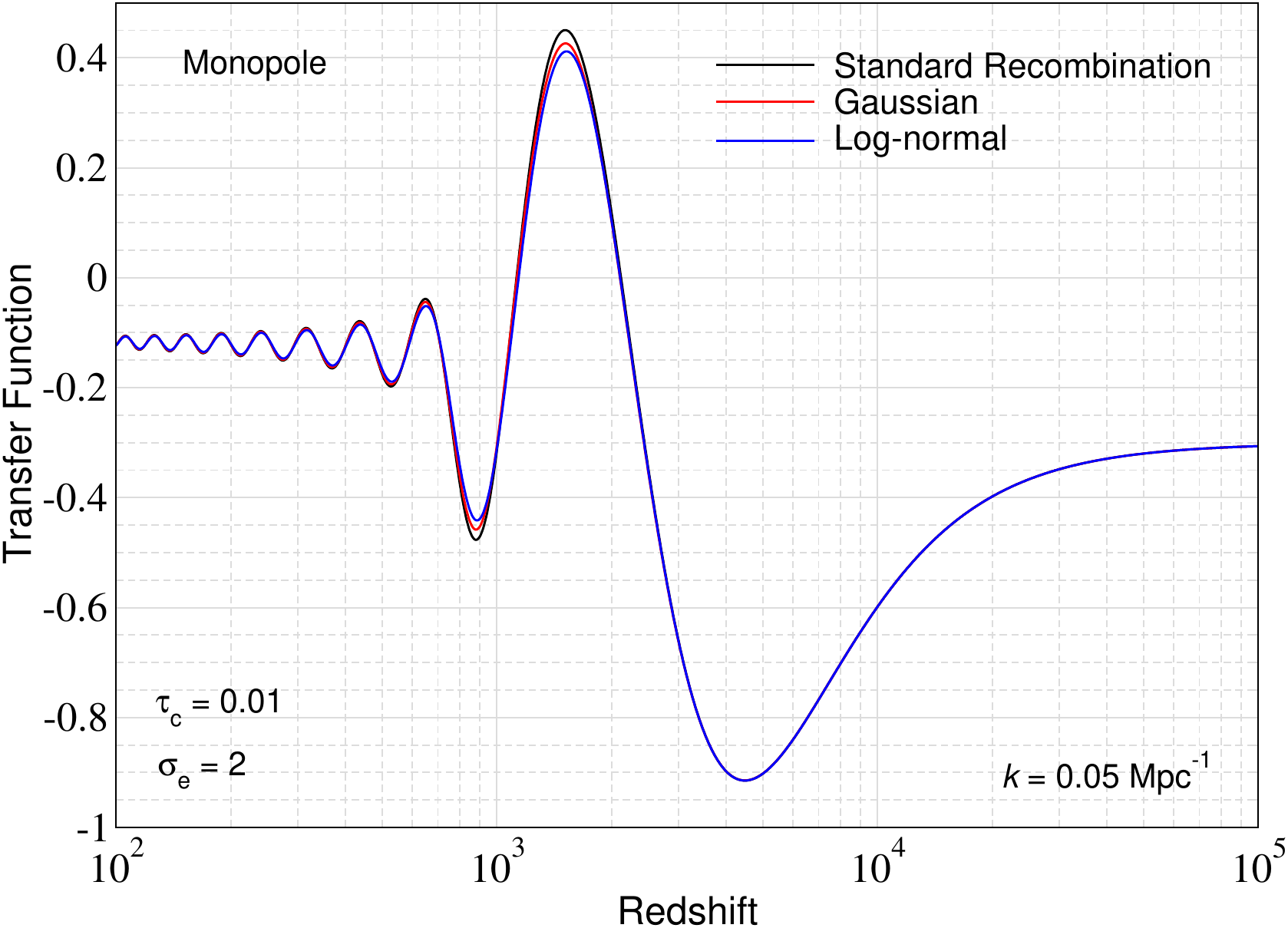}
\\[2mm]
\includegraphics[width=\columnwidth]{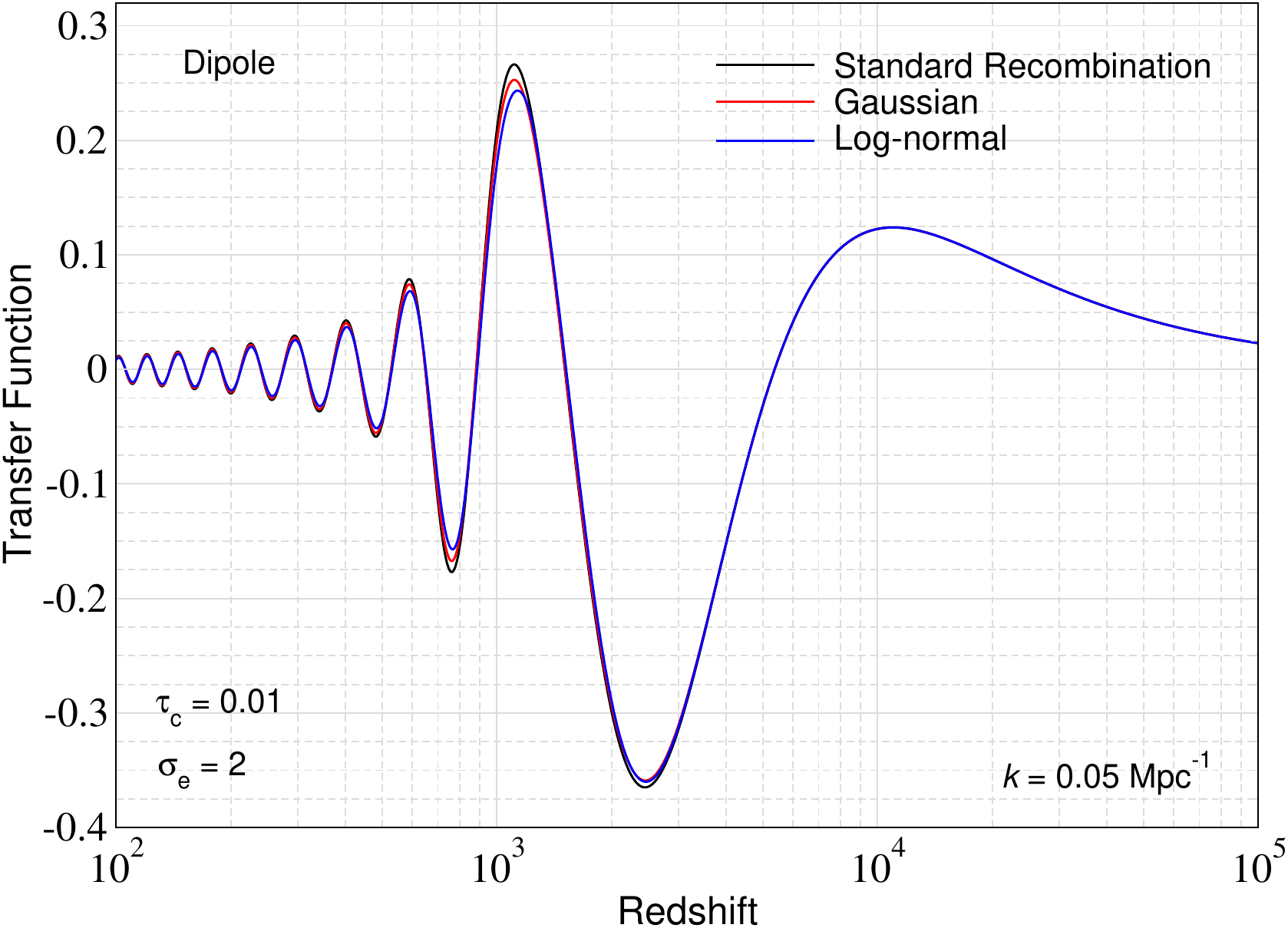}
\\[2mm]
\includegraphics[width=\columnwidth]{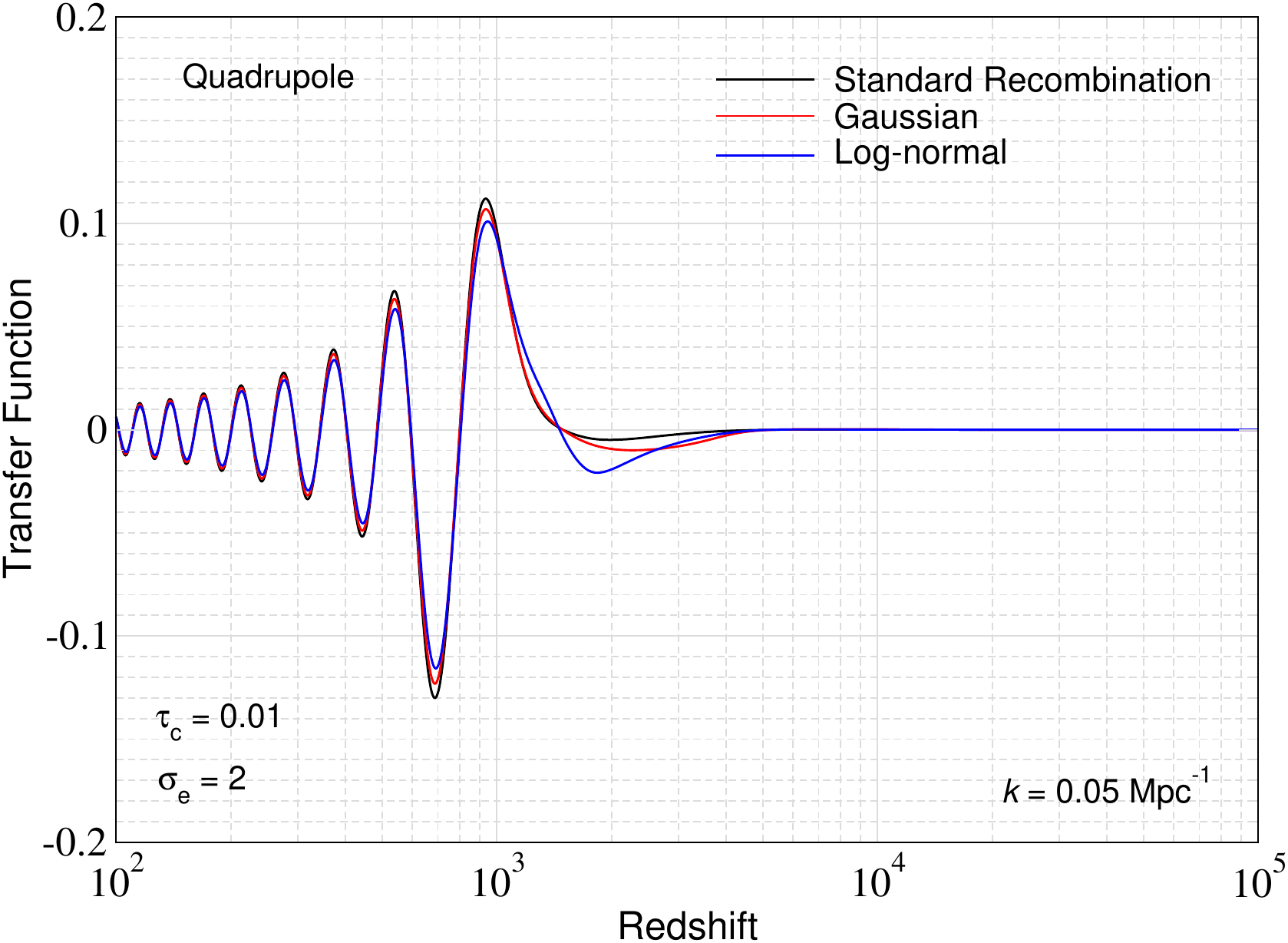}
\\
\caption{Photon transfer functions for the temperature monopole, dipole and quadrupole for $k=0.05\,\Mpc^{-1}$. The standard \LCDM result is compared with the Gaussian and log-normal treatments for $\tauc=0.01$ and $\sigmae=2$ at $z=1100$. For the computations we used the simplified BH.}
\label{fig:Transfer_example_1}
\end{figure}

In Fig.~\ref{fig:Transfer_example_1} we illustrate the changes to the photon transfer functions, $\Theta_\ell(\eta, k)$, of the monopole, dipole and quadrupole at wavenumber $k=0.05\,\Mpc^{-1}$ in the simplified BH approach (Sect.~\ref{sec:Boltz_simp}). We used $\ell_{\rm max}=50$ for all species and included polarization effects. The parameters have been normalized at $z=1100$ and in all cases we use the \LCDM recombination history for $\Neb$. We left $\sigmae$ constant, while the optical depth across the coherence time was scaled as 
\begin{align}
\label{eq:tauc_scaling}
\tauc(z)=\frac{\tauc(z=1100)}{\tau_{\rm s}(z=1100)}\,\tau_{\rm s}(z),
\end{align}
where $\tau_{\rm s}(z)=\Gammab(z)\,r_{\rm s}(z)$, meaning that $\tauc(z)$ increases as $\propto (1+z)$ in the pre-recombination era. This implies a coherence scale that remains a constant fraction of the sound horizon.

\begin{figure}
\centering
\includegraphics[width=\columnwidth]{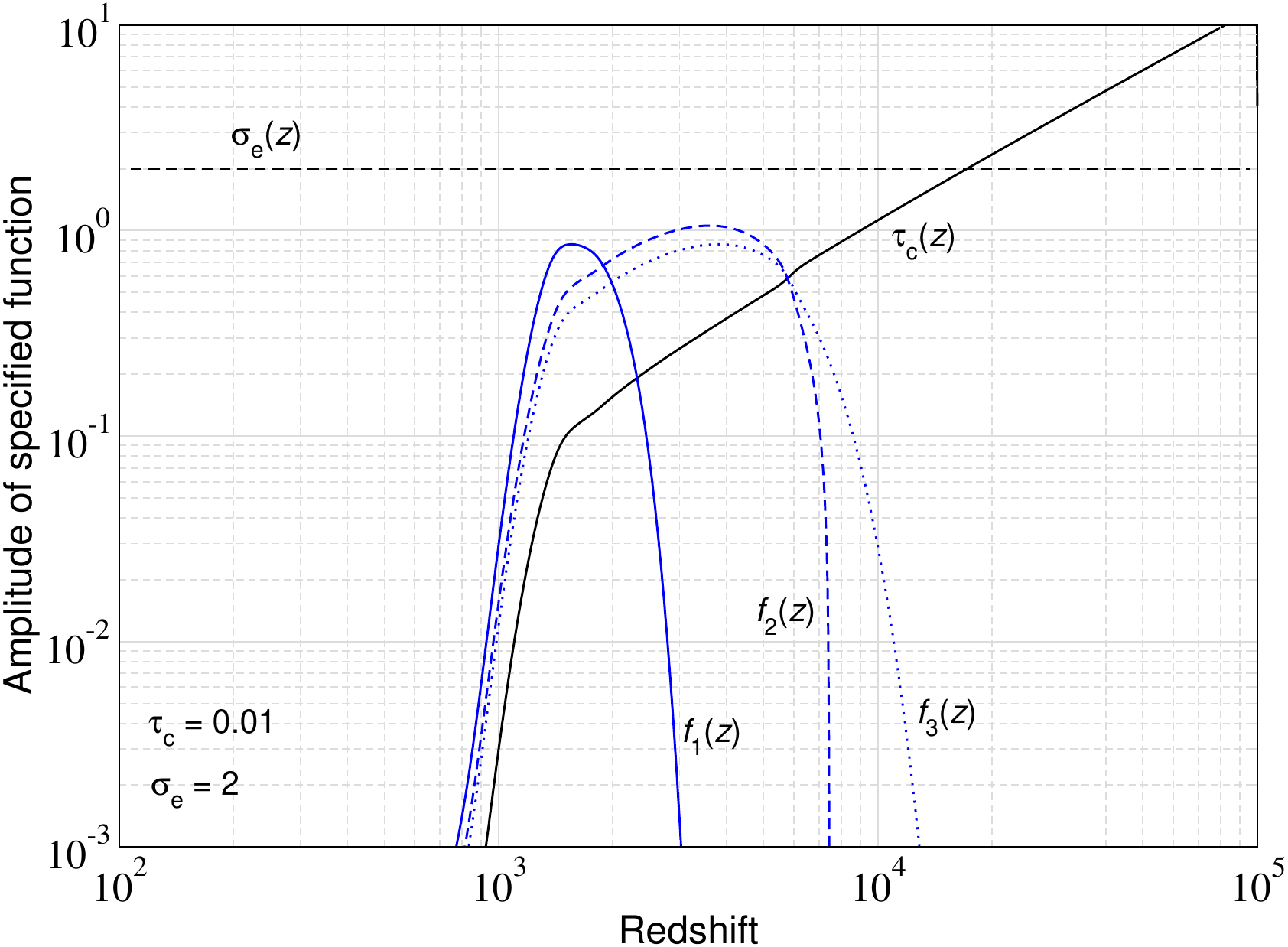}
\\[2mm]
\includegraphics[width=\columnwidth]{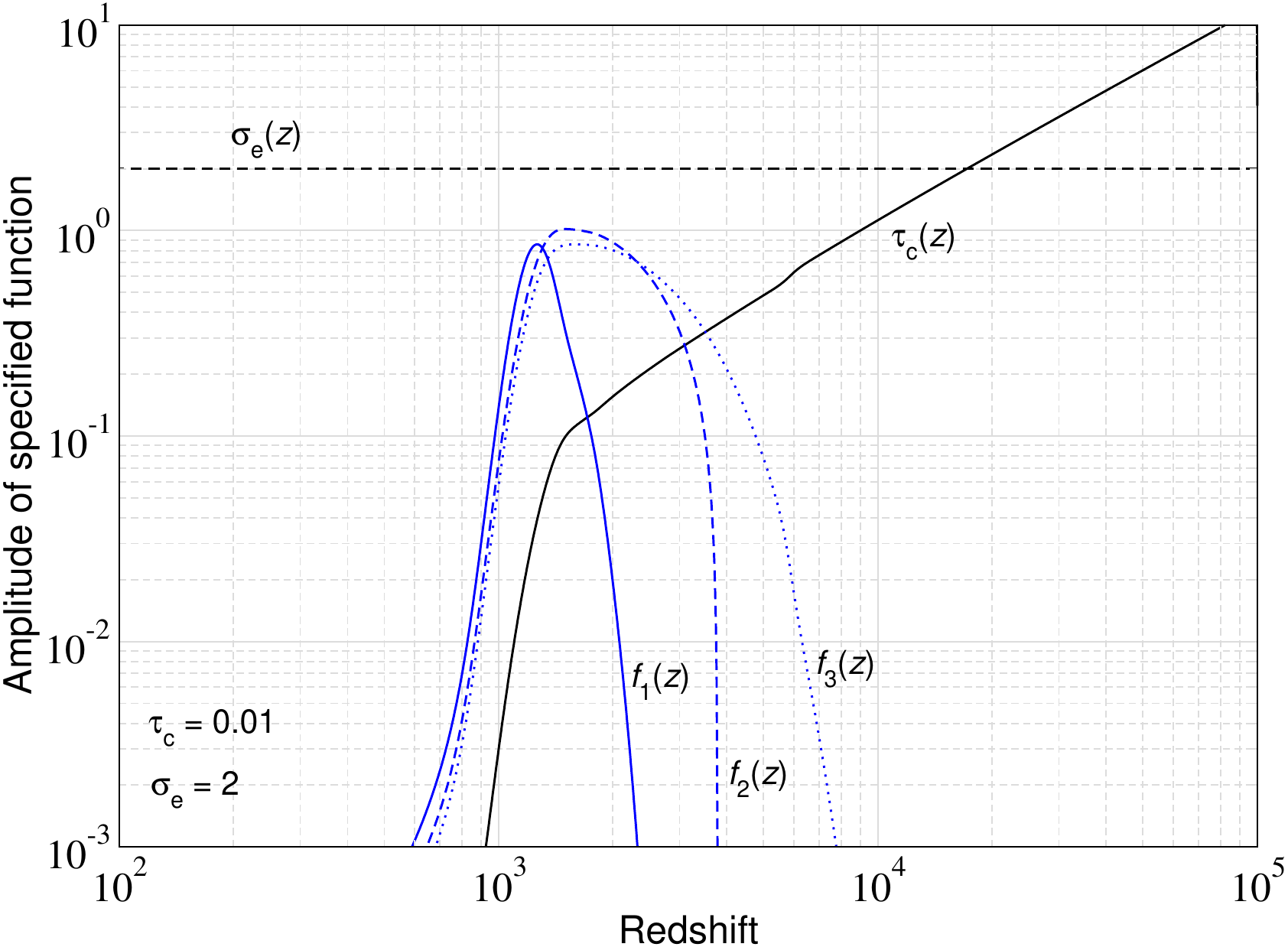}
\\
\caption{Model variables $\tauc$, $\sigmae$ and $f_i$ for the examples shown in Fig.~\ref{fig:Transfer_example_1}. The upper panel is for the modified BH treatment with Gaussian driver while the lower is for the log-normal setup.}
\label{fig:fi_example_1}
\end{figure}

As Fig.~\ref{fig:Transfer_example_1} shows, the amplitudes of the transfer functions are diminished around recombination, as already anticipated from the changes to the damping scale.
In addition, we notice a small phase-shift towards higher redshifts.
The effects are slightly more pronounced for the log-normal treatment, which stems from the differences in the scaling of the $f_i$ (see Fig.~\ref{fig:fi_example_1}). For instance, we can notice a large effect in the photon quadrupole transfer function related to the higher amplitude of $f_1$ and $f_2$ around $z\simeq 2000$. 

From Fig.~\ref{fig:fi_example_1} we can also see that the corrections are small at high redshifts due to the exponential suppression with increasing optical depth. The leading order terms alone would instead lead to (unphysical) exponential growth of the photon transfer functions since $\zetae>1$ at $z \simeq 2000-3000$, highlighting the importance of the higher order terms as already anticipated. Overall, Fig.~\ref{fig:Transfer_example_1} demonstrates that one can expect noticeable effects on the photon transfer functions due to the propagation through a clumpy medium.

\begin{figure}
\centering
\includegraphics[width=\columnwidth]{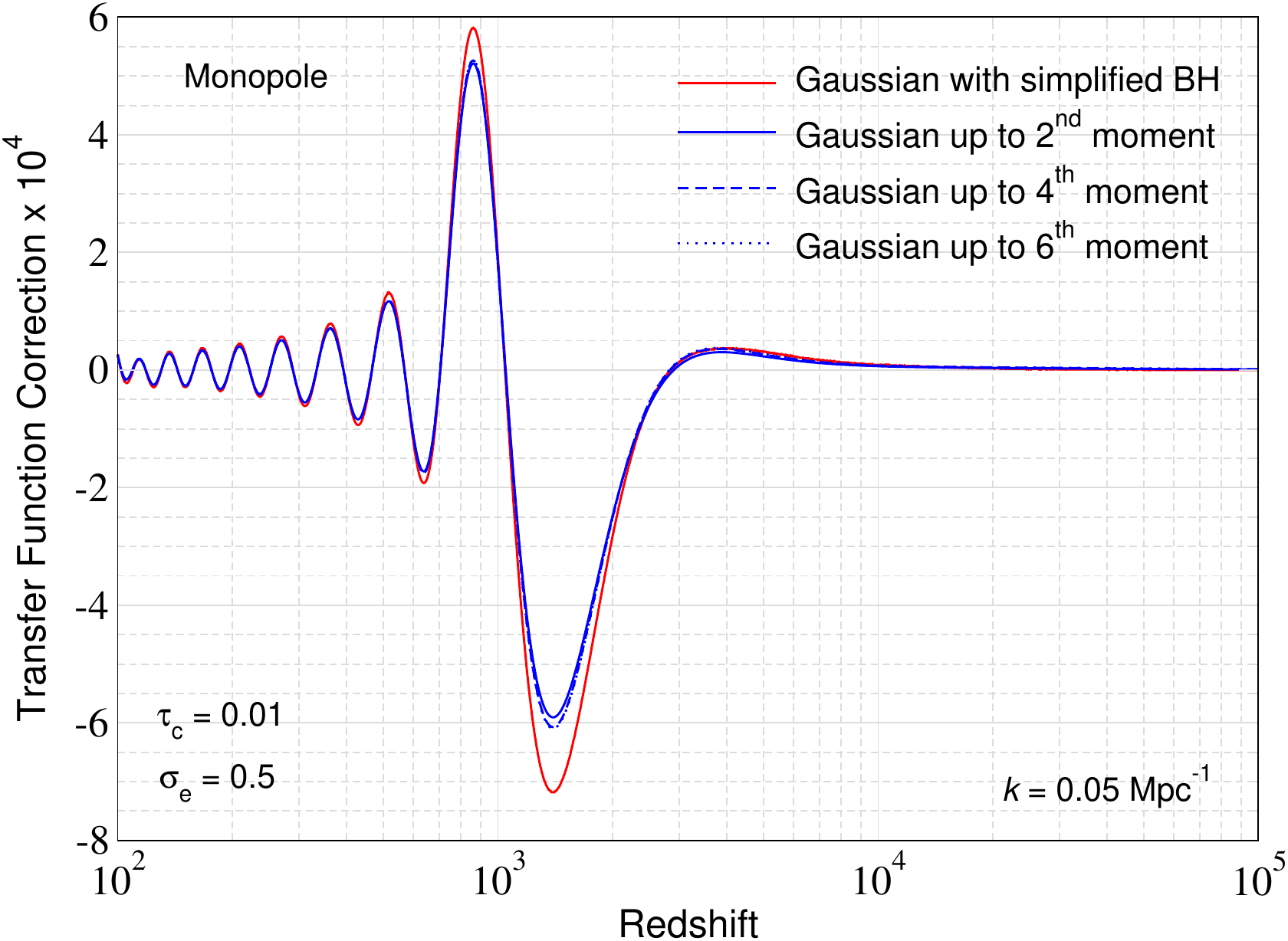}
\\[1mm]
\includegraphics[width=\columnwidth]{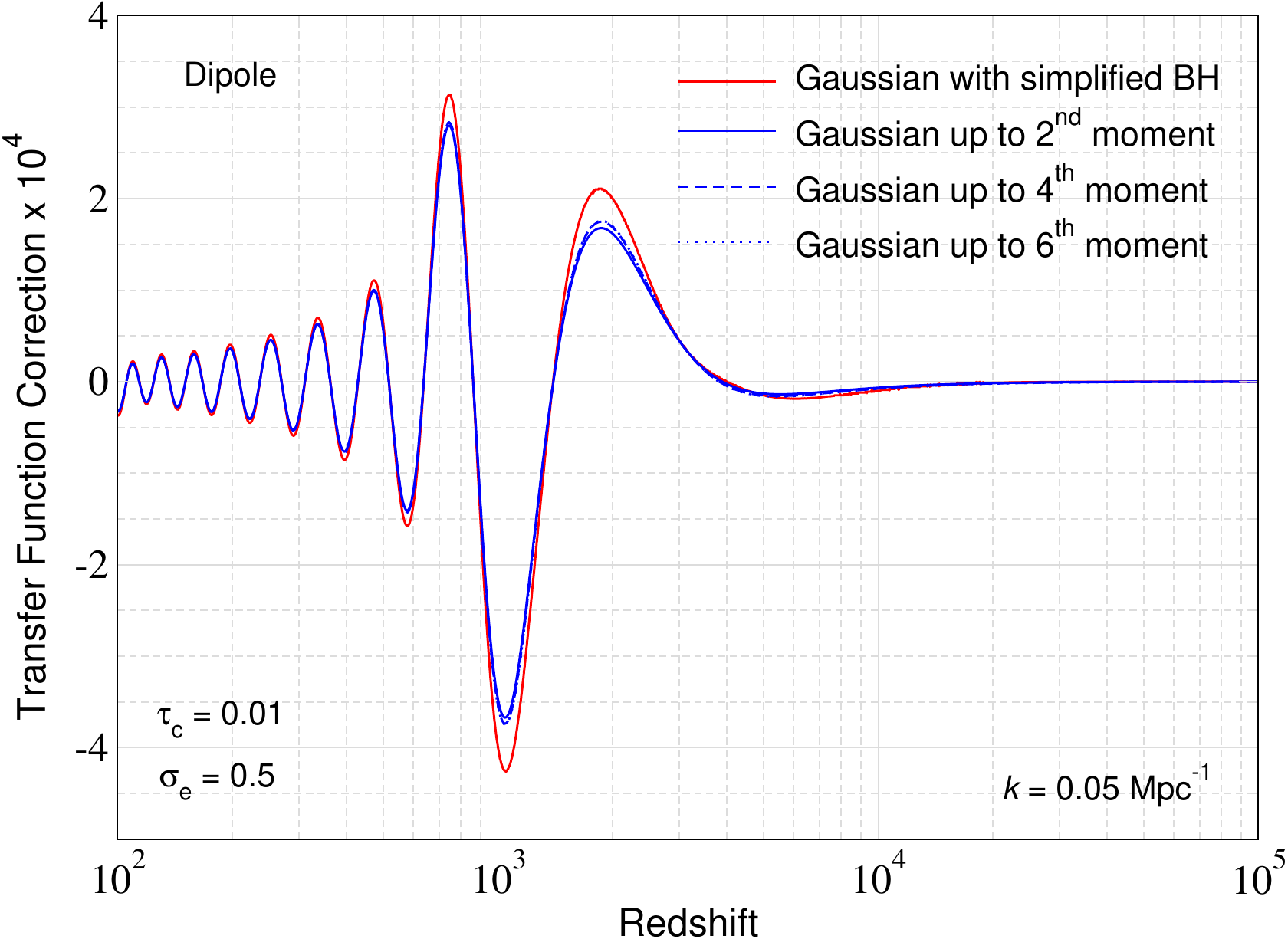}
\\[1mm]
\includegraphics[width=\columnwidth]{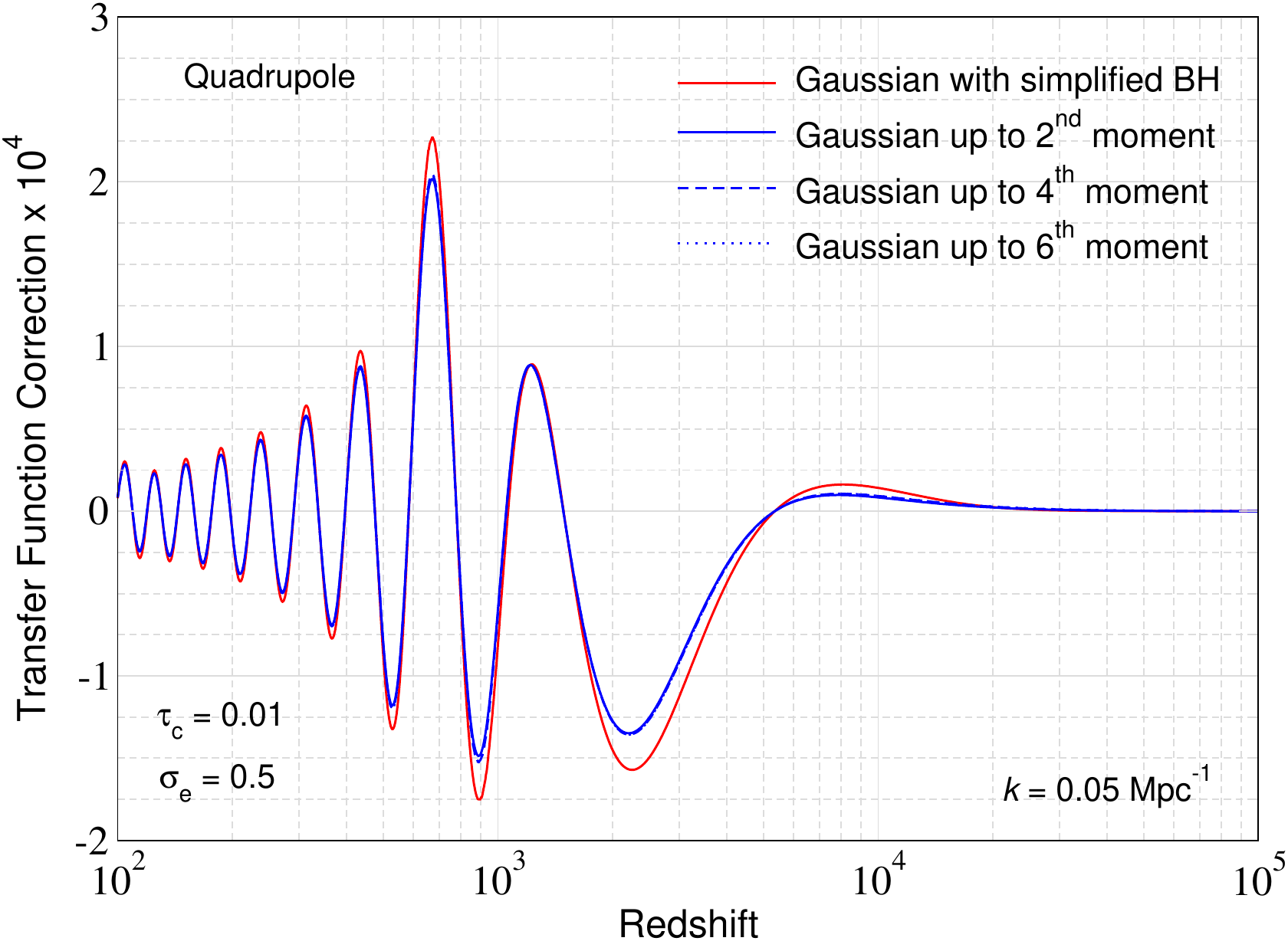}
\\
\caption{Photon transfer function corrections (with respect it the \LCDM solution) for the temperature monopole, dipole and quadrupole at wavenumber $k=0.05\,\Mpc^{-1}$. We assumed the Gaussian scenario with $\tauc=0.01$ and $\sigmae=0.5$ at $z=1100$ and compare the simplified BH approach with the moment approach up to 6$^{\rm th}$ order. Note that the lines for the 6$^{\rm th}$ order moment solutions practically coincide with the curves for the 4$^{\rm th}$ order solutions.}
\label{fig:Transfer_example_2}
\end{figure}

\begin{figure}
\centering
\includegraphics[width=\columnwidth]{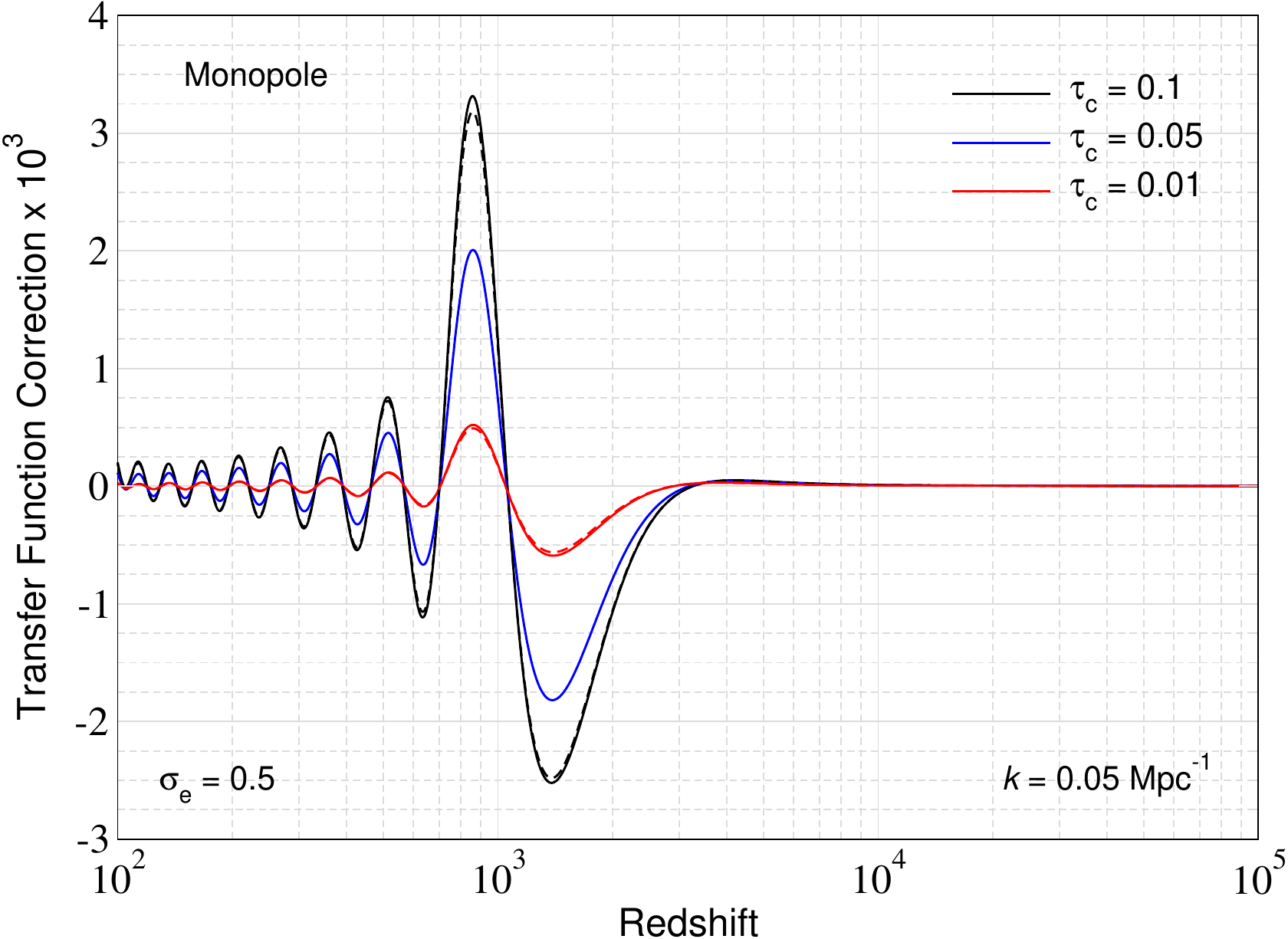}
\\[2mm]
\includegraphics[width=\columnwidth]{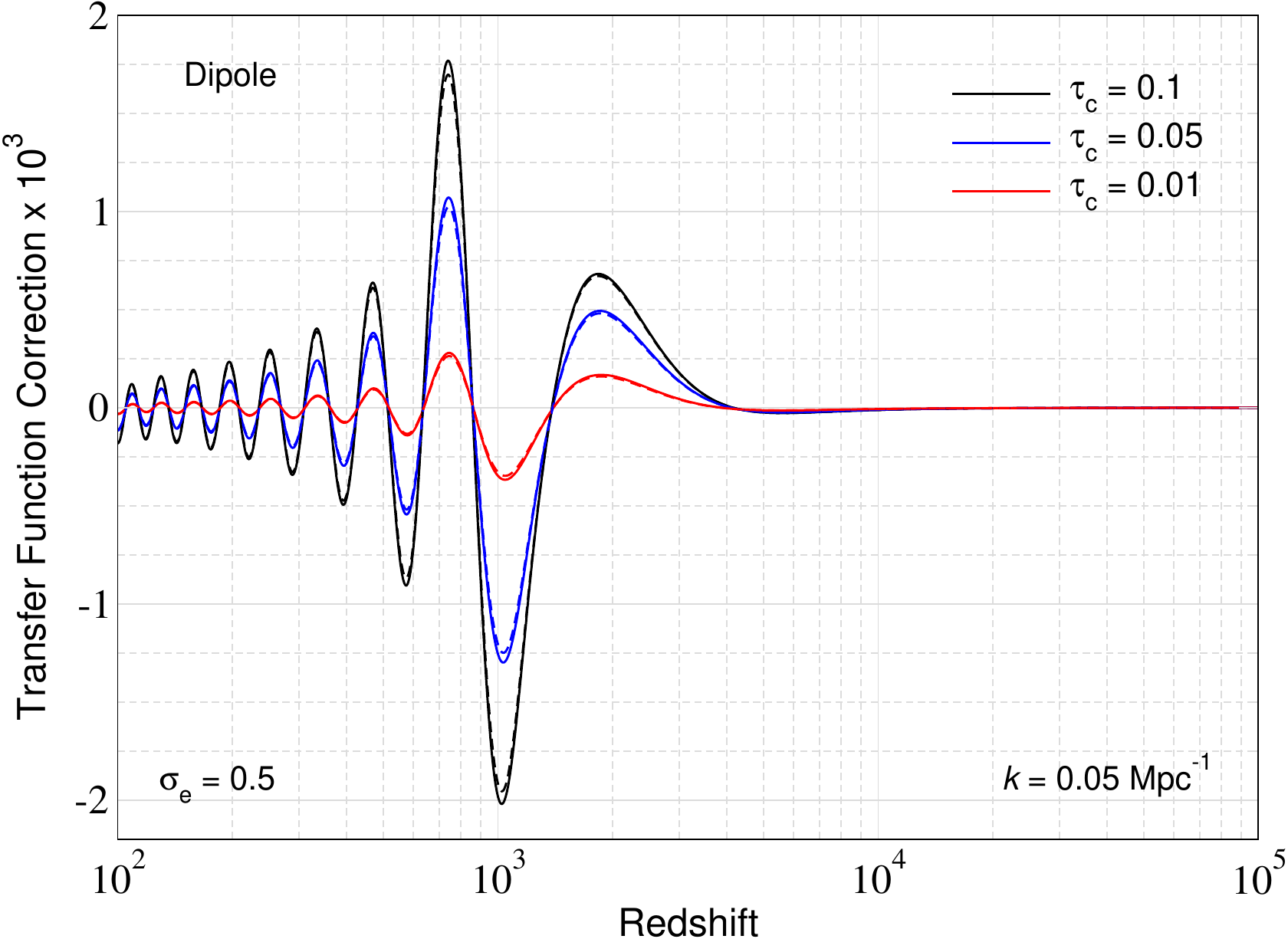}
\\
\caption{Photon transfer function corrections for the temperature monopole and dipole at $k=0.05\,\Mpc^{-1}$. We used $\sigmae=0.5$ for varying values of $\tauc$ all normalized at $z=1100$. In each line group, we show the results from the Gaussian (solid line) and log-normal (dashed line) setups including up to 2$^{\rm nd}$ order moments.}
\label{fig:Transfer_LN_G_compare}
\end{figure}

\subsubsection{Comparison with moment treatment}
\label{sec:effects_Transfer_moments}
The simplified BH treatment is an approximation to the problem. It is therefore important to test its validity. A few challenges appear: First, solving the moment hierarchy is a lot more computationally challenging due to the increase in the number of coupled equations with moment order. Second, a truncated moment hierarchy is bound to exhibit numerical issues, meaning we have to limit the values of $\sigmae$ to remain convergent.
Third, a perturbative treatment in $\tauc$ has its limitations which will become more severe at high redshifts, when $\tauc$ can exceed unity due to the increasing average density of the Universe. This is expected to lead to noticeable differences with the simplified BH treatment that also make the system more stiff and harder to solve numerically. 

To illustrate these aspects, in Fig.~\ref{fig:Transfer_example_2} we show the transfer function corrections for the monopole, dipole and quadrupole for a more moderate choice of parameters in the Gaussian scenario at $k=0.05\,\Mpc^{-1}$ (and $\ell_{\rm max}=25$). The results of the simplified BH treatment agree well (to within $\simeq 10\%-20\%$) with the more rigorous moment treatment. For the chosen example, the latter converges once 4 order moments or more are included. We can notice that the moment treatment predicts a slightly smaller effect. 
Overall, the main cause of the difference is due to the dependence of the corrections on $\tauc$. By decreasing $\tauc$ we confirmed that all the treatments agree, as also anticipated from the analytical discussion. 
We confirmed numerically that around recombination the difference is roughly consistent with an additional suppression of the mode amplitude by a factor of $\simeq 1/(1+\tauc)$. However, this did not improve the agreement at $z\gtrsim 2000$ such that we did not follow up on this further. 
We also confirmed that the Gaussian and log-normal treatments yield very similar results for the chosen example. This is illustrated in Fig.~\ref{fig:Transfer_LN_G_compare} for the photon monopole and varying values of $\tauc$.

We highlight that the moment treatment does not allow an arbitrary exploration of the possible parameter space. For instance, we could not numerically solve the problem for the cases illustrated in Fig.~\ref{fig:Transfer_example_1} when using the moment method. While increasing $\tauc$ seems to be handled quite well, increasing $\sigmae$ can quickly lead to numerical instabilities. This is expected especially in the Gaussian setup, which will inevitably explore regimes of $1+\deltae<0$ when large variance is permitted. For the log-normal setup, the moment truncation similarly prohibits too large values of $\sigmae$. However, it seems that the zones of numerical instability are in regimes that are less motivated such that we postpone a more detailed study to future work. For model-exploration, the modified BH treatment is more robust and can serve as a first step.

\subsubsection{Exploring the high-redshift dependence}
\label{sec:high_redshift_dependence}
So far we kept $\sigmae$ constant while we assumed that the coherence length was a fixed fraction of the sound horizon. This was for illustration only and by looking at Fig.~\ref{fig:fi_example_1} we can understand that this is not expected to lead to a significant effect early on. Naively, one expects the largest effects when increasing both $\tauc$ and $\sigmae$; however, the significant exponential suppression, e.g., $f_3 \simeq \tauc \sigmae^2 \expf{-\tauc^2 \sigmae^2}$, of the scattering rate corrections with $\tauc$ renders this choice less interesting. Indeed the largest effect is expected when $\zetae=\tauc \sigmae^2<1$ remains roughly constant and $\tauc(z)<1$. In this scenario, the photon transfer functions could also be modified in the high-redshift domain, leading to changes of the CMB power spectra at small scales. 

\begin{figure}
\centering
\includegraphics[width=0.99\columnwidth]{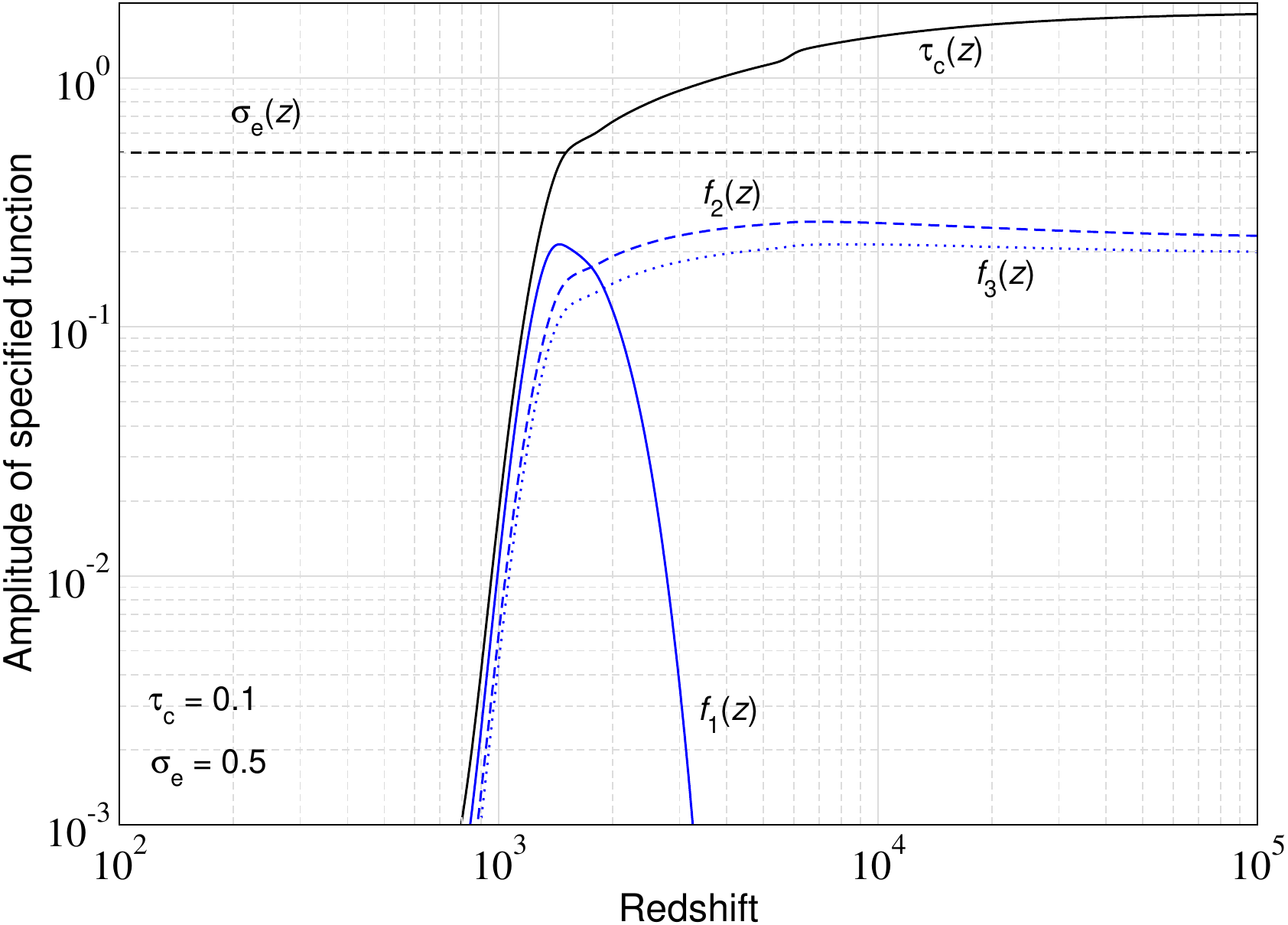}
\caption{Model variables $\tauc(z)$, $\sigmae$ and $f_i$ for $\sigmae=0.5$ and $\tauc=0.1$ at $z=1100$. We assumed the redshift scaling of $\tauc$ according to Eq.~\eqref{eq:tauc_scaling_M1} with $z_{\rm s}=1500$ and $\gamma_{\rm s}=1$.}
\label{fig:fi_scalings_M1}
\end{figure}

\begin{figure}
\centering
\includegraphics[width=0.99\columnwidth]{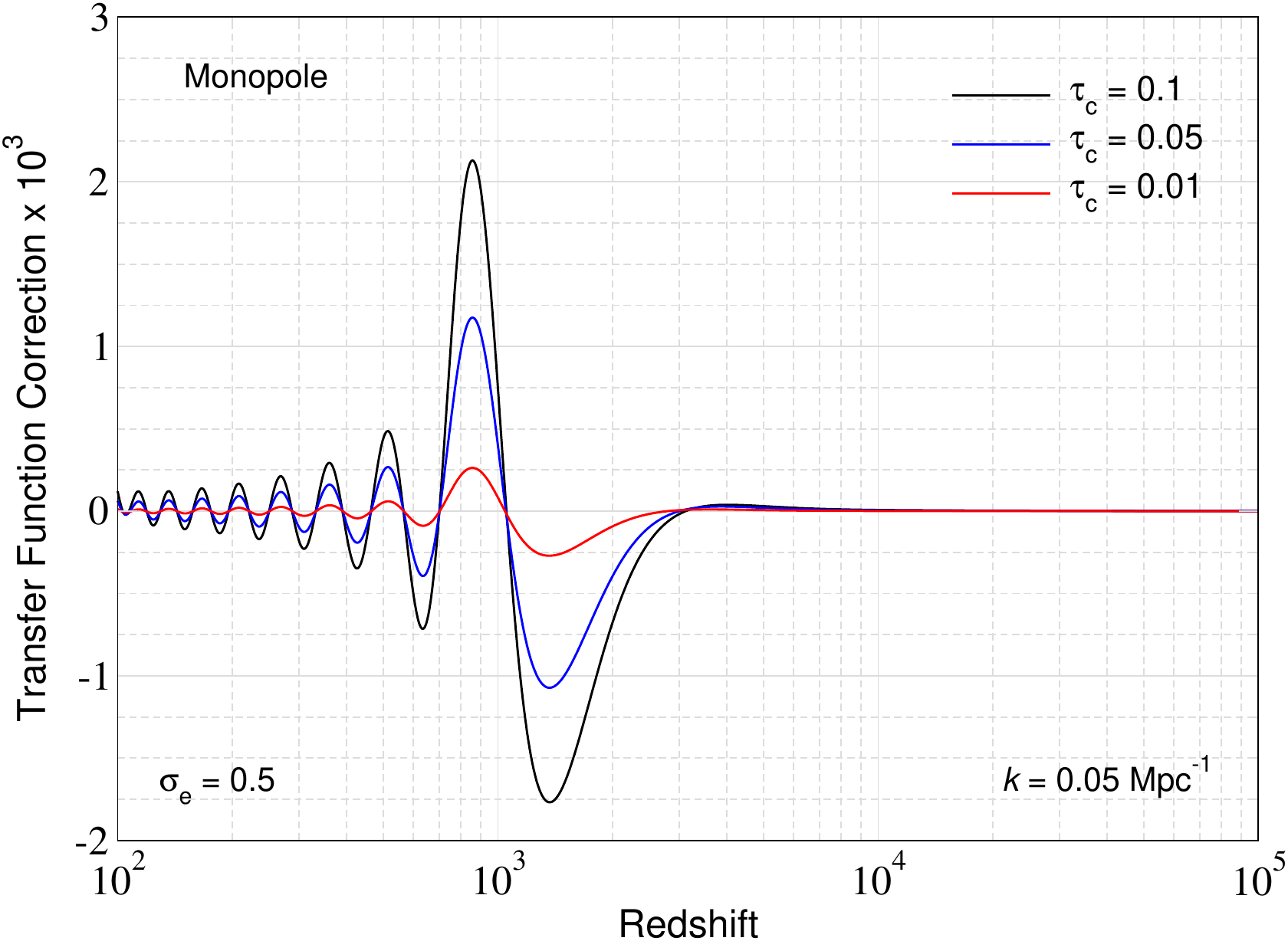}
\\
\includegraphics[width=0.99\columnwidth]{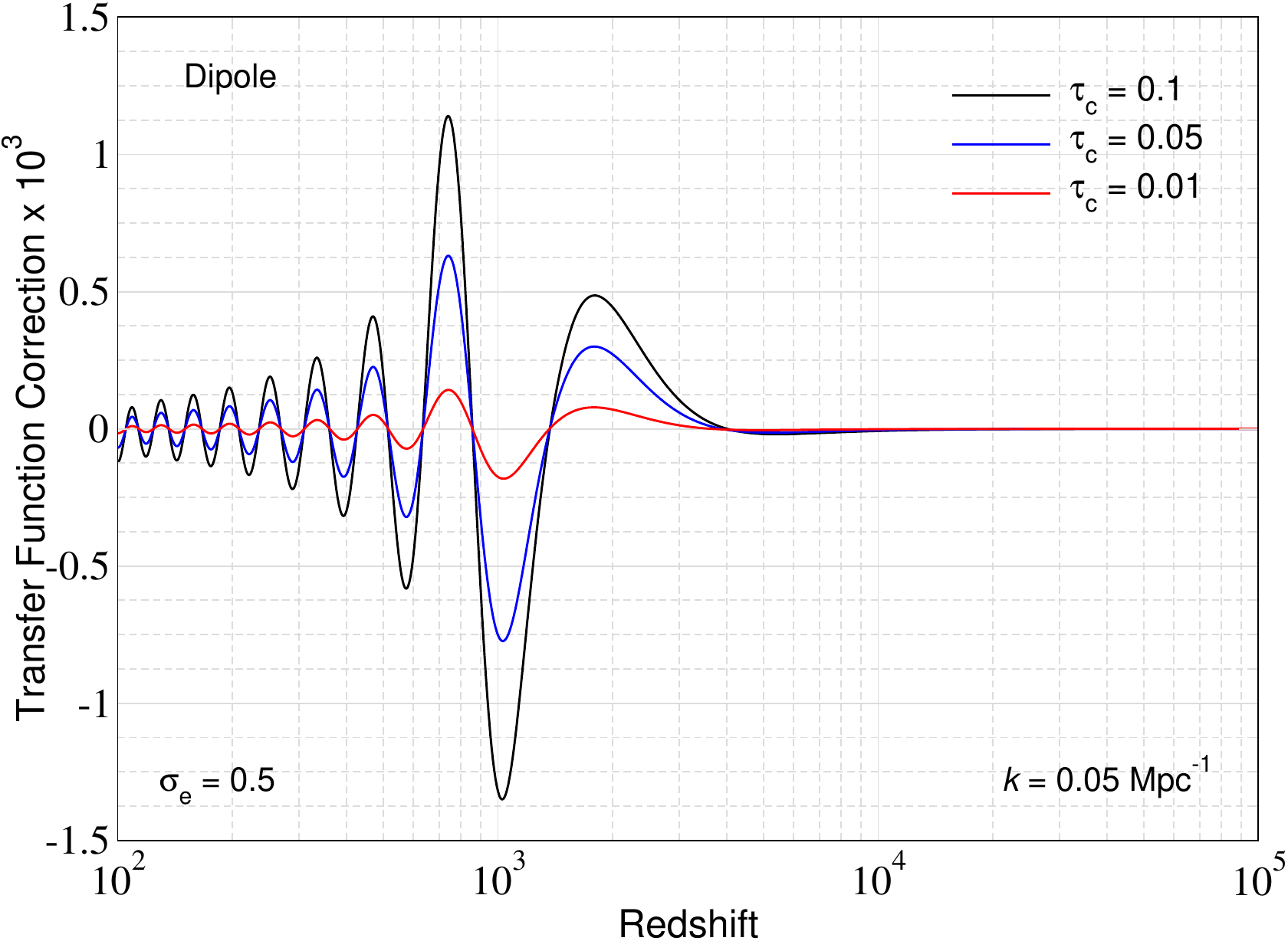}
\\
\caption{Photon transfer function corrections for the temperature monopole and dipole at $k=0.05\,\Mpc^{-1}$. We used $\sigmae=0.5$ for varying values of $\tauc$ all normalized at $z=1100$ with a scaling according to Eq.~\eqref{eq:tauc_scaling_M1}. The results were obtained with a 4$^{\rm th}$ order Gaussian moment setup.}
\label{fig:Transfer_scaling_M1}
\end{figure}

Given that in the previous examples $\tauc \propto (1+z)$ at early times (cf., Fig.~\ref{fig:fi_example_1}), one can simply assume that the coherence length decreases as $\Delta \eta_{\rm c} \propto r_{\rm s}(z)/(1+z)$ leaving $\tauc \sigmae^2 \simeq {\rm const}$. In contrast, scaling $\sigmae\propto (1+z)^{-1/2}$ does not fully avoid the exponential suppression since one would still obtain $\tauc^2 \sigmae^2\propto (1+z)$ for $\Delta \eta_{\rm c} \propto r_{\rm s}$.

For illustration, we assume a scaling like
\begin{align}
\label{eq:tauc_scaling_M1}
\tauc(z)=\frac{\tauc(z=1100)}{\tau_{\rm s}(z=1100)}\,\frac{\tau_{\rm s}(z)}{1+\left(\frac{1+z}{1+z_{\rm s}}\right)^{\gamma_{\rm s}}},
\end{align}
with pivot redshift $z_{\rm s}$ and power-law index $\gamma_{\rm s}$. This means around recombination the corrections can remain similar to the cases discussed before but at high redshifts a larger correction is expected. This can be anticipated from Fig.~\ref{fig:fi_scalings_M1}, where we see that both $f_2$ and $f_3$ have a large high-$z$ tail. The function $f_1$ still drops exponentially since the extra factor of $(1+R)/R\simeq (1+z)$ in $f_1$ is not suppressed [see Eq.~\eqref{eq:equations_sim_Gb}]. This means that the tight coupling solution is not affected much and that the corrections in this case mainly appear as a suppression of the overall amplitude (Fig.~\ref{fig:Transfer_scaling_M1}).

\begin{figure}
\centering
\includegraphics[width=\columnwidth]{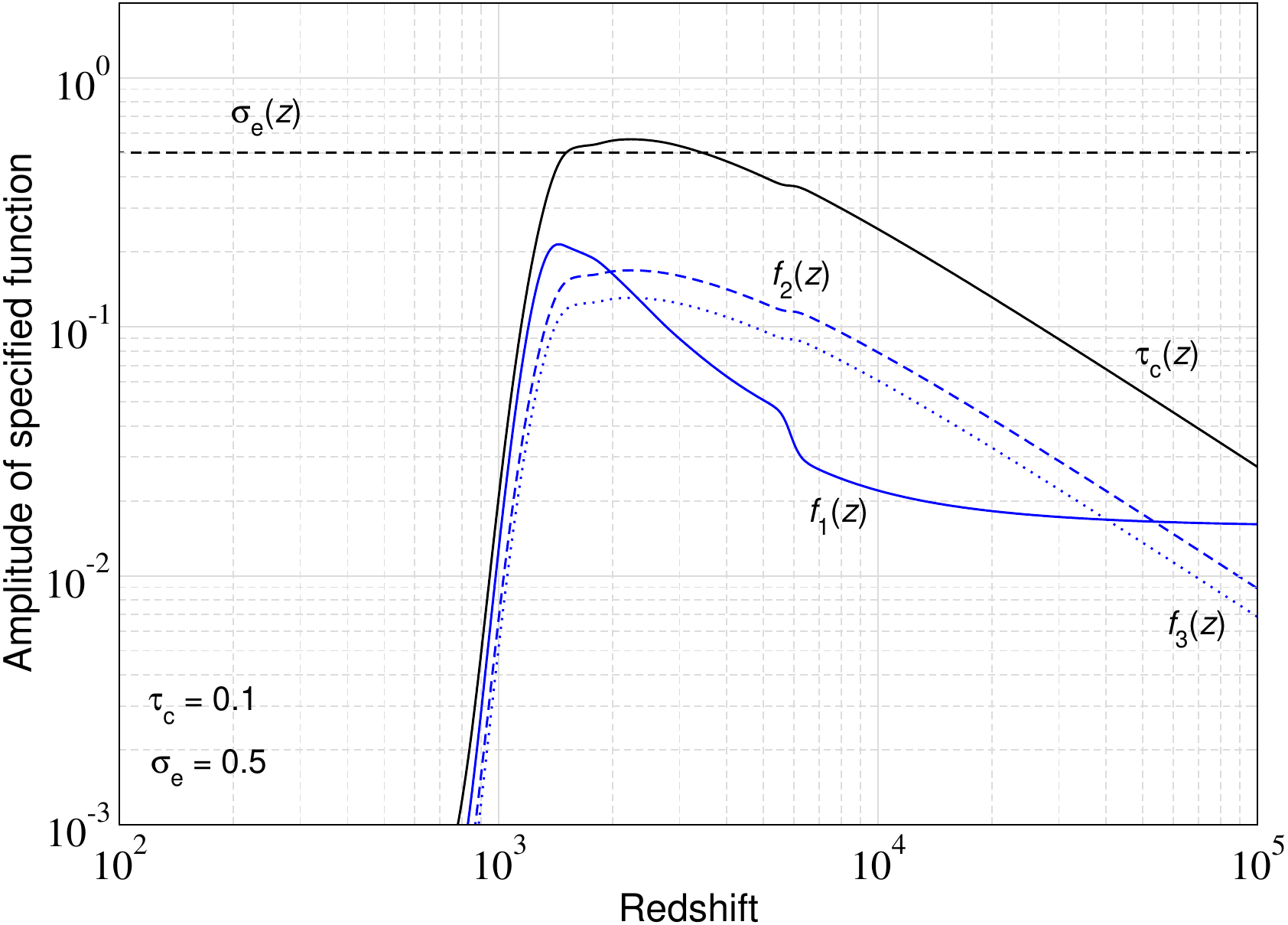}
\\[2mm]
\includegraphics[width=\columnwidth]{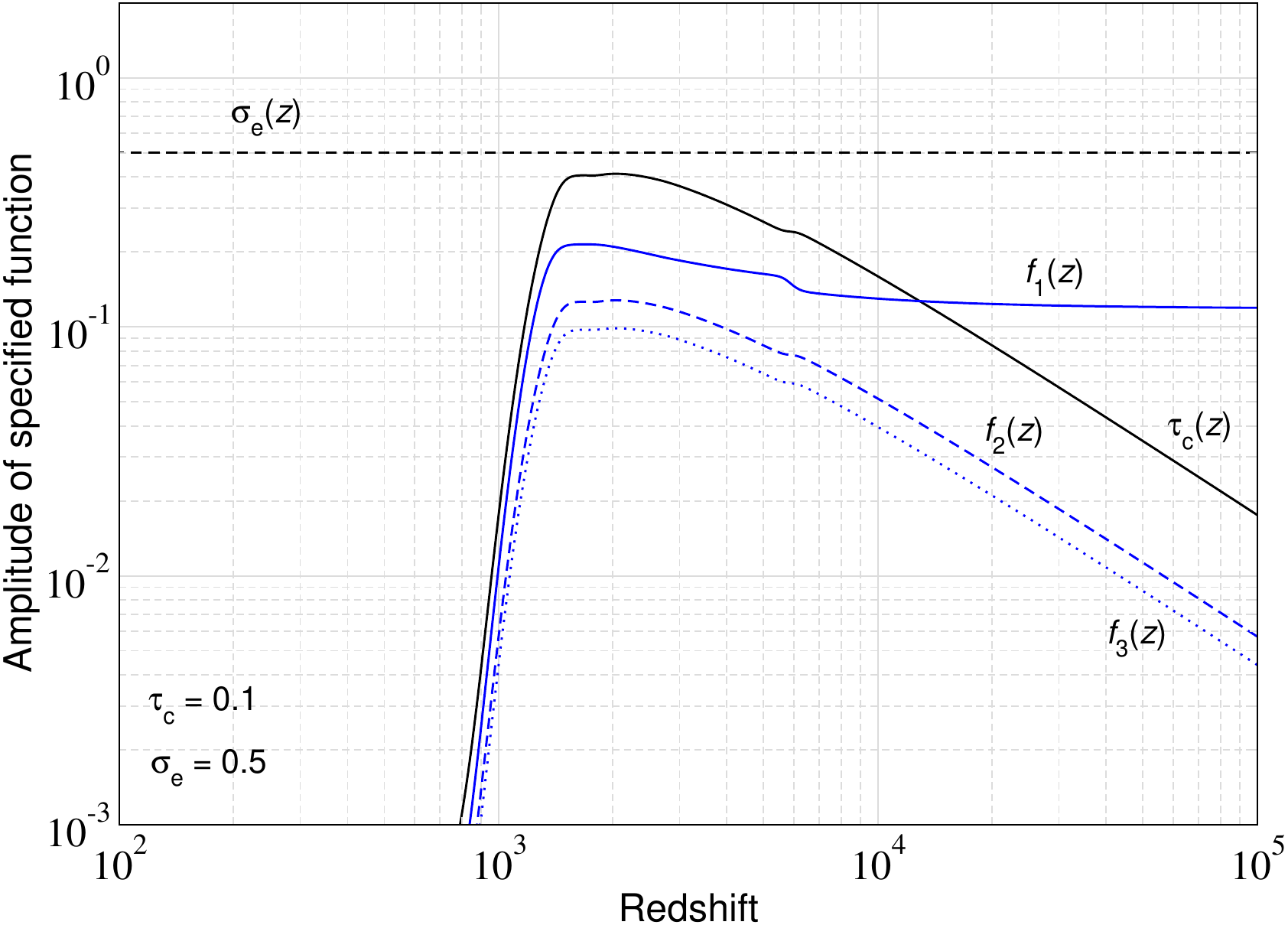}
\caption{Model variables $\tauc(z)$, $\sigmae$ and $f_i$ for $\sigmae=0.5$ and $\tauc=0.1$ at $z=1100$. We assumed the redshift scaling of $\tauc$ according to Eq.~\eqref{eq:tauc_scaling_M1} with $z_{\rm s}=1500$ (upper panel) and $z_{\rm s}=1200$ (lower panel) for $\gamma_{\rm s}=2$. We note that the features at $z\simeq 6000$ are due to the first recombination of helium}
\label{fig:fi_scalings_M2}
\end{figure}

\begin{figure}
\centering
\includegraphics[width=0.99\columnwidth]{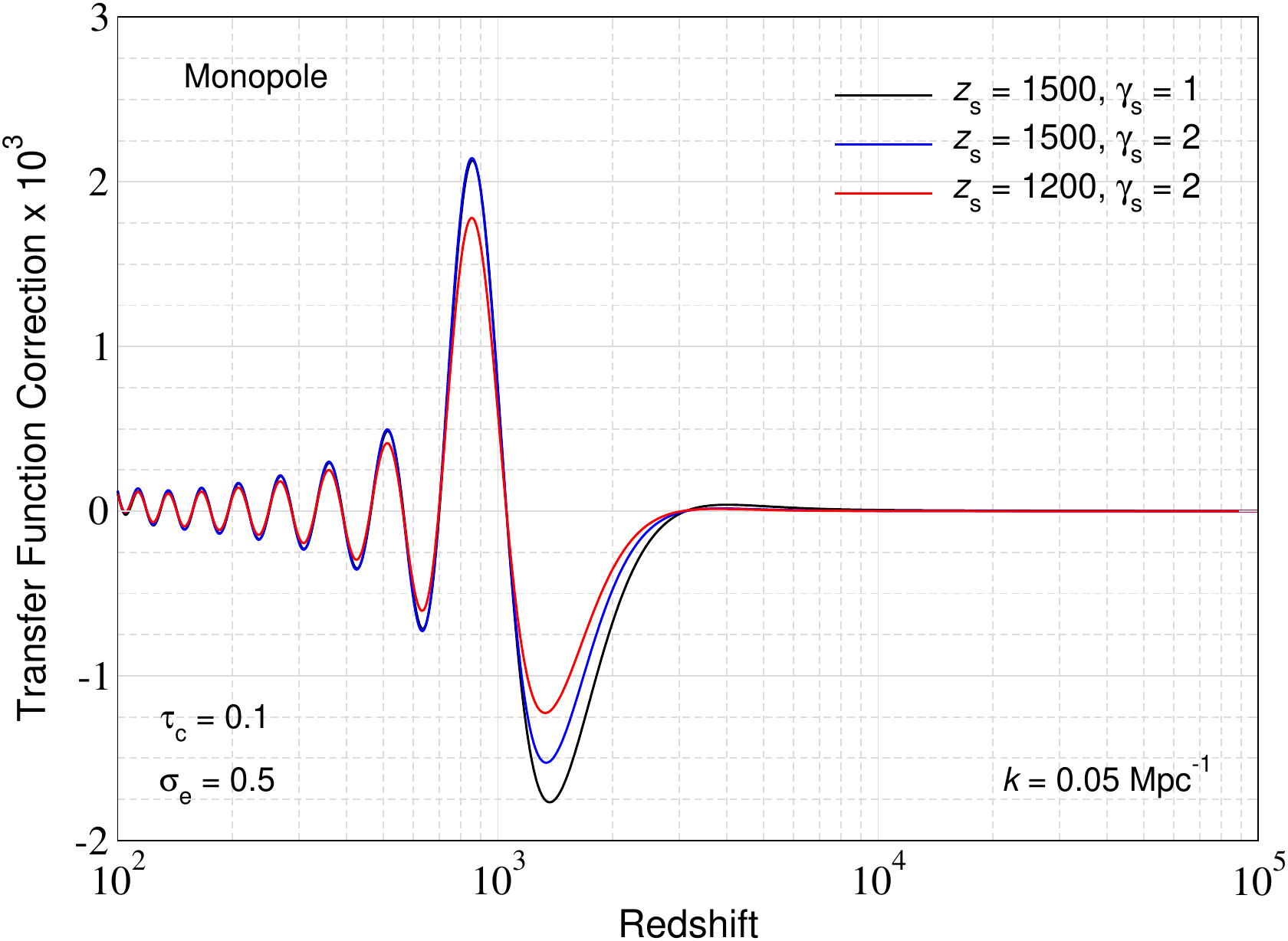}
\\[2mm]
\includegraphics[width=0.99\columnwidth]{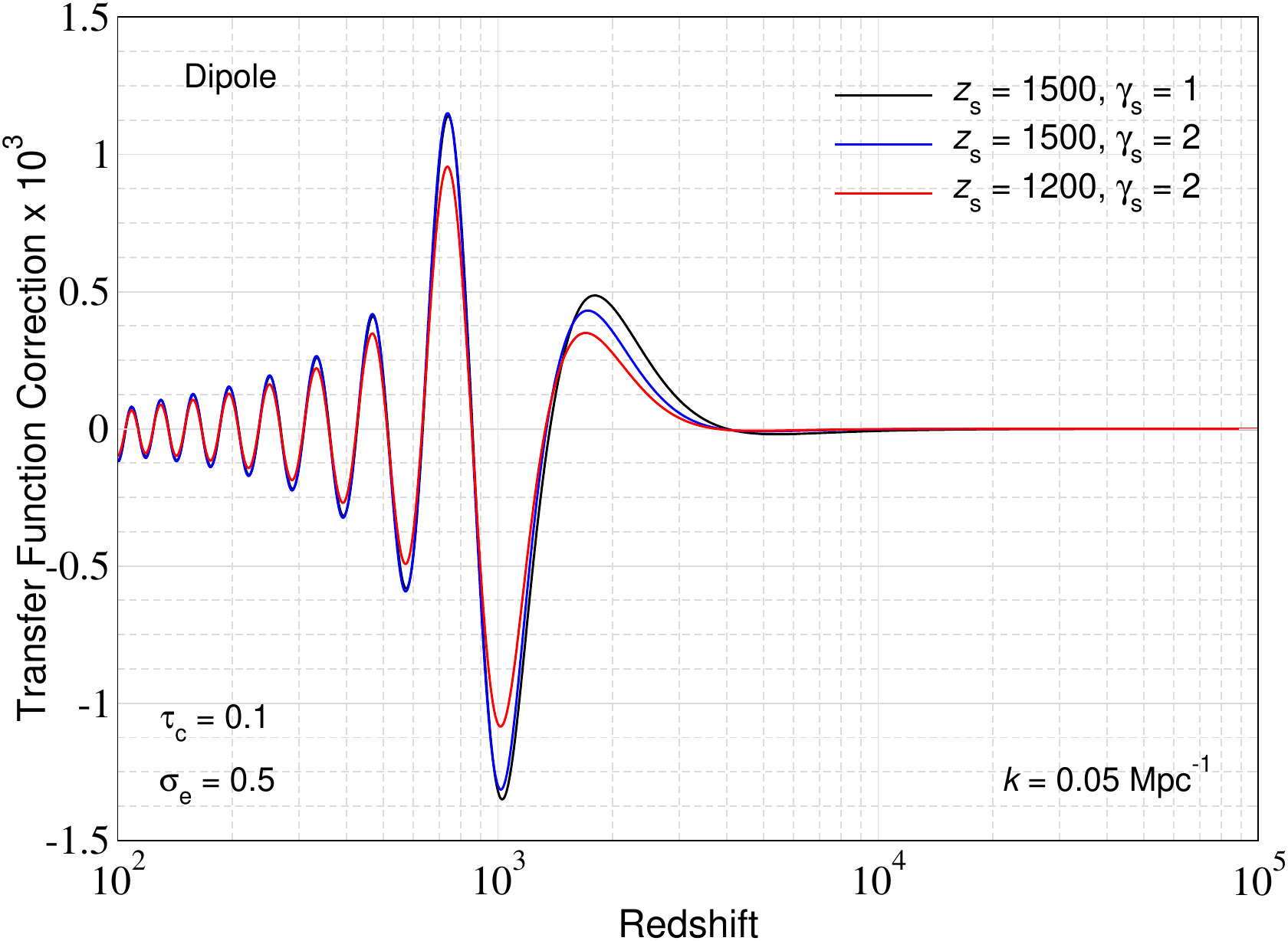}
\\[2mm]
\includegraphics[width=0.99\columnwidth]{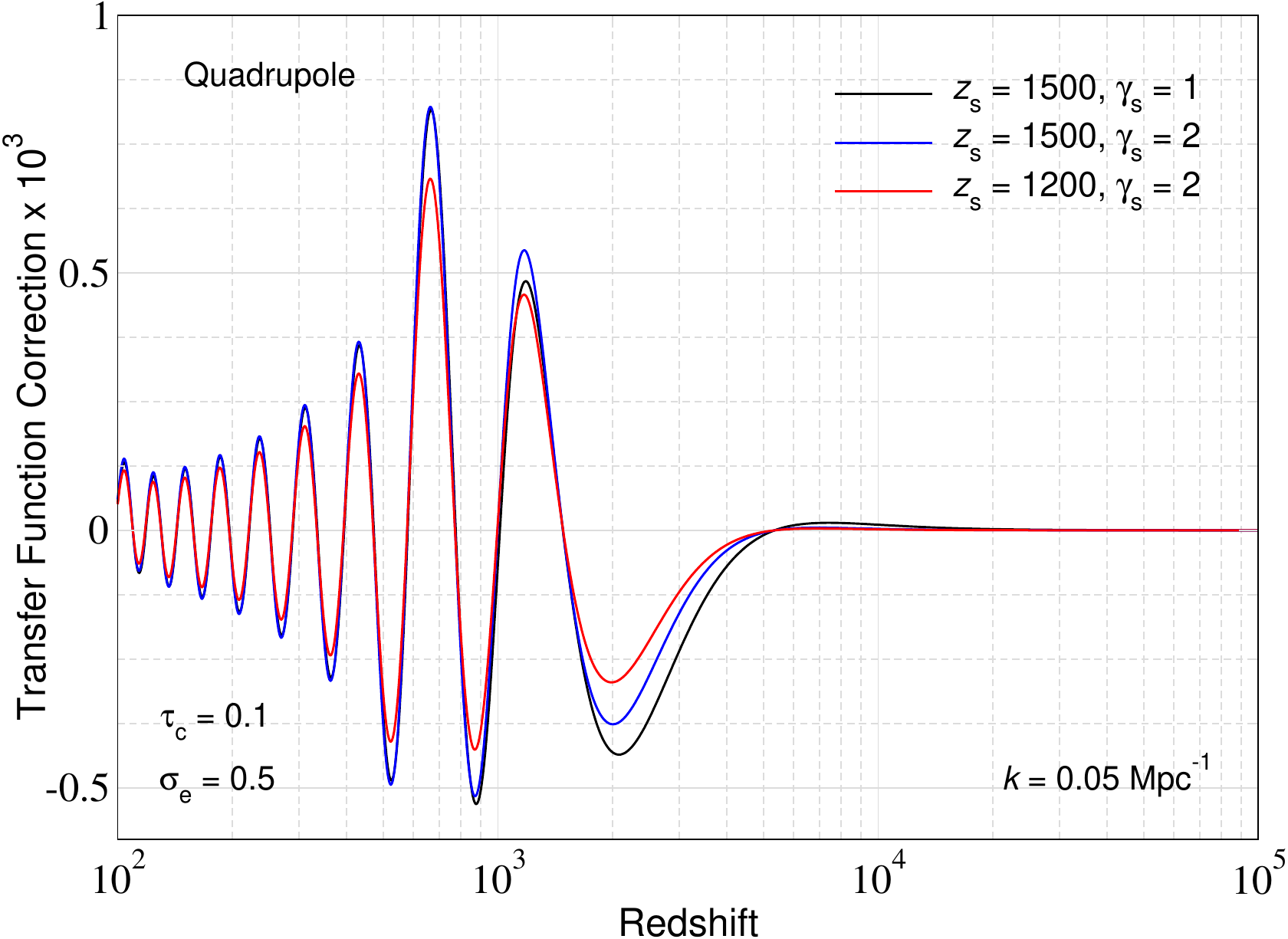}
\\
\caption{Photon transfer function corrections for the temperature monopole, dipole and quadrupole at $k=0.05\,\Mpc^{-1}$. We used $\sigmae=0.5$ normalized at $z=1100$ with a scaling according to Eq.~\eqref{eq:tauc_scaling_M1} and model parameters as labeled. A 4$^{\rm th}$ order Gaussian moment setup was used.}
\label{fig:Transfer_scaling_M2}
\end{figure}
In Figs.~\ref{fig:fi_scalings_M2} and \ref{fig:Transfer_scaling_M2} we illustrate the dependence on the values of $\gamma_{\rm s}$ and $z_{\rm s}$. Setting $\gamma_{\rm s}=2$ means that even $f_1$ maintains a high-redshift tail and hence implies early time corrections to the transfer function. Decreasing $z_{\rm s}$ further enhances the effect as can be anticipated from the redshift dependencies of the functions $f_i$. We highlight that for $\gamma_{\rm s}=2$ we see a noticeable phase shift in the monopole transfer function prior to recombination, an effect that is also visible in the CMB power spectra a small scales (see Fig.~\ref{fig:CMB_PS_example_1}).

\vspace{-3mm}
\subsubsection{Exploring the low-redshift dependence}
\label{sec:low_redshift_dependence}
We now focus on how the late-time scalings enter the problem. For all the cases considered so far we see that $\tauc$ drops sharply around $z\simeq 10^3$ simply because $\Gammab\propto \Neb$ decreases quickly. It is hard to imagine a physical process that could overcome this scaling unless the recombination process is locally delayed, e.g., by the injection of some ionizing radiation \citep{Chen2004, Slatyer2009, Chluba2010a}. A slightly faster growth of the coherence length in the free electrons could be anticipated given that dense clumps recombine faster leaving the large scale (lower density) free electrons exposed, but this would likely have a power-law scaling.

In addition, from our discussion of the recombination history in the separate Universe approach (Sect.~\ref{sec:LN_delta_b_fluctuations} and Fig.~\ref{fig:Moment2}) we anticipate that assuming $\sigmae\approx {\rm const}$ at $z\lesssim 10^3$ is rather unrealistic. However, the results shown in Fig.~\ref{fig:Moment2} assumed constant baryon density variance, $\sigma_{\rm b}$, when one could expect an increase towards lower redshifts due to the onset of structure formation (and similarly a change of $\sigmae$ in the post- and pre-recombination eras). Overall this shows that significant uncertainties exist in the choices of these scaling.

\begin{figure}
\centering
\includegraphics[width=0.99\columnwidth]{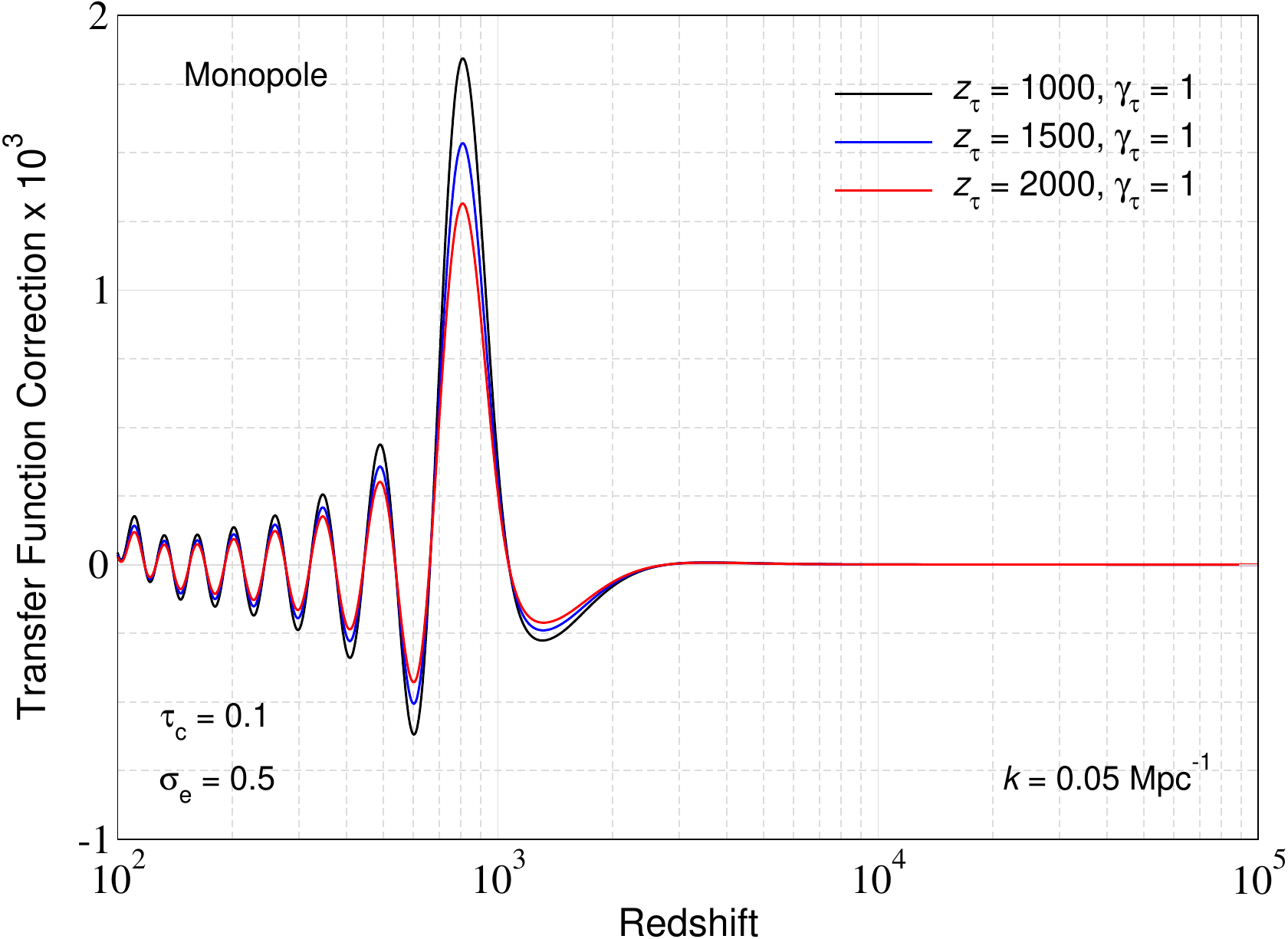}
\\[2mm]
\includegraphics[width=0.99\columnwidth]{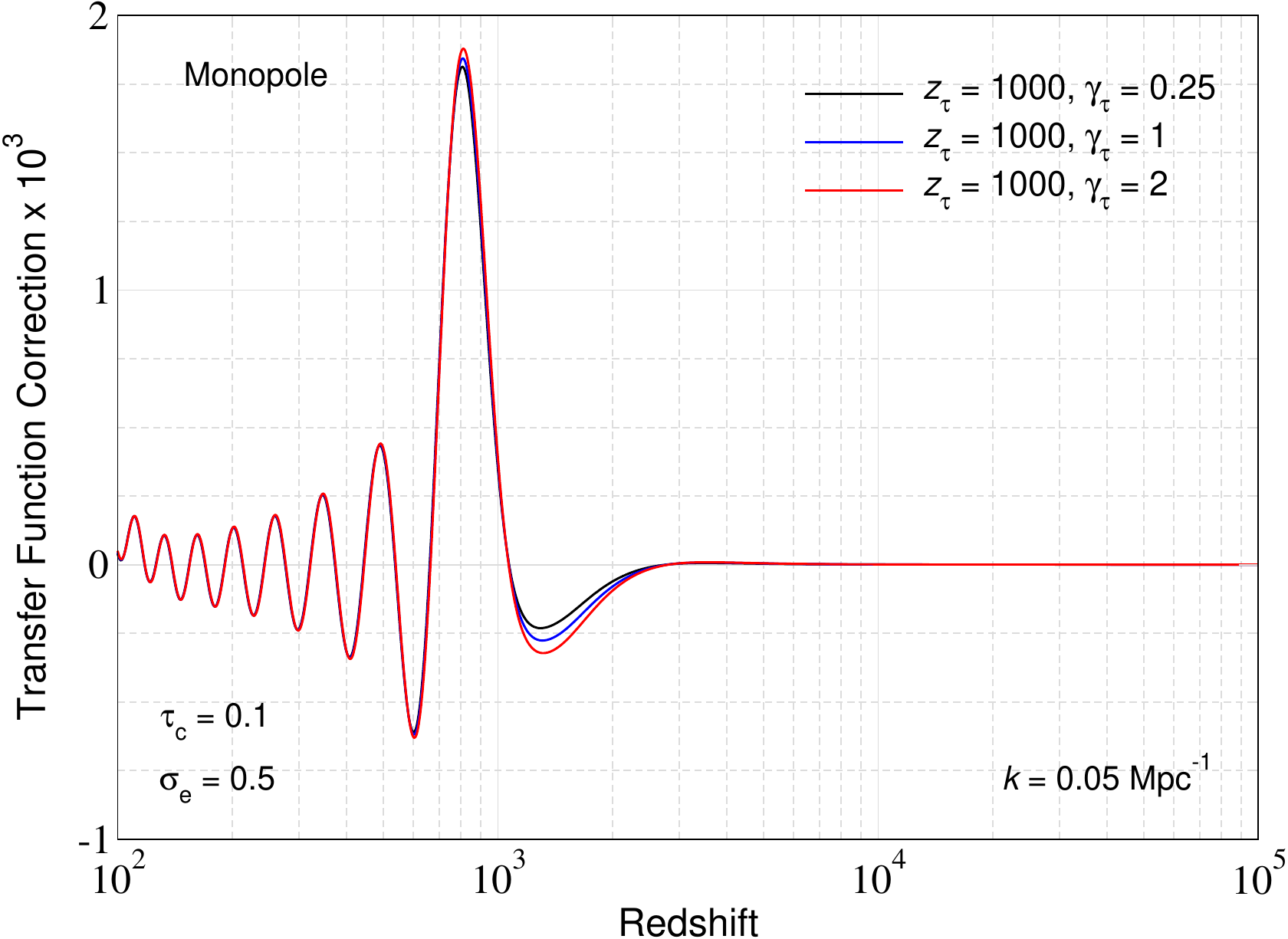}
\\
\caption{Photon transfer function corrections for the temperature monopole at $k=0.05\,\Mpc^{-1}$. We used $\sigmae(z)=0.5$ and $\tauc(z=10^4)=0.1$ with a scaling according to Eq.~\eqref{eq:tauc_sigmae_scaling_late_M1} for values of $\gamma_\tau$ and $z_\tau$ as labeled. The results were obtained with a 4$^{\rm th}$ order Gaussian moment setup.}
\label{fig:Transfer_scaling_M1-lowz}
\end{figure}
For illustration, one can assume that $\tauc$ and $\sigmae$ both scale as
\bsub
\begin{align}
\label{eq:tauc_sigmae_scaling_late_M1}
\tauc(z)=\frac{\tauc(z=10^4)}{1+\left(\frac{1+z}{1+z_{\tau}}\right)^{-\gamma_{\tau}}},
\\
\sigmae(z)=\frac{\sigmae(z=10^4)}{1+\left(\frac{1+z}{1+z_{\sigma}}\right)^{-\gamma_{\sigma}}},
\end{align}
\esub
meaning $\tauc(z) \approx {\rm const}$ and $\sigmae(z) \approx {\rm const}$ early on while they may decay as power-laws at late times. 

For simplicity, we can assume $\sigmae={\rm const}$ (i.e., $z_{\sigma}\rightarrow 0$). Cases with varying $\sigmae$ can be mapped back to modifications of $\tauc(z)$ at least in the regime for which $\tauc^2 \sigmae^2\ll 1$.
For fixed values of $\gamma_{\tau}$ we expect that increasing $z_{\tau}$ will lead to a reduction of the amplitude of the modifications while leaving the redshift dependence of the correction rather unchanged. In the upper panel of Fig.~\ref{fig:Transfer_scaling_M1-lowz} we illustrate this aspect for the monopole. 
If in contrast we vary $\gamma_{\tau}$ we see a redshift-dependent correction (lower panel of Fig.~\ref{fig:Transfer_scaling_M1-lowz}).
If in addition we varied $\sigmae$ round recombination we find a stronger response than for varying $\tauc$ since the leading order term scales quadratically in $\sigmae$.
Overall, these illustrations highlight a complicated interplay between various redshift-dependent effects that one will have to explore more carefully in the future.

\subsection{Effects on the CMB power spectra}
\label{sec:effects_PS}
We now briefly illustrate the possible changes to the CMB power spectrum for a few models. This is mainly as a demonstration of the effects and we therefore use the modified BH treatment to run the computations in all cases. A more detailed parameter study and model comparison is left for the future.

\begin{figure}
\centering
\includegraphics[width=\columnwidth]{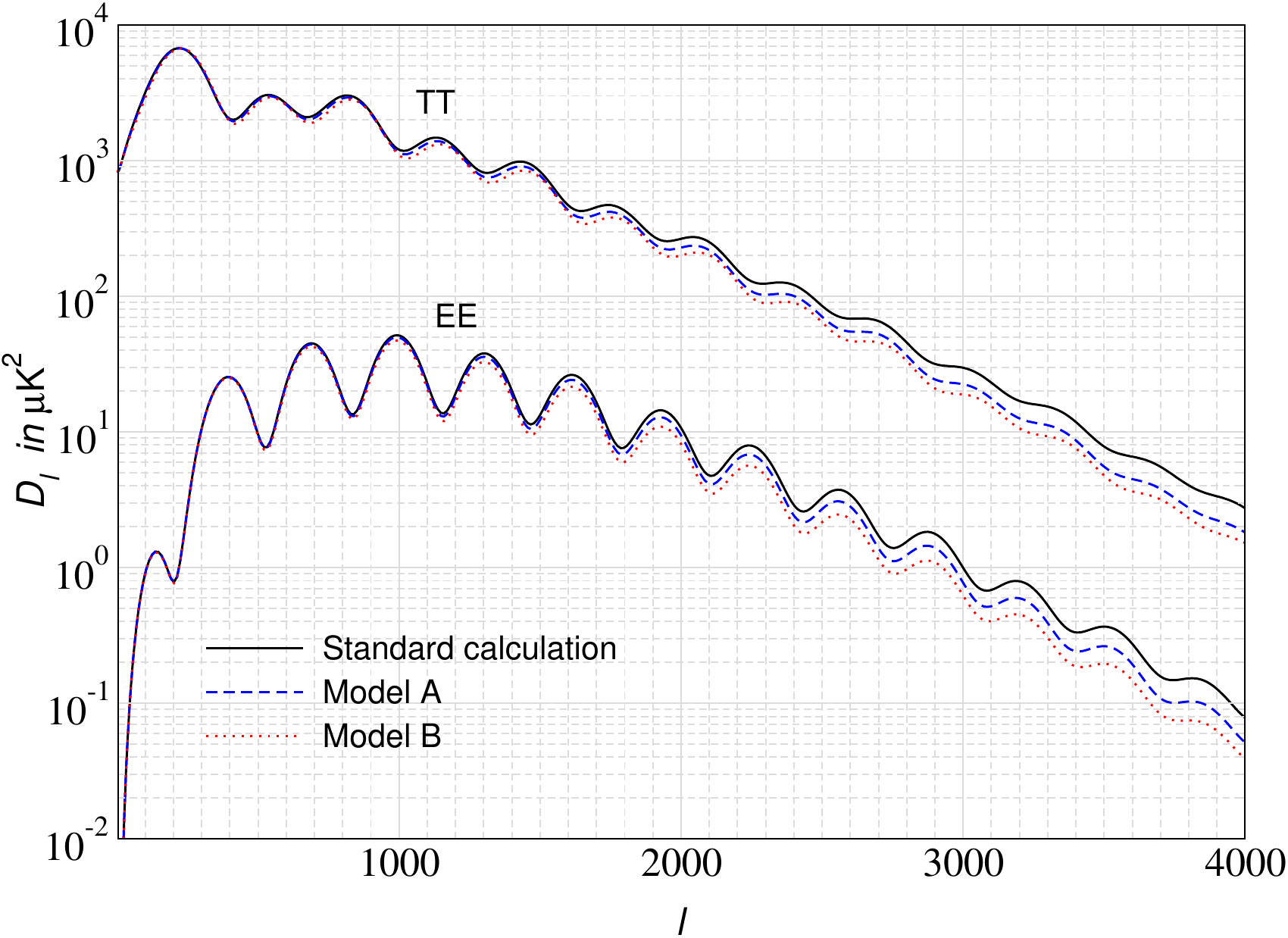}
\\
\caption{CMB $TT$ and $EE$ power spectra, $D_\ell=\ell(\ell+1) C_\ell/2\pi$. The solid black lines show the standard \LCDM computation. The other cases where computed using the log-normal modified BH setup. For Model A, we use a setup similar to Fig.~\ref{fig:Transfer_example_1} with $\sigmae(z)=1$ and $\tauc=0.01$ at $z=1100$ and scaling Eq.~\eqref{eq:tauc_scaling}. Model B is similar to the scenario shown in Fig.~\ref{fig:Transfer_scaling_M2} with $\sigmae(z)=1$ and $\tauc=0.1$ at $z=1100$ with scaling Eq.~\eqref{eq:tauc_scaling_M1} for $z_{\rm s}=1200$ and $\gamma_{\rm s}=2$.}
\label{fig:CMB_PS_example_1}
\end{figure}

In Fig.~\ref{fig:CMB_PS_example_1} we show two examples in comparison to the \LCDM cosmology. The non-standard scenarios are similar to the setups illustrated in Figs.~\ref{fig:Transfer_example_1} and \ref{fig:Transfer_scaling_M2}. We {\it did not} modify the average recombination history in the computations to focus on the new physical effect from modified scattering rates. In both models we see significant extra damping at small scales ($\ell \gtrsim 2000$), as anticipated from the previous discussion. For Model B we also see a clear phase shift especially in the $EE$ power spectrum. This already indicates that non-trivial effects can be expected once allowing for modified recombination histories in a clumpy Universe. The expected range of models is quite rich given that in general two new functions, $\tauc(z)$ and $\sigmae(z)$, have to be specified. However, a comprehensive exploration of this high-dimensional parameter space is beyond the scope fo this paper and will therefore be considered elsewhere.

\begin{figure}
\centering
\includegraphics[width=\columnwidth]{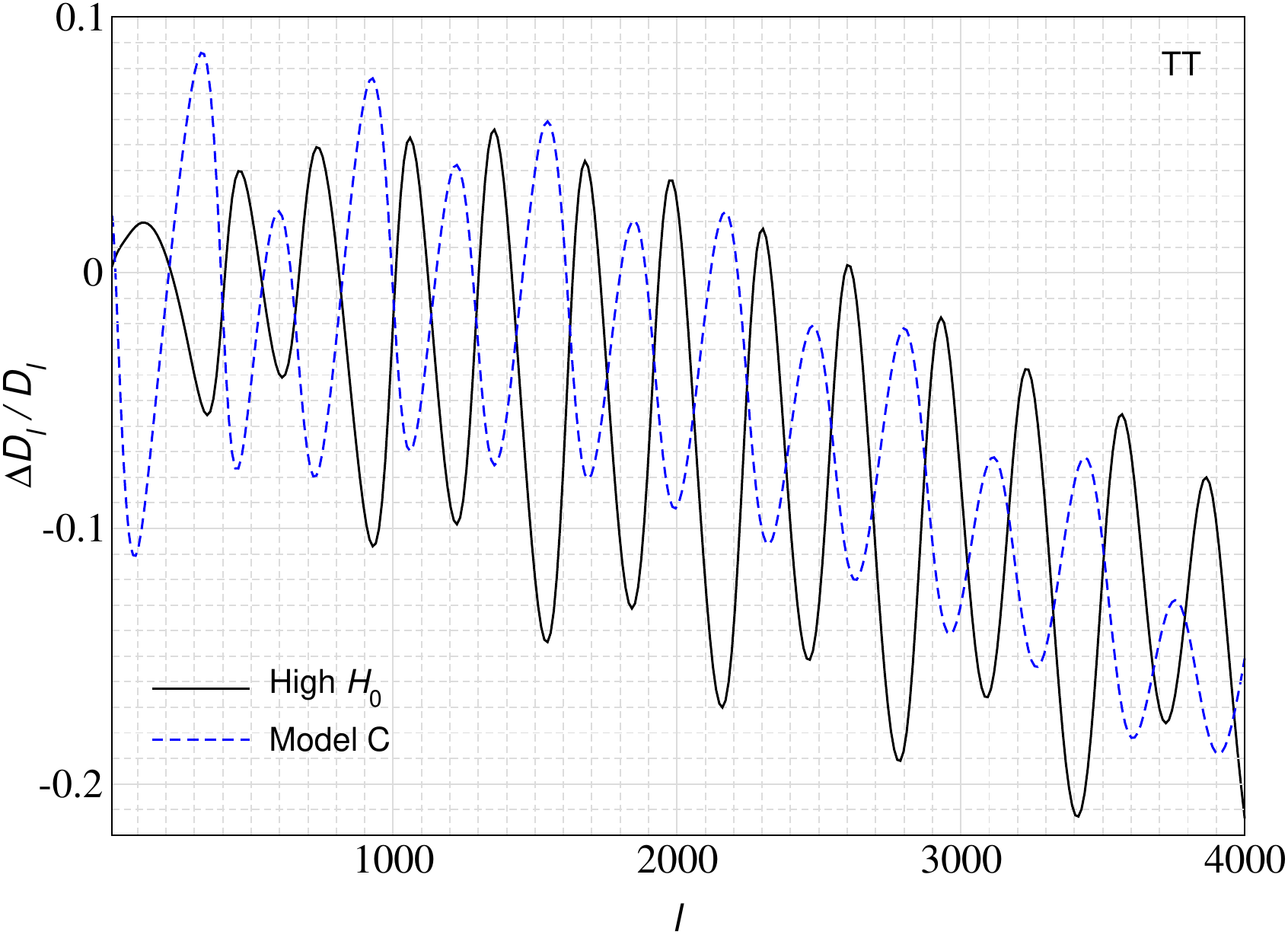}
\\
\includegraphics[width=\columnwidth]{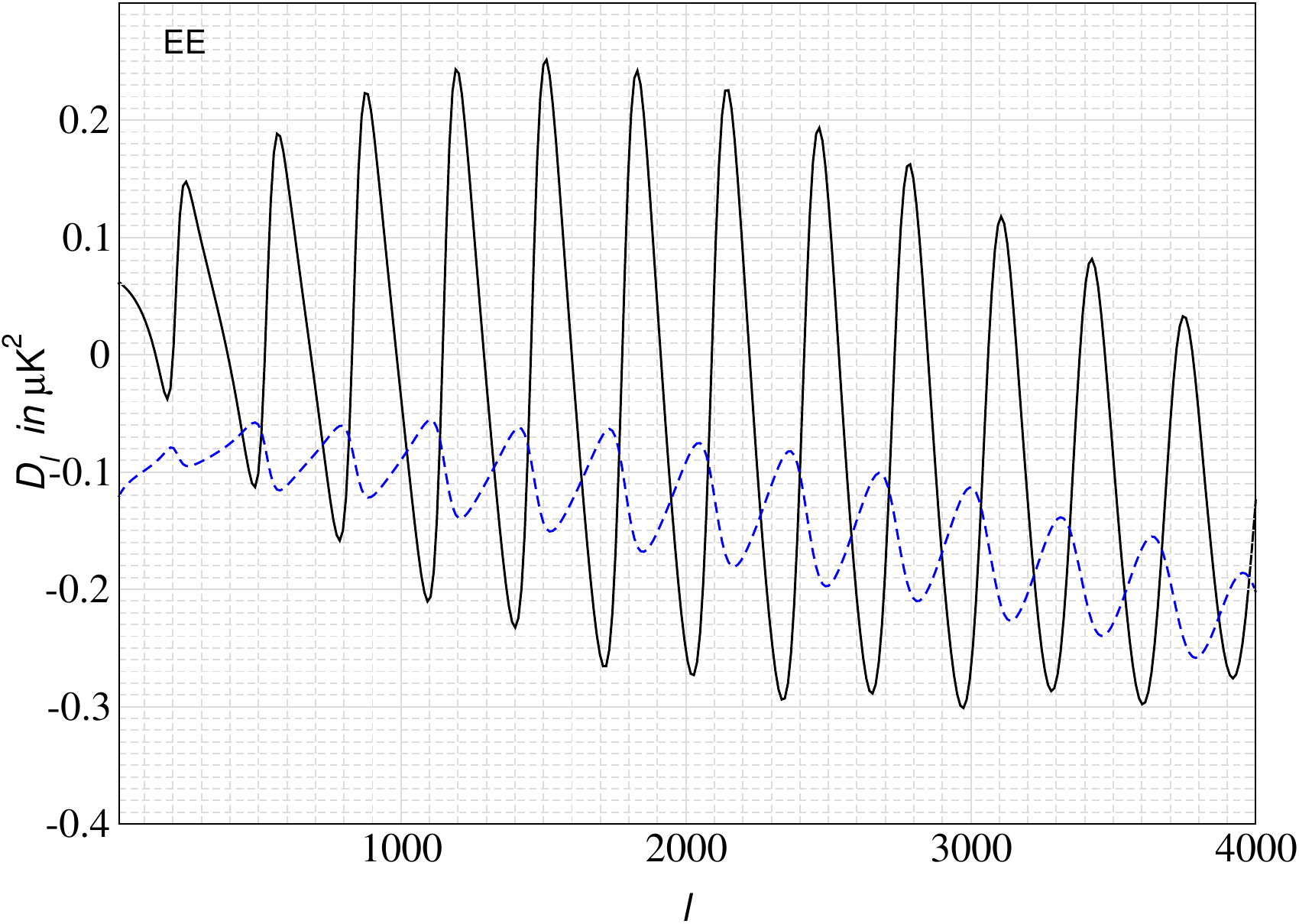}
\\
\caption{Relative change of CMB $TT$ and $EE$ power spectra with respect to the \LCDM model. The solid black lines show the difference cause by using a high value of $H_0$. Model C was computed using the modified BH with unchanged average recombination history for a fixed value of $\zetae(z)=0.1$ and $\sigmae(z)=0.75$, meaning that $\tauc=\zetae/\sigmae^2\approx 0.18$.}
\label{fig:CMB_PS_example_H0}
\end{figure}

The non-standard scenarios shown in Fig.~\ref{fig:CMB_PS_example_1} are mainly for illustration and unlikely to agree with existing measurements. However, we can attempt a search for models that come close to mimicking the changes that are required to solve the Hubble tension. For this, we show the change of the CMB $TT$ and $EE$ power spectrum when increasing the value of $H_0$ from $67.4\,{\rm km\,s^{-1}\,\Mpc^{-1}}$ to $73\,{\rm km\,s^{-1}\,\Mpc^{-1}}$ in Fig.~\ref{fig:CMB_PS_example_H0}. The changes are similar to shift of the peak positions towards lower values of $\ell$ (larger scales), which in spirit is akin to a decrease of the distance to the last scattering surface. 

We found that with modified scattering rates due to electron clumping for $D_\ell^{TT}$ one can indeed mimic a change to the effective redshifting for models with fixed $\zetae=\tauc \sigmae^2$. This is also illustrated in the upper panel of  Fig.~\ref{fig:CMB_PS_example_H0} for a case that closely mirrors the change induced by increased $H_0$ in the CMB $TT$ power spectrum. Although it is unclear if models with $\zetae\approx {\rm const}$ can be realized by physical scenarios (especially in the post-recombination era), this avenue looks very promising. However, for the model under consideration, the changes to $D_\ell^{EE}$ do not mimic those of changes to $H_0$, suggesting that a one-parameter modification is unlikely to work. 

Our simple examples highlight the importance of performing a detailed exploration of the new parameter space in light of existing cosmological datasets. A combination of effects when also changing the average recombination history will be crucial in this. Differences between the effects on CMB temperature and polarization terms will also play an important role in identifying possible solutions, however, this is beyond the scope of this work.

\vspace{-4mm}
\section{Discussion and conclusions}
\label{sec:conclusion}
In this paper, we developed a new framework to incorporate the effects of small-scale electron density fluctuations on the CMB power spectra. Using It\^o calculus (see Sect.~\ref{sec:ito_formalism}), we obtained ensemble-averaged Boltzmann hierarchies (BHs) that depend on the variance, $\sigmae^2(\eta)$, of the electron density fluctuations and the corresponding Thomson optical depth across the coherence length, $\tauc(\eta)$. The leading order correction terms were confirmed using a Langevin-type approach to the problem (Sect.~\ref{sec:pert_sol}). We furthermore derived an approximate photon BH (Sect.~\ref{sec:Boltz_simp}) for efficient exploration of the new parameter space in future analyses.

In addition to modifying the average recombination history, we have shown that the small-scale electron density fluctuations reduce the effective scattering rate of the medium if time-correlations are taken into account. The related corrections should be negligible in \LCDM (even if they are certainly present) but could become important in cosmologies with an onset of structure formation during or before the recombination era. \changeJ{In our model, the} changes to the average recombination history are mainly controlled by the electron density variance, $\sigmae^2(\eta)$, as we illustrated in Sect.~\ref{sec:LN_delta_b_fluctuations} (see Fig.~\ref{fig:Av_Xe}). In contrast, the main new scattering effects identified here depend on $\zetae(\eta)=\tauc \sigmae^2$ (see Sect.~\ref{sec:estimate_Gamma} for more detailed discussion). 

The functional forms of $\sigmae^2(\eta)$ and $\zetae(\eta)$ are currently unknown, opening up a large parameter space for exploration, as we illustrate in Sect.~\ref{sec:effects_T_Cell}.
Importantly, our results suggest that solutions to the Hubble tension (HT) may be linked to phenomena arising in the presence of small-scale electron density fluctuations (see Fig.~\ref{fig:CMB_PS_example_H0}). However, here we have performed the most naive demonstration without fully accounting for all the effects relevant to the problem (e.g., the related change to the average recombination history). With a data-driven reconstruction of the recombination history, it has already been shown that an early recombination is indeed preferred and alleviates the HT \citep{Lynch2024I, Lynch2024II}. Importantly, a modified recombination history alone cannot capture the new effects found here. The new terms in the BH therefore open new degrees of freedom that have to be carefully studied. We anticipate that a combination of temperature and polarization data may become crucial in distinguishing models due to the differing effects the new scattering terms have.

In this work, we have only taken a first step towards a better understanding of the effects that a clumpy Universe could have on the CMB anisotropies. Our approach was purely statistical, without directly specifying a physical scenario. This is motivated by our lack of understanding of the small-scale Universe in early phases and also by the fact that non-linearities inevitably scramble the link to primordial physics. 
However, it will be extremely important to perform realistic simulations in the presence of non-standard initial conditions at small scales. For example, significantly enhanced small-scale power could lead to the formation of the seeds of early structures \citep[e.g.,][]{Dekker2025, Kruijf2025}. Small-scale isocurvature perturbations \citep[e.g.,][]{Han2025, Buckley2025} could furthermore play an important role in this context, as could causally-produced perturbation from cosmic defects \citep[e.g.,][]{Vilenkin2000,Battye2020, Cyr2025DomainWalls} and scale-dependent primordial non-Gaussianity. 
\changeJ{In this, it will be crucial to carefully include the full dynamics of the perturbations while accounting for the coupling to the photons that will tend to suppress density perturbations. However, since the ionization fraction is determined by a non-linear coupling to the local properties of the medium, this will be a challenging computation in detail.}

No matter which physical process may be causing significant early electron density fluctuations, it will be extremely important to understand if these effects currently bias our cosmological inference. For instance, discussions about evolving dark energy in light of DESI data \citep{DESI-DR1-2024, DESI-DR2-2025} could be significantly hampered without confirming that the standard recombination history is valid \citep{Lynch2024II, Mirpoorian2025HTModRec}. Similarly, questions about neutrino physics and other extensions to the standard cosmological model may be biased without marginalizing over uncertainties in the recombination scenario. A data-driven assessment of this question is therefore a crucial next step.
A combination with future data from Pulsar Timing \citep[e.g.,][]{NANOGrav2023}, Gravitational Waves \citep{Bailes2021GWs} and CMB spectral distortions \citep{Chluba2019BAAS, Chluba2021Voyage}, especially using the cosmological recombination radiation \citep{Hart2020CRR, Lucca2023} could allow us to further constrain various scenarios.

With a clearer understanding of the viable parameter space, it will also be important to refine the framework developed here. For instance, we limited ourselves to two-point time correlations in the stochastic driver field, when one could expect higher order correlators to become important. The log-normal setup includes some higher order contributions, however, extending the framework (possibly guided by dedicated numerical simulations) to explicitly model these would be helpful. In addition, one should investigate the effects of mode-coupling in regimes where a scale-separation is not warranted. This may even open up new ways to directly study the properties of the small-scale electron density fluctuations and their imprints with future experiments like CMB-HD \citep{Sehgal2019CMBHD}.
%

The formalism developed in this work could have further applications. For instance, a refined treatment of the recombination history computation with It\^o calculus could identify additional corrections to the average recombination history that our simple separate Universe approach did not capture. 
Secondly, in the post-recombination era one expects a highly variable Rayleigh scattering rate if the small-scale Universe is indeed very clumpy. This will imprint new frequency-dependent effects on the CMB anisotropies that are not captured in existing treatments \citep{Yu2001, Lewis2013Rayleigh, Coulton2021Rayleigh}, potentially opening another way to constrain the presence of non-standard small-scale density fluctuations with upcoming experiments such as CCAT \citep{CCATp2018, CCAT2023}.
And finally, the late reionization process is highly inhomogeneous, possibly further affecting large-scale CMB and 21 cm signals in ways that could be treated using It\^o calculus. 
We leave an exploration of these ideas to future work.

%

{
\vspace{-3mm}
\section*{Acknowledgments}
The authors thank Gabriel Lynch, Lloyd Knox and Subodh Patil for valuable discussion of the problem and comments on the manuscript.

\vspace{-3mm}
\section*{Data availability}
The data underlying this article are available in this article and can further be made available on request.
}

{\small
\vspace{-3mm}
\bibliographystyle{mn2e}
\bibliography{Lit-2025}

\appendix

\vspace{-3mm}
\section{Constant density spheres}
\label{sec:example_spheres}
We illustrate the fact that $\zetae$ can in principle exceed unity for a simple two-zone model. We assume that overdensities are in spheres of a radius $R$. We shall set the parameter $F>1$ to define the electron overdensity, such that $\rho=F \,\bar{\rho}$, where $\bar{\rho}$ is the mean density. The volume filling factor is $f_{\rm V} \propto R^3$. This means that we have the volume factor of $1-f_{\rm V}$ for the (amorphous) underdense regions. We demand that the average density is $\bar{\rho}\equiv f_{\rm V} F \bar{\rho} + (1-f_{\rm V}) F_{\rm u}\,\bar{\rho}$, which implies 
\begin{align}
F_{\rm u}=\frac{1-f_{\rm V} F}{(1-f_{\rm V})}.
\end{align}
We require $0<F_{\rm u}\leq 1$ and $0<f_{\rm V}<1$ which means $1<F<1/f_{\rm V}$. For large overdensity, this means that the radius of the region, $R$, has to be small.
We then also have the density variance of the field as 
\begin{align}
\sigma^2=\left<\left(\frac{\rho}{\bar{\rho}}-1\right)^2\right>&=f_{\rm V} F^2+(1-f_{\rm V}) F_{\rm u}^2-1
=\frac{f_{\rm V}(F-1)^2}{1-f_{\rm V}},
\end{align}
meaning that $0<\sigma^2<(1-f_{\rm V})/f_{\rm V}$. For small $f_{\rm V}$ we can have large variance and also allow large density enhancements without changing the average properties of the medium.

We can use the exact results for the two-point correlation function of overlapping spheres \citep[Chapter 5.1.1 in][]{TorquatoBook}, which is given by 
\bsub
\begin{align}
\xi(r)&=\left[f_{\rm V}\right]^{\kappa(r)}-f_{\rm V}^2,
\\
\kappa(r)&=
\begin{cases}
1+\frac{3}{4}\,\frac{r}{R}-\frac{1}{16}\,\frac{r^3}{R^3}&\text{for}\qquad r<2R
\\
2 &\text{for}\qquad r\geq 2R.
\end{cases}
\end{align}
\esub
Here, $\sigmae^2=\xi(0)=f_{\rm V}-f_{\rm V}^2=f_{\rm V}(1-f_{\rm V})$. We can evaluate the integral $C=R \int \xi(r) \id r$ numerically as a function of $f_{\rm V}$. At $f_{\rm V}\lesssim 0.2$ we find $C\approx 0.6 \,R f_{\rm V}^{1.15}$. It has a maximum of $C\simeq 0.17 \,R$ at $f_{\rm V}\simeq 0.55$. No matter what $f_{\rm V}$ and $R$ is selected, it will always be possible to make $\zetae = \Gammab C(R)>1$.

\vspace{-3mm}
\section{Approximate problem in tight-coupling regime}
\label{app:TC_derivation}
In the tight coupling regime we expect all photon transfer functions for $\ell>2$ to vanish. The
simplified photon hierarchy the reads
\bsub
\begin{align}
\label{eq:Theta_equations_simpl}
\Theta_0'&=-k \Theta_1-\Phi',
\\
\Theta_1'&=\frac{k}{3}\left[\Theta_0+\Psi\right]
-\frac{2 k}{3}\Theta_2
-\Gamma\left[\Theta_1-\frac{\varv_{\rm b}}{3}\right],
\\
\Theta'_2&=\frac{2k}{5}\Theta_1-\frac{9}{10}\Gamma\,\Theta_2,
\\
\varv_{\rm b}'&= k\Psi-\mathcal{H} \varv_{\rm b}
+\frac{3\Gamma}{R}\left[\Theta_1-\frac{\varv_{\rm b}}{3}\right],
\end{align}
\esub
where $\Gamma=\Gammab(1+\deltae)$. As the first step we use 
\begin{align}
\label{eq:Theta2_sol}
\Theta_2&\approx \frac{4}{9\Gamma} k \Theta_1 = -\frac{4}{9 \Gamma }[\Theta_0+\Phi]'.
\end{align}
The coefficient $\alpha_2=4/9 \rightarrow 8/15$ when polarization terms are included. This yields the modified dipole equation
\begin{align}
\label{eq:Theta1_mod}
\Theta_1'&=\frac{k}{3}\left[\Theta_0+\Psi\right]
-\frac{k^2\alpha_2}{3\Gamma}\Theta_1
-\Gamma\left[\Theta_1-\frac{\varv_{\rm b}}{3}\right]
\end{align}
for later use. Next we eliminate the dependence on $\Delta_1=\Theta_1-\frac{\varv_{\rm b}}{3}$. By combining with the dipole equation, we rewrite the baryon velocity equation as 
\begin{align}
\label{eq:Delta1_equation}
\Delta_1'&= \frac{k}{3}\Theta_0+\left[\mathcal{H}-\frac{k^2\alpha_2}{3\Gamma}\right] \Theta_1 - \mathcal{H}\Delta_1
-\frac{\Gamma(1+R)}{R}\Delta_1.\end{align}
We now define $\epsilon = R/[\Gamma (1+R)]\ll 1$ to write 
\bsub
\begin{align}
\Theta_1'&=\frac{k}{3}\left[\Theta_0+\Psi\right]
-\frac{k^2\alpha_2}{3 f_R}\,\epsilon\Theta_1
-\frac{f_R}{\epsilon}\Delta_1,
\\
\Delta_1&\approx
\epsilon\left\{\frac{k}{3}\Theta_0+\mathcal{H} \Theta_1- \mathcal{H}\Delta_1-\partial_\eta\Delta_1\right\}
-\epsilon^2\,\frac{k^2\alpha_2}{3 f_R}\Theta_1.
\end{align}
\esub
with $f_R=R/(1+R)$. To have the result for $\Theta_1$ to order $\epsilon$, we need $\Delta_1$ to second order in $\epsilon$. Since $\epsilon$ depends on time, we also have to include corrections $\epsilon'$. Inserting the definition, we find
$$\frac{\epsilon'}{\epsilon}=\frac{\mathcal{H}}{1+R}-\frac{\Gamma'}{\Gamma}.$$
Inserting the series Ansatz $\Delta_1=\Delta_1^{(0)}+\epsilon \Delta_1^{(1)}+\epsilon^2 \Delta_1^{(2)}$ we then obtain 
\begin{align}
\nonumber
\Delta_1^{(0)}&=0
\\
\Delta_1^{(1)}&= \frac{k}{3}\Theta^{(0)}_0+\mathcal{H} \Theta^{(0)}_1
\approx \frac{k}{3}\Theta_0+\mathcal{H} \Theta_1-
\epsilon \left(\frac{k}{3}\Theta^{(1)}_0+\mathcal{H} \Theta^{(1)}_1\right)
\\ 
\nonumber
\Delta_1^{(2)}&=\frac{k}{3}\Theta^{(1)}_0+\mathcal{H} \Theta^{(1)}_1-\frac{k^2\alpha_2}{3 f_R}\Theta^{(0)}_1
-\left[\mathcal{H}+\frac{\epsilon'}{\epsilon}\right]\Delta_1^{(1)}-\partial_\eta\Delta_1^{(1)}
\\ 
\nonumber
&\approx \frac{k}{3}\Theta^{(1)}_0+\mathcal{H} \Theta^{(1)}_1-\frac{k^2\alpha_2}{3 f_R}\Theta_1
-\left[\mathcal{H}\,\frac{2+R}{1+R}-\frac{\Gamma'}{\Gamma}\right]\left(\frac{k}{3}\Theta_0+\mathcal{H} \Theta_1\right)
\\ 
\nonumber
&\qquad-\partial_\eta\left(\frac{k}{3}\Theta_0+\mathcal{H} \Theta_1\right),
\end{align}
where we added higher order terms to rewrite $\Theta^{(0)}_0$ in terms of $\Theta_0$ and so forth. Using
\begin{align}
\partial_\eta\left(\frac{k}{3}\Theta_0+\mathcal{H} \Theta_1\right)
&=-\frac{k}{3}\left[k\Theta_1+\Phi'\right]+\mathcal{H}'\Theta_1
+\mathcal{H}\Theta_1'
\\ \nonumber 
&\approx -\frac{k}{3}\left[k\Theta_1+\Phi'\right]+\mathcal{H}'\Theta_1
+\frac{k}{3}\,\mathcal{H}\left[\Theta_0+\Psi\right]
\end{align}
we can then find
\begin{align}
\label{eq:Delta1_equation_sol}
\Gamma \Delta_1&\approx f_R \left(\frac{k}{3}\Theta_0+\mathcal{H} \Theta_1\right)
-\frac{k^2\alpha_2 f_R}{3 \Gamma}\Theta_1
+\frac{f^2_R}{\Gamma}\frac{k}{3}\left[k\Theta_1+\Phi'\right]
\nonumber \\
&\qquad
-\frac{f^2_R}{\Gamma}\left[\mathcal{H}\,\frac{2+R}{1+R}-\frac{\Gamma'}{\Gamma}\right]\left(\frac{k}{3}\Theta_0+\mathcal{H} \Theta_1\right)
\nonumber \\
&\qquad\qquad
-\frac{f^2_R}{\Gamma}\left[
\frac{k}{3}\,\mathcal{H}\left(\Theta_0+\Psi\right)+\mathcal{H}'\Theta_1\right].
\end{align}
Inserting back into the dipole equation and using the sound speed square as $c_{\rm s}^2=1/[3(1+R)]$, then yields
\begin{align}
\label{eq:Theta1_mod_B}
\Theta_1'+&\left\{\frac{R'}{1+R}+\frac{k^2c_{\rm s}^2}{\Gamma}\left[\alpha_2+\frac{R^2}{(1+R)}\right]\right\}\Theta_1- k c_{\rm s}^2 \kappa \,\Theta_0
\nonumber\\
&\;
-\frac{f_R^2\mathcal{H}}{\Gamma}\left\{\frac{\mathcal{H}'}{\mathcal{H}}+
\,\frac{2+R}{1+R}-\frac{\Gamma'}{\Gamma}\right\} \Theta_1
-\frac{k f_R^2}{3 \Gamma}\left[2\mathcal{H}-\frac{\Gamma'}{\Gamma}\right] \Theta_0
\nonumber \\
&\qquad\qquad=
\frac{k}{3}\left[\kappa \Psi-\frac{R^2}{\Gamma (1+R)^2}\Phi'
\right],
\end{align}
where $\kappa = 1 + \mathcal{H} f_R^2 /\Gamma$.
Dropping all terms $\propto 1/\Gamma$ one recovers the common textbook solution without scattering effects \citep{DodelsonBook}.
The terms $\mathcal{O}(f_R^2 \mathcal{H}/\Gamma)$ are furthermore all minor corrections. 
Dropping these terms, we obtain
\begin{align}
\label{eq:Theta1_mod_C}
\Theta_1'+&\left\{\frac{R'}{1+R}+\frac{k^2c_{\rm s}^2}{\Gamma}\left[\alpha_2+\frac{R^2}{(1+R)}\right]\right\}\Theta_1- k c_{\rm s}^2 \Theta_0
\nonumber \\
&\qquad\qquad=
\frac{k}{3}\left[\Psi-\frac{R^2}{\Gamma (1+R)^2}\Phi'
\right].
\end{align}
We can identify the time-derivative of the diffusion damping scale, 
\begin{align}
\label{eq:kD_derivative}
\frac{\partial}{\partial_\eta} \frac{k^2}{k^2_{\rm D}}&=\frac{k^2 c_{\rm s}^2}{2\Gamma}\left[\alpha_2+\frac{R^2}{(1+R)}\right].
\end{align}
Taking the time-derivative of the monopole equation and combining all expressions we find
\begin{align}
\label{eq:Theta0_mod_B}
&\Theta_0''+\Phi''
+\left\{\frac{R'}{1+R}+\frac{k^2 c_{\rm s}^2}{\Gamma}\left[\alpha_2+\frac{R^2}{(1+R)}\right]\right\}\left[\Theta'_0+\Phi'\right]
\nonumber\\
&\qquad +k^2 c_{\rm s}^2[\Theta_0+\Phi]
=k^2 c_{\rm s}^2 \Phi+\frac{k^2}{3}\left[\frac{R^2}{\Gamma(1+R)^2}\Phi'-\Psi\right],
\end{align}
which includes all scattering related damping effects. 

}

\end{document}